\documentclass[twocolumn]{aastex631}

\usepackage{newtxtext,newtxmath}
\usepackage{ae,aecompl}

\usepackage{graphicx}	
\usepackage{amsmath}	
\newcommand\msun{{\,M_\odot}}

\newcommand\zsun{{\rm \,Z_\odot}}

\newcommand{\Ang}{~\mbox{\AA}}

\newcommand{\cMpc}{~\mbox{comoving}~\mbox{Mpc}}

\newcommand{\cmci}{~\mbox{cm}^{-3}}

\newcommand{\K}{~\mbox{K}}

\newcommand{\traphic}{{\sc traphic}}

\newcommand{\gadgetthree}{{\sc p-gadget3}}

\newcommand{\astraeus}{{\sc astraeus}}

\newcommand{\invs}{~\mbox{s}^{-1}}

\newcommand{\Msun}{~\mbox{M}_{\odot}}
\newcommand{\Msunyri}{~\mbox{M}_{\odot}~\mbox{yr}^{-1}}
\newcommand{\Msuni}{~\mbox{M}_{\odot}^{-1}}
\newcommand{\Zsun}{~Z_{\odot}}

\newcommand{\kmsMpc}{~\mbox{km~s}^{-1}~\mbox{Mpc}^{-1}}

\newcommand{\music}{{\sc music}}
\newcommand{\cloudy}{{\sc cloudy}}
\newcommand{\rockstar}{{\sc rockstar}}
\newcommand{\fsps}{{\sc fsps}}
\newcommand{\fspspython}{{\sc fsps-python}}
\newcommand{\hyperion}{{\sc hyperion}}
\newcommand{\yggdrasil}{{\sc yggdrasil}}
\newcommand{\st}{{\sc starburst99}}
\newcommand{\bpass}{{\sc bpass}}

\newcommand{\nspace}{{ }}

\begin{document}

\title{Simulating high-redshift galaxies: Enhancing UV luminosity with star formation efficiency and a top-heavy IMF}

\author[0009-0000-8108-6456]{Tae Bong Jeong}
\affiliation{School of Space Research, Kyung Hee University, 1732 Deogyeong-daero, Giheung-gu, Yongin-si, Gyeonggi-do 17104, Republic of Korea}

\author[0000-0001-6529-9777]{Myoungwon Jeon}
\affiliation{School of Space Research, Kyung Hee University, 1732 Deogyeong-daero, Giheung-gu, Yongin-si, Gyeonggi-do 17104, Republic of Korea}
\affiliation{Department of Astronomy \& Space Science, Kyung Hee University, 1732 Deogyeong-daero, Yongin-si, Gyeonggi-do 17104, Republic of Korea}
\correspondingauthor{Myoungwon Jeon}
\email{myjeon@khu.ac.kr}

\author[0000-0002-4362-4070]{Hyunmi Song}
\affiliation{Department of Astronomy \& Space Science, Chungnam National University, 99 Daehak-ro, Yoosung-gu, Daejeon 34134, Republic of Korea}

\author[0000-0003-0212-2979]{Volker Bromm}
\affiliation{Department of Astronomy, The University of Texas at Austin, Austin, TX, 78712, USA}
\affiliation{Weinberg Institute for Theoretical Physics, University of Texas at Austin, Austin, TX 78712, USA}

\received{18 Sep 2024}
\revised{4 Nov 2024}
\accepted{20 Dec 2024}
\submitjournal{ApJ}

\begin{abstract}
Recent findings from photometric and spectroscopic JWST surveys have identified examples of high-redshift galaxies at $z \gtrsim 10$. These high-$z$ galaxies appear to form much earlier and exhibit greater UV luminosity than predicted by theoretical work. In this study, our goal is to reproduce the brightness of these sources by simulating high-redshift galaxies with virial masses $M_{\rm vir} = 10^{9} - 10^{10} \msun$ at $z > 10$. To achieve this, we conduct cosmological hydrodynamic zoom-in simulations, modifying baryonic sub-grid physics, and post-process our simulation results to confirm the observability of our simulated galaxies. Specifically, we enhanced star formation activity in high-redshift galaxies by either increasing the star formation efficiency up to 100\% or adopting a top-heavy initial mass function (IMF). Our simulation results indicate that both increasing star formation efficiency and adopting a top-heavy IMF play crucial roles in boosting the UV luminosity of high-redshift galaxies, potentially exceeding the limiting magnitude of JWST surveys in earlier epochs. Especially, the episodic starburst resulting from enhanced star formation efficiency may explain the high-redshift galaxies observed by JWST, as it evacuates dust from star-forming regions, making the galaxies more observable. We demonstrate this correlation between star formation activity and dust mass evolution within the simulated galaxies. Also, adopting a top-heavy IMF could enhance observability due to an overabundance of massive stars, although it may also facilitate rapid metal enrichment. Using our simulation results, we derive several observables such as effective radius, UV slope, and emission line rates, which could serve as valuable theoretical estimates for comparison with existing spectroscopic results and forthcoming data from the JWST NIRSpec and MIRI instruments.
\end{abstract}


\section{Introduction}

Before the launch of the James Webb Space Telescope (JWST), galaxies from the era of cosmic dawn---when the universe was first illuminated in starlight---were too distant and faint to be observed, and only small samples of candidates, such as GN-z11, had been detected by surveys using the Hubble Space Telescope (HST) during this era (e.g., \citealp{Ellis2013, Oesch2016}). However, since the initial data release in July 2022, studies utilizing early survey data from JWST NIRCam have unveiled a broad window into the cosmic dawn and pushed the redshift frontier up to $z \sim 16$ (e.g., \citealp{Naidu2022, Tacchella2022, Furtak2023, Harikane2023, Finkelstein2022, Finkelstein2023, Donnan2023b, Donnan2023a, Bouwens2023a, Bouwens2023b, Adams2023, Tacchella2023, Donnan2024, Finkelstein2024, Harikane2024}). For instance, using the early released observation results from the Cosmic Evolution Early Release Science Survey (CEERS) project, \citet{Finkelstein2024} identified 88 galaxy candidates at high-$z$, probing the evolution of the UV luminosity function (UVLF) across redshifts ranging from $z \sim 8.5$ to $z \sim 14.5$. Interestingly, unlike the theoretical predictions, the UVLF at high redshift appears to evolve less steeply, suggesting that the early Universe was over-abundant in UV-luminous galaxies.

\par
Additionally, combining data from the COSMOS/UltraVISTA survey,  \citet{Donnan2023b, Donnan2023a} suggested that the shape of the UVLF at high redshift is better described by a double power-law function, which features a flatter slope at the bright end, compared to the Schechter function. \citet{Donnan2024} reinforced the double power-law characterization of the UVLF at high redshift using a sample of 2,548 galaxies from several major JWST Cycle-1 photometric surveys, including PRIMER, JADES, and NGDEEP, confirming their UV-luminous nature. Follow-up studies using spectroscopic data from JWST have confirmed the redshifts of high-$z$ galaxies more accurately (e.g., \citealp{Curtis-Lake2023, Harikane2024, ArrabalHaro2023}). Although such spectroscopic confirmation efforts are still ongoing, \citet{Harikane2024} found that one of the extreme high-$z$ galaxy candidates (CR2-z16-1) turned out to be a lower-redshift interloper ($z_{\rm spec} = 4.912$). However, the remaining spectroscopically confirmed galaxies are indeed at high-$z$ and thus UV luminous. This indicates that the "Early Galaxy Problem," as described by \citet{Steinhardt2023} might persist, showing that the early Universe was more rapidly evolved and bluer (UV luminous) than theoretical expectations, suggesting a possible tension with the standard $\rm \Lambda CDM$ cosmological model (e.g., \citealp{Boylan-Kolchin2023}).

\par
\par
To address the early galaxy problem, various ideas have been proposed, ranging from adopting extreme assumptions to suggesting that there is no tension within the $\rm \Lambda CDM$ cosmology framework. For instance, some studies have considered somewhat exotic scenarios such as the acceleration of the formation of small-scale structures facilitated by primordial black holes formed immediately after inflation (e.g., \citealp{Liu2022, Liu2023, Zhang2024}), or modifications to the $\rm \Lambda CDM$ power spectrum on small scales (e.g., \citealp{Padmanabhan2023, Hirano2024}). More moderate suggestions within the $\rm \Lambda CDM$ cosmology framework include altering the dominant physical processes that regulate star formation. This could be achieved by adopting enhanced star formation efficiency (e.g., \citealp{Inayoshi2022, Mason2023}) or a top-heavy initial mass function (IMF) in high-$z$ galaxies (e.g., \citealp{Inayoshi2022, Cameron2024, Finkelstein2023, Yung_2023, Lu2024}), or a time-dependent IMF, progressing from top-heavy to normal (e.g., \citealp{Trinca2023}). Accounting for the contribution of Active Galactic Nuclei (AGN) in high-$z$ galaxies could also help to explain the UVLF (e.g., \citealp{Ono2023, Trinca2023, Hegde2024}).

\par
Another possibility is the diminished impact of dust attenuation on high-$z$ galaxies. Based on observational data from ALMA, \citet{Ferrara2023}, \citet{Ziparo2023} and \citet{Bakx2023} proposed that the over-abundance of the bright end of the UVLF at high redshift could be attributed to reduced dust attenuation. They suggested that dust ejection by radiative pressure, causing a segregation of the spatial distribution of dust from UV-emitting regions, likely played a role in diminishing the dust attenuation effect, resulting in brighter high-$z$ galaxies. Similarly, using a semi-analytic model (SAM) to explain high-$z$ galaxies observed by JWST and ALMA, \citet{Tsuna2023} emphasized the importance of dust opacity, indicating that dust ejection due to bursty star formation is required to account for the brightness of these galaxies. Alternatively, stellar feedback might not have operated effectively in high-$z$ galaxies, a scenario known as feedback-free starburst (\citealp{Dekel2023, Li2023}). Specifically, \citet{Dekel2023} argued that the free-fall time of gas clouds could be shorter than the time required for stellar feedback to take effect. This feedback-free scenario is applicable only to massive halos ($M_{\rm vir} \gtrsim 10^{10.8} \msun$). However, through this regulated feedback process, they concluded that massive galaxies could form in high-$z$ environments, coinciding with highly efficient star formation.

\par
Lastly, several studies employing numerical simulations suggest that there is "no tension" between observations and $\rm \Lambda CDM$ cosmology. Utilizing pre-existing data from various cosmological simulation projects (e.g., EAGLE, Illustris), \citet{Keller2023} proposed that previous cosmological simulations can reproduce the high-$z$ galaxies observed by JWST. By extrapolating results from the Renaissance simulation (\citealp{O'Shea2015}), for instance, \citet{McCaffrey2023} also demonstrated that high-redshift galaxies could have been situated in "rare-peak" regions of the early Universe, naturally explaining the observations. Recently, \citet{Sun2023a, Sun2023b} suggested that bursty star formation in high-$z$ galaxies might be the key to boosting the UV luminosity without altering sub-grid physics. Specifically, using results from the FIRE-2 simulation (\citealp{Ma2018a, Ma2018b}), they demonstrated that the luminosity of high-redshift galaxies might fluctuate in and out of the detection limit. This fluctuation is closely related to the star formation rate (SFR) and its evolution, giving rise to a selection effect in the observation of high-redshift galaxies.
\par

On the contrary, using simulated data from SERRA (\citealp{Pallottini2022}), \citet{Pallottini2023} concluded that stochastic variability in the star formation process might not be the key to resolving the discrepancy. They found that the predicted SFR could not account for the necessary UVLF boost in high-$z$ galaxies, suggesting that other physical mechanisms are needed to align with the observational results. Moreover, the currently suggested studies are based on cosmological simulations conducted over large volumes, which have relatively coarse resolutions (e.g., $m_{\rm DM} = 2.0 \times 10^4 \Msun$ for {\sc Renaissance} simulations, $m_{\rm DM} \geq 4.5 \times 10^5 \Msun$ for Illustris) due to the high computational cost. While these resolutions are sufficient for investigating galaxies of a similar size to the Milky Way (MW), they introduce significant uncertainty in the physical properties of high-$z$ galaxies, which are considerably smaller. This uncertainty particularly affects the SFR within the first galaxies, which has a substantial impact on their luminosity.

\par
\par

\par
As mentioned earlier, various sub-grid physics have been proposed to explain the discrepancy, but these scenarios require validation with high-resolution simulations, as modifying sub-grid physics can indeed impact other physical properties. For instance, adopting a top-heavy IMF could enhance the UV luminosity of each galaxy, but it would also increase the population of massive stars that end their lives as supernovae (SNe). As a result, this could significantly boost metal enrichment in the early Universe, potentially creating another inconsistency between theoretical predictions and observational data. In this work, we present cosmological radiation hydrodynamic zoom-in simulations with high resolution to form high-redshift galaxies comparable to those observed by JWST. Our goal is to test the various scenarios suggested by previous studies and validate their implications on the physical properties of high-redshift galaxies.

\par
Here, we are focusing on the moderate scenarios that do not violate the $\rm \Lambda CDM$ cosmology but introduce variations in the sub-grid physics and associated parameters. Specifically, among the possible scenarios for explaining the discrepancy, we are focusing on mechanisms that can alter the star formation process, such as enhancing star formation efficiency and adopting a top-heavy IMF not only for Population~III (Pop~III) stars but also for the second generation of Population~II (Pop~II) stars. One distinguishing aspect of this work is our adoption of a top-heavy IMF for Pop II stars, which appear to be the predominant population in high-redshift galaxies (e.g., \citealp{Jeon_2019, Jaacks2019}). Adopting a top-heavy IMF for Pop II stars offers the possibility of boosting UV fluxes due to their small mass-to-luminosity ratio for a substantial amount of time, considering the short duration of Pop III star formation.
\par

It has been suggested that by delaying the transition from Pop III to Pop II star formation, high-$z$ galaxies could be dominated by Pop III stars (e.g., \citealp{Liu2020MNRAS, Nakajima2022, Inayoshi2022, Trussler2023, Cameron2024, Ventura2024MNRAS}). However, metals ejected from a few Pop~III SNe can easily enrich nearby gas clouds above the critical metallicity threshold ($\rm Z_{thr} \lesssim 10^{-5.5} \Zsun$; e.g., \citealp{Safranek-Shrader2016}), leading to an immediate transition from Pop III to Pop II star formation (e.g., \citealp{Jeon_2014, Katz2023}). For instance, \citet{Yajima2023} confirmed that the mass fraction of Pop III stars decreases as stellar mass increases, representing less than 10\% for galaxies with $M_{\star} \lesssim 10^5 \msun$. This suggests that massive galaxies at $z > 10$ observed by JWST are likely dominated by Pop~II stars (but see \citealp{Venditti2024}). We will provide physical justifications for a possible top-heavy IMF for Pop II stars in Section \ref{subsubsec:p2} below.

\par
Based on our simulated results, we derive synthetic observational properties to be compared with JWST observations. These synthetic observational results are further post-processed using a dust radiative transfer code combined with a photoionization module for emission line properties. Several previous theoretical studies have focused solely on the continuum $L({1500\Ang})$, which can be calculated based on the relationship between SFR and UV luminosity applicable to the local Universe. Although this relationship is calibrated by observations, calculating the UV luminosity is highly model-dependent because various physical properties are involved. For example, the properties of gas and dust and their spatial distribution affect UV luminosity through extinction and emission processes.
Consequently, while using this simple relationship might work well, more elaborate processes are required to accurately explain galaxies in the early Universe. In this work, by comparing our synthetic predictions from the sophisticated post-processing pipelines with actual observations, we are exploring the possibilities of various sub-grid physics to narrow the discrepancy without violating the $\rm \Lambda CDM$ cosmology.

\par
This paper is organized as follows. In Section 2, we describe the numerical methodology, followed by a presentation of the detailed results from our simulation sets in Section 3. 
In Section 4, we discuss the observable properties derived from our simulations, along with the caveats and limitations of our work. We compare our findings with those of similar studies and observational data in Section 5. Finally, we provide a summary and conclusions in Section 6. For consistency, all distances are given in physical (proper) units, unless otherwise noted.

\section{Numerical methodology}
\label{Sec:Method}

\subsection{Simulation Setup}
We have performed radiation hydrodynamic zoom-in simulations using a modified version of the N-body/TreePM Smoothed Particle Hydrodynamics (SPH) code \gadgetthree \nspace (\citealp{2001NewA....6...79S}, \citealp{Springel_2005}). For the cosmological parameters, we adopt a matter density parameter $\Omega_{\rm m} = 1 - \Omega_{\rm \Lambda}= 0.265$, baryon density $\Omega_{\rm b}=0.0448$, Hubble constant $H_{\rm 0} = 71 \kmsMpc$, and normalization parameter $\sigma_{\rm 8} = 0.8$  (\citealp{Planck2016}). The multi-scale initial conditions are generated using the cosmological initial condition code \music \nspace (\citealp{Hahn_2011}). To identify the target halos at $z \approx 10$, we initially perform dark matter-only simulations at a lower resolution using $128^3$ particles within a box of $L_{\rm box}=6.25 h^{-1} \cMpc$. The target halos are then identified using the halo-finder code \rockstar \nspace (\citealp{Behroozi_2013}). After selecting the target halos, we conduct successive refinements on the particles within 3 $R_{\rm vir}$ at $z \approx 10$, where $R_{\rm vir}$ denotes the virial radius of the halos. This process results in the refined DM and SPH particle masses of $m_{\rm DM} \approx 2500\Msun$ and $m_{\rm SPH} \approx 500\Msun$, respectively, corresponding to an effective resolution of $2048^3$.

\par
Our simulation starts at $z = 125$ and terminates within the range $8 \lesssim z \lesssim 10$, to compare with the high-redshift galaxies observed by JWST. The adopted softening length in our simulations is $\epsilon_{\rm soft} \sim 30$ physical pc, which is kept constant across all redshifts for both DM and baryonic particles. Also, similar to \citet{Jeon_2017}, we use adaptive softening lengths for the gas particles, where the softening length is proportional to the SPH kernel length with a minimum value of $\epsilon_{\rm gas, min} = 2.8 \rm \, pc$. At each timestep, we solve the non-equilibrium rate equations for the primordial chemistry involving nine atomic and molecular species, H, H$^+$, H$^{-}$, H$_2$, H$^{+}_{2}$, He, He$^+$, He$^{++}$, e$^{-}$, D, and D$^+$. In addition to primordial cooling, our simulations incorporate metal cooling processes for elements such as carbon, oxygen, silicon, magnesium, neon, nitrogen, and iron. The cooling rates for these elements are determined using the photoionization package \cloudy \nspace (\citealp{1998PASP..110..761F}). 

\subsection{Star formation}
We consider the formation of Pop III stars, which are the metal-free first generation of stars, and the subsequent formation of Pop II stars that are formed from gas clouds enriched with metals from the SN explosions of Pop III stars. Stars are formed when the hydrogen number density of a gas particle exceeds the density threshold of $n_{\rm H, thr} = 100 \cmci$. We implement a stochastic conversion of gas particles into star particles, where the conversion rate follows the Schmidt law (\citealp{Schmidt1959}). Specifically, gas particles are converted into stars according to $\dot{\rho}^{*} = \rho / \tau_{\star}$, where $\rho$ represents the gas density, and $\tau_{\star}$ is the star formation timescale, which is defined as $\tau_{\star} = \tau_{\rm ff}/ \epsilon_{\rm ff}$. Here, $\epsilon_{\rm ff}$ denotes the star formation efficiency, and the free-fall time $\tau_{\rm ff}$ is given by $\tau_{\rm ff} = \left[3\pi/(32G\rho) \right]^{1/2}$. During each numerical time interval of the simulation, $\Delta t$, gas particles are converted into star particles only if a randomly generated number between 0 and 1 is less than $\rm min(\Delta t / \tau_{\star}, 1)$.

The star formation time scales for both Pop~III and Pop~II stars depend on the gas number density, $n$ for star formation, as follows:
\begin{equation}
    \tau_{\star} = \frac{\tau_{\rm ff}(n)}{\epsilon_{\rm ff}} = 201.2 \, \rm Myr \, \Big( \frac{\epsilon_{\rm ff}}{0.01} \Big)^{-1} \it \, \Big(\frac{n}{\rm 500  \cmci}\Big)^{\rm -1/2}.
\end{equation}
We use the same star formation efficiency for both Pop III and Pop II stars, $\epsilon_{\rm ff}=\epsilon_{\rm ff, Pop III}=\epsilon_{\rm ff, Pop II}=0.01$ in the default set of our simulations, which is a typical value observed in the local Universe (e.g., \citealp{Leroy2008}). When a gas particle fully satisfies these conditions, it is converted into a collisionless star particle with an initial mass of $m_{\star}=500\Msun$. Due to the limited numerical resolution, we treat our star particles as single stellar clusters (SSP), with masses extracted from the assumed IMFs, rather than as individual stars. For more detailed descriptions of the star formation recipes and associated stellar feedback, we refer readers to \citet{Jeon_2017}.

\subsubsection{Pop III stars}
\label{subsubsec:p3}
Pop III stars form in metal-free gas clouds, where the primary cooling mechanism is through molecular hydrogen ($\rm H_{2}$) cooling. This process is relatively inefficient, allowing temperatures to drop only to about $T \approx 200$K, increasing the probability of Pop III stars being more massive ($\gtrsim 100\msun$) (e.g., \citealp{Bromm2013, Klessen2023}). Some theoretical studies, which incorporate radiative feedback from protostars and disk fragmentation, proposed the formation of multiple stellar systems with less massive stars reaching up to several tens of solar masses (e.g., \citealp{Clark2011, Stacy2016, Sugimura2020, Latif2022}). Nonetheless, the mass spectrum of Pop III stars remains broad, ranging from $10\msun$ to over $1000\msun$ (e.g., \citealp{Susa2014, Hirano2015, Hosokawa2016}). Given this uncertainty, we assume a top-heavy initial mass function (IMF) for Pop III stars, described by the functional form $\phi_{\rm PopIII}(m) = dN/d\log m\approx m^{-\alpha}$, with a slope of $\alpha=1.0$ across the mass range $[m_{\rm min},m_{\rm max}]=[10\msun,150\msun]$.

Following the suggestions by several studies indicating that star formation in the early Universe might have been significantly more efficient than in the local Universe to explain the JWST high-$z$ galaxies (e.g., \citealp{Inayoshi2022, Furtak2023, Mason2023, Harikane2024}), one way to achieve this is by simply increasing $\epsilon_{\rm ff}$ in our simulations. Specifically, we increase $\epsilon_{\rm ff}$ to a range of 0.3 to 1.0, which is 30 to 100 times higher than the value used in our default settings. We expect that this increase in $\epsilon_{\rm ff}$ would result in more frequent star formation activities, thereby increasing the stellar mass of galaxies and proportionally boosting their UV luminosity.

\subsubsection{Pop II stars}
\label{subsubsec:p2}
When massive Pop III stars reach the end of their life cycles, those with masses in the range of $11\msun \lesssim m \lesssim 40\msun$ explode as conventional core-collapse SNe, while those with masses between $140\msun \lesssim m \lesssim 260\msun$ undergo pair-instability SNe (PISNe) (e.g., \citealp{Heger2002, Heger2010}). These explosive events eject metals into the surrounding pristine gas clouds and when the metallicity of this enriched gas exceeds a certain threshold, $\rm Z_{thr}=10^{-5.5}\zsun$, Pop II stars begin to form (e.g., \citealp{Omukai2000}; \citealp{Schneider2010}; \citealp{Safranek-Shrader2016}). Once a gas particle satisfies the above conditions, we form a star particle with an initial mass of $m_{\star}=500\msun$, adopting the Chabrier IMF (\citealp{Chabrier2003}) within the mass range of $[m_{\rm min}, m_{\rm max}] = [0.1\msun, 100\msun]$ in the default cases of our simulations. To generate highly UV-luminous galaxies in our simulations, we also increased $\epsilon_{\rm ff}$ to a range of 0.3 to 1.0 for Pop II stars, similar to the approach used for Pop III stars.
\par
\begin{figure}
    \centering
    \includegraphics[width = 85mm]{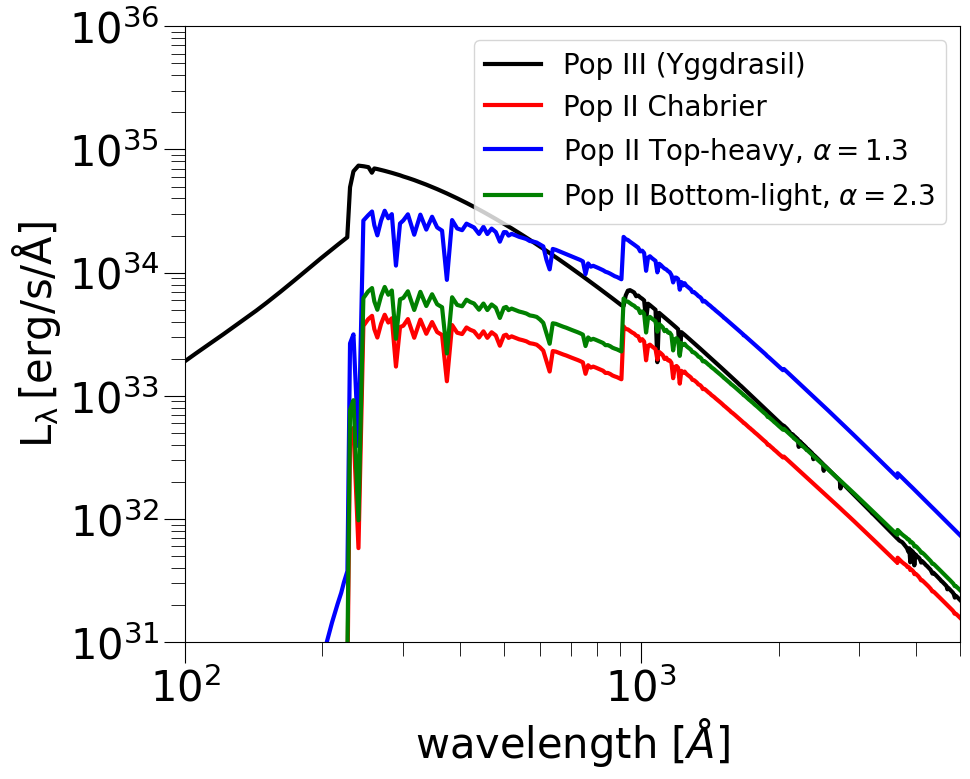}
    \caption{Generated synthetic SEDs display the differences arising from the use of various IMFs. The black solid line represents the SED of a Pop III star modeled with Yggdrasil (\citealp{Zackrisson2011}), while the red solid line illustrates the SED of a Pop II star generated using FSPS (\citealp{Conroy2010}). Both SEDs are shown at $t_{\star} = 0.01 \rm Myr$, with $m_{\star} = 1 \Msun$, adopting the Chabrier IMF. On the other hand, the blue and green solid lines depict the SEDs of Pop II stars using a top-heavy IMF, with different slope values, $\rm \alpha = 2.3$ for the green line and $\rm \alpha = 1.3$ for the blue line. Notably, the SED luminosity using the top-heavy IMF with $\rm \alpha = 1.3$ is almost 1 dex higher compared to the Chabrier IMF for Pop II stars, exhibiting values similar to the Pop III SED.}
    \label{fig:spectrum_by_IMF}
\end{figure}
\par
Another potential way to boost luminosity in our simulations is by considering a top-heavy IMF for Pop II stars (e.g., \citealp{Finkelstein2023, Harikane2023, Inayoshi2022, Yung_2023, Harvey2024}).
The IMF of Pop II stars is generally assumed to be bottom-heavy, following distributions such as the Salpeter IMF (\citealp{Salpeter1955}) or the Chabrier IMF (\citealp{Chabrier2003}). However, gas clouds in high-$z$ environments might experience higher temperature floors due to factors such as low-metallicity environments and the radiative coupling to the cosmic microwave background (CMB), which has a redshift-dependent temperature of $T_{\rm CMB}=2.73(1+z) \rm K$ (\citealp{Larson1998, Schneider2010, Safranek-Shrader2014, Jerabkova2018, Haslbauer2024, Kroupa2024}), leading to the formation of more massive stars. Similarly, hydrodynamical simulations by \citet{Chon2022} suggest that even for metallicities higher than our $\rm Z_{thr}$ value, stellar mass functions of star clusters at $z \gtrsim 10$ tend to exhibit top-heavy components, predominately resulting from CMB heating. To implement this idea, we use a modified version of the Salpeter IMF (\citealp{Salpeter1955}), adopting different values for $\alpha$, $m_{\rm min}$, and $m_{\rm max}$. The functional form of a top-heavy IMF for Pop II stars can be expressed as follows,
\begin{equation}
\label{eq2}
\xi(m) = \frac{dN}{dm} = \xi_{0} \, m^{-\alpha} \qquad  \begin{cases} \alpha=1.3 \\ 1\msun \leq m \leq 100 \msun \end{cases} .
\end{equation}

\par
In Figure \ref{fig:spectrum_by_IMF}, we compare the synthetic spectral energy distributions (SEDs) of Pop III and Pop II stars, illustrating the differences that arise from adopting various IMFs. To generate these SEDs, we use Yggdrasil (\citealp{Zackrisson2011}) and Flexible Stellar Population Synthesis (FSPS, \citealp{Conroy2010}) with different IMFs at $t_{\star} = 0.01 \rm Myr$ for a stellar mass of $m_{\star} = 1 \Msun$. The black line corresponds to the SED of Pop III stars, constructed using Yggdrasil (\citealp{Zackrisson2011}), whereas the colored lines represent the SEDs of Pop II stars utilizing FSPS (\citealp{Conroy2010}) with varied IMF shapes. Specifically, the red line shows the result of using the Chabrier IMF (\citealp{Chabrier2003}), with $0.1 \msun \leq m \leq 100 \msun$, which is the default Pop II star IMF adopted in our simulations. The blue line represents the result of a top-heavy IMF as expressed in Eq. \eqref{eq2}, while the green line uses the same IMF as the blue line but adopts a different value for the slope ($\alpha=2.3$). The comparison shows that using a top-heavy IMF (blue line) results in an almost 1 dex higher UV-boosted SED compared to the default SED using the Chabrier IMF (red line). This demonstrates that considering the boosted UV luminosity from a top-heavy IMF for Pop II stars could be a key factor in reducing the discrepancy between simulations and the JWST observations by producing brighter galaxies.

\begin{table}
\centering
\begin{tabular}{c|c|c|c}
\hline
Name & ${M_{\rm halo}}$ at $z=10\nspace[{\Msun}] $ & $\epsilon_{\rm ff}$ & Pop II IMF \cr
\hline
{\sc HM9-E001} & $4.2 \times 10^{9}$ & $0.01$ & Chabrier \cr
{\sc HM9-E030} & $4.2 \times 10^{9}$ & $0.3$ & Chabrier \cr
{\sc HM9-E100} & $4.2 \times 10^{9}$ & $1.0$ & Chabrier \cr
{\sc HM9-T001} & $4.2 \times 10^{9}$ & $0.01$ & Top-heavy \cr
{\sc HM9-T030} & $4.2 \times 10^{9}$ & $0.3$ & Top-heavy \cr
\hline
{\sc HM10-E001} & $1.3 \times 10^{10}$ & $0.01$ & Chabrier \cr
{\sc HM10-E030} & $1.3 \times 10^{10}$ & $0.3$ & Chabrier \cr
\hline
\end{tabular}
\caption{Summary of the simulation sets. Column (1): Name of run. Column (2): Halo mass at $z=10$.  Column (3): Star formation efficiency, $\epsilon_{\rm ff}$. Column (4): IMF choice for Pop~II.}
\label{tab:set}
\end{table}

\subsubsection{Summary of simulation sets}
To explore high-$z$ galaxies, we conduct simulations with variations in two elements of the sub-grid physics across two initial conditions. Specifically, we examine high-$z$ galaxies with halo masses of $\rm M_{halo} \approx 4.2 \times 10^9 \msun$ ({\sc HM9}) and $\rm M_{halo} \approx 1.3 \times 10^{10} \msun$ ({\sc HM10}) formed at $z = 10$. Table \ref{tab:set} presents information about the seven different sets of simulation combinations, incorporating different sub-grid physics as described in \ref{subsubsec:p3} and \ref{subsubsec:p2}, along with the two different initial conditions. Specifically, Table \ref{tab:set} summarizes each set, including the names of each run, the final halo masses $M_{\rm halo}$ at $z = 10$, the modified $\rm \epsilon_{ff}$, indicating $ \epsilon_{\rm ff} = \epsilon_{\rm ff, Pop III} = \epsilon_{\rm ff, Pop II}$, and the choice of the IMF for Pop II stars in each column.

\subsection{Stellar feedback}
\subsubsection{Photoionization feedback}
\label{RT}
Once the stars are created, they start to emit ionizing photons into the interstellar medium (ISM). We employ the radiative transfer (RT) module \traphic \nspace (\citealp{Pawlik_2008}; \citealp{Pawlik2011TRAPHHIC}) to address the equations governing the photoionization feedback process. To summarize briefly, when the photon packets are emitted from the radiation source, they propagate along the spatially adaptive, unstructured grid traced out by the SPH particles in a photon-conserving manner. In our simulations, the radiation sources emit photon packets in $N_{\rm EC} = 8$ random directions per RT step. After calculating absorption and scattering within the SPH particles, those containing the emitted photon packets transfer the remaining energy to neighboring SPH particles in $N_{\rm TC} = 32$ random directions. If these cones do not contain SPH particles, virtual particles (ViPs) are introduced into the cones to facilitate the transfer of photon packets. To enhance the sampling of the volume, photon packets are emitted $N_{\rm em} = \Delta t_{\rm r} / \Delta t_{\rm em}$ times by randomly rotating the orientation of the cones. The emission time step is $\Delta t_{\rm em} = 5 \times 10^{-3}$ Myr, and the radiative time step is $\Delta t_{\rm r} = {\rm min}(10^{-2} \rm Myr, \Delta t_{\rm hydro})$, where $\Delta t_{\rm hydro}$ is the minimum time step of the SPH particles in the simulations. For more details on \traphic, we refer readers to \citet{Pawlik_2008, Pawlik_2011}.
\par
For the Pop III star clusters, the total ionizing photons emitted per second from a single Pop III cluster are calculated by integrating the contributions from individual stars, weighted by the assumed IMF. The exact expression is as follows,
\begin{equation}
\label{eq3}
    \dot{N}_{\rm ion, tot}(t) = \int^{m_{\rm up}}_{m_{\rm low}}\dot{N}_{\rm ion}(m)\phi_{\rm Pop III}(m)\eta(m)dm, 
\end{equation}
using the same Pop III IMF parameters as in Sec.~\ref{subsubsec:p3}. Here, the contribution function, $\eta(m)$, is set to 1 when the Pop III cluster is younger than the Pop III star lifetime, $t_{\rm Pop III}$. Once the lifetime of a cluster surpasses $t_{\rm Pop III}$, we assume the star particle has left the main sequence and set $\eta(m) = 0$. For simplicity, we assume a lifetime of $t_{\rm Pop III} = 3$ Myr for all Pop III clusters in our simulations, based on the assumption that Pop III stars are massive and short-lived (\citealp{Bromm2001, Schaerer2003}). To determine the ionizing photon rates of individual Pop III stars, we use polynomial fits from \cite{Schaerer2003}, $\log_{10} \dot{N}_{\rm ion} = 43.61 + 4.90 x - 0.83 x^2$, where $x = \log_{10} (m / \msun)$ represents the initial mass of a star. For example, the integrated ionizing photon rate on the zero-age main sequence is $\dot{N}_{\rm ion, tot}(\rm t = 0) = 10^{47.9} \invs \Msuni$. 

For Pop II stars, we do not consider photoionization feedback and only account for SN feedback, as the ionizing photon rate of Pop II stars is negligible compared to that of Pop~III stars, and the corresponding feedback is less significant than the impact of Pop~II SNe (e.g., \citealp{Bromm2001, Schaerer2003}). To account for the absence of photoionization feedback from Pop II stars, we simply assume that massive Pop II stars undergo SN explosions immediately after their formation. This approach omits the SN delay period dictated by the stellar evolution time necessary before an explosion, during which photoionization heating would take place, thus allowing us to reduce computational costs. We acknowledge that the choice of delay time for SN feedback is crucial, as it can significantly influence galaxy properties such as stellar mass and the burstiness of star formation in high-$z$ galaxies with short dynamical timescales (e.g., \citealp{Faucher-Giguere2018, Furlanetto2022}). Consequently, our results should be considered as an upper limit, particularly for stellar masses, which would likely be lower if photoionization heating and radiation pressure were included (e.g., \citealp{Wise2012}). We plan to conduct similar simulations in future work, incorporating all relevant feedback, including photoionization and photoheating feedback from all stars, and we will compare these results with the current work.

\subsubsection{Supernova feedback}
\label{Supernova feedback}
SN feedback is one of the primary feedback mechanisms within galaxies and is highly effective at disrupting the dense gas in the ISM, thereby regulating star formation activity. This feedback is implemented as a thermal energy mechanism, where the rest-mass energy of the dying stellar cluster is converted into thermal energy and transferred to neighboring SPH particles. However, there is a well-known issue associated with the thermal energy scheme, known as the over-cooling problem. In this scenario, gas particles heated by a SN explosion radiate their energy too quickly, rendering the SN feedback ineffective. To avoid this problem, we adopt the method proposed by \citet{DallaVecchia2012}, which guarantees a temperature increase of more than $10^{7.5} \rm K$ by limiting the number of neighboring particles that receive the SN thermal energy. To implement this approach, we reduce the number of gas particles receiving the SN energy to $N_{\rm ngb} = 1$. This strategy helps prevent the overcooling problem by concentrating the energy on fewer particles.

\par
The total SN energy per unit solar mass, $\epsilon_{\rm SN}$, is calculated using the adopted IMF for both Pop III and Pop II stars, with the assumption that each SN releases $10^{51} \rm erg$ of energy. Thus, this can be expressed as $\epsilon_{\rm SN} = n_{\rm SN} \times 10^{51} \rm erg$, where $n_{\rm SN}$ is the number of SNe per unit mass, calculated by integrating the IMF, $\phi(m)$, over the mass range from $m_{\rm min}$ to $m_{\rm max}$. Here, $m_{\rm min} = 8 \msun$ and $m_{\rm max} = 40 \msun$ represent the lowest and highest initial masses of stars that can undergo a SN. For Pop III stars, the resultant value is $\epsilon_{\rm SN, PopIII} = 5.56 \times 10^{49} \rm erg  \Msun^{-1}$, while for Pop II stars, it is $\epsilon_{\rm SN, PopII} = 1.73 \times 10^{49} \rm erg \Msun^{-1}$.

\par
We also consider the chemical enrichment contributed by the winds from asymptotic giant branch (AGB) stars, and the explosions of core-collapse supernovae (CCSNe) and Type Ia SNe for Pop II stars, as well as CCSNe and PISNe for Pop III stars, using the methods described in \citet{Wiersma2009}. At each timestep in the simulations, we calculate the masses of nine individual elements (H, He, C, N, O, Si, Mg, Ne, and Fe) produced by the dying stars and release them into the neighboring ISM and IGM. For the Pop III stars, we adopt nucleosynthetic metal yields and remnant masses from \citet{Heger2010} for CCSNe and from \citet{Heger2002} for PISNe.

\par
For Pop II stars, we calculate the yields of elements and evolutionary tracks of each star using the metallicity-dependent tables within the range $\rm Z = 0.0004 \Zsun$ to $\rm Z = 1.0 \Zsun$ (\citealp{Portinari1998}). Additionally, the mass loss of intermediate-mass stars ($0.8 \msun \lesssim m \lesssim 8 \msun$) during the AGB phase is calculated using \citet{Marigo2001} and we use an empirical delay time function expressed in terms of e-folding times (e.g., \citealp{Tonry2006}) for Type~Ia SNe due to uncertainties in the detailed evolution of Type Ia SNe. We disperse the ejected metals from dying stars into neighboring gas particles ($N_{\rm ngb} = 48$), and then transport these metals to the IGM and ISM by solving the diffusion equation (e.g., \citealp{Greif2009}). For more details about SN feedback in our simulations, we refer readers to \citet{Kim2023} and \citet{Lee2024}.

\subsection{Post-processing}
\par
\begin{figure}
    \centering
    \includegraphics[width = 85mm]{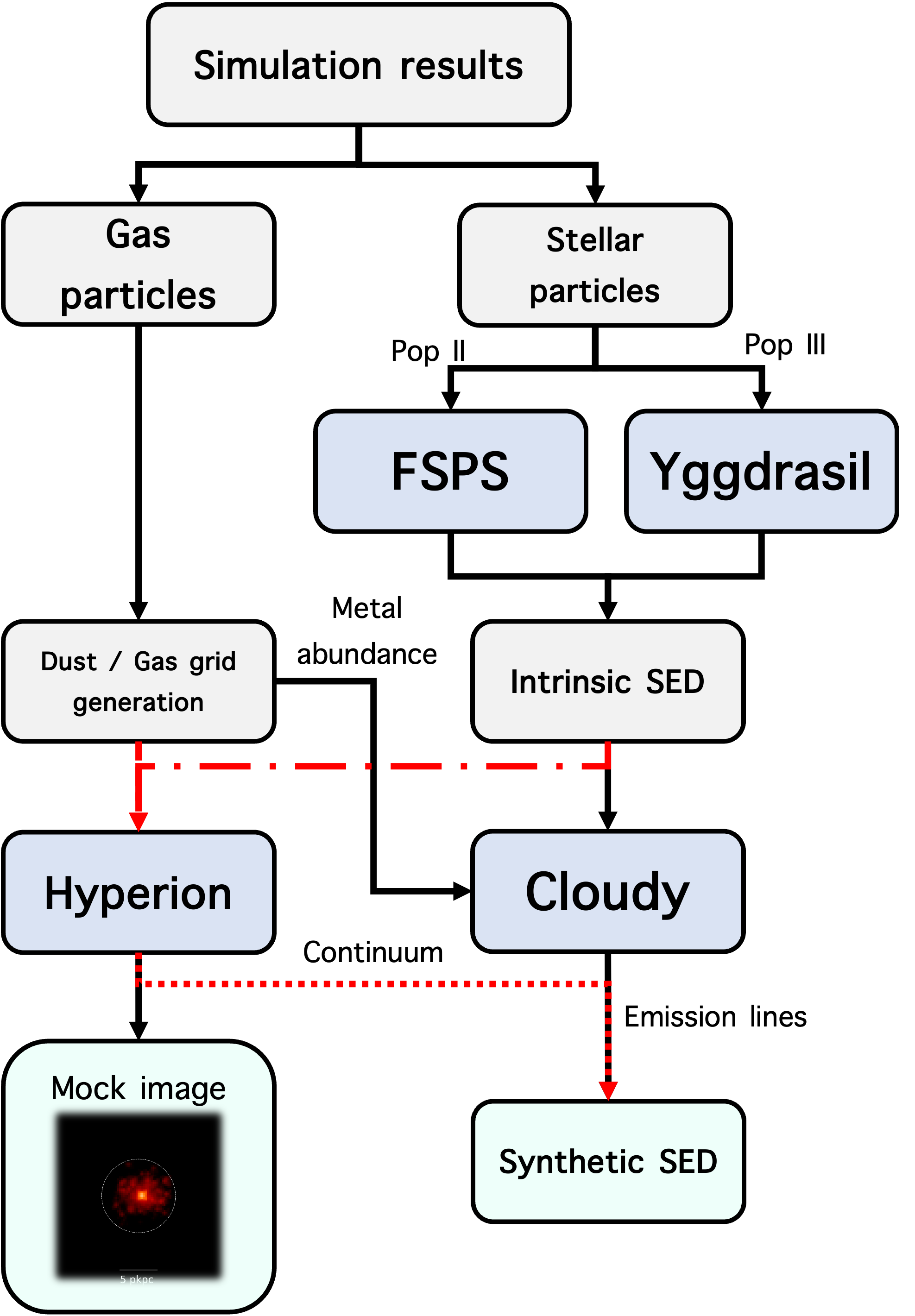}
    \caption{The diagram illustrates the basic flow for conducting post-processing. In the diagram, gray squares represent raw data from simulation snapshots and preliminary results from each step. Shaded blue squares indicate the modules utilized in the process, while shaded cyan squares denote the final synthetic observational results, such as mock images and synthetic SEDs.}
    \label{fig:pipeline for post-processing}
\end{figure}

After running the simulation sets, we post-process the results to derive synthetic observations for comparison with actual observational data. To achieve this, we follow the basic pipeline frameworks described in \citet{Barrow2017}, which integrate the stellar synthetic library code \fsps \nspace (\citealp{Conroy2010}), the dust radiative transfer code \hyperion \nspace (\citealp{Robitaille2011}), and the photoionization code \cloudy \nspace (\citealp{Chatzikos2018}). It is important to note that the SED of Pop III stars is not considered in \citet{Barrow2017}, and \fsps \nspace does not provide synthetic libraries for Pop III stars as well. Therefore, we utilize SED data for Pop III stars from \yggdrasil \nspace (\citealp{Zackrisson2011}) and calculate their bolometric luminosities. Figure \ref{fig:pipeline for post-processing} illustrates the basic workflow of our post-processing pipelines. The gray boxes represent raw data from simulation snapshots and preliminary results obtained during post-processing. The blue boxes denote the modules used to process the raw data and preliminary results. Finally, the cyan boxes indicate the end results of the post-processing procedure, such as mock photometry results and the resultant synthetic SED.
\par
\par
To briefly summarize, we first produce the intrinsic stellar spectra and bolometric luminosity for each star particle extracted from a simulation snapshot. This is done using \fspspython \nspace (\citealp{ben_johnson_2024_12447779}) for Pop II stars and \yggdrasil \nspace for Pop III stars. Each star particle is treated as a SSP, meaning they are considered as stellar clusters consisting of stars with the same age and metallicity. Next, we generate grids to describe the dust and gas density distribution by using the metallicity and neutral atomic hydrogen fraction extracted from our simulation results. Stellar particles are then superimposed onto these dusty and gaseous grids. We assume that dust contributes 7\% of the metal elements and use the MW dust model from \citet{Draine2003} ($R_{V} = 3.1$) for dust extinction (\citealp{Barrow2017}). Additionally, for gas extinction, we assume that neutral atomic hydrogen is the primary species contributing to gas opacity.
\par

Finally, we apply gas and dust extinction and absorption using \hyperion, and derive emission line strengths with \cloudy. For calculating emission lines, we reuse the gas grid enriched with metal-abundant gas particles and treat star particles as single source points at the center of each cell. Using these processed results, we combine the emission line data with the continuum to obtain the total SEDs for the target galaxies. The emission lines are then adjusted based on the ratio of the intrinsic SED to the post-processed continuum. As described in \citet{Barrow2017}, we only construct emission line data for the SEDs, which are not included in the mock images. Recent studies have suggested the potential significance of a strong nebular continuum arising from the surrounding ISM of massive stars (\citealp{Cameron2024, Trussler2023}). However, in this work, we do not include the nebular continuum in either the mock images or the SEDs. Considering the IMF assumed for Pop II stars in this study, the ratio of massive stars ($m_{\rm \star} \gtrsim 50 \msun$) to those with masses in the range of $5 \msun \lesssim m_{\rm \star} \lesssim 50 \msun$ is about three times lower than the values reported by \citet{Cameron2024}. This suggests that the nebular continuum may have a less significant impact on our results. Note that we only focus on the observabilities of our simulated galaxies at $9 \leq z \leq 13$, taking into account gas extinction by the ISM within galaxies, while neglecting the effect of the IGM due to the substantial uncertainty in its variations. For the IGM effect, we explicitly cut off the post-processed SED of simulated galaxies below the Lyman limit ($\lambda_{\rm rest} \leq 912 \Ang$), setting $f_{\lambda\leq912\Ang} = 0$.
\par
\begin{figure*}
    \centering
    \includegraphics[width = 170mm]{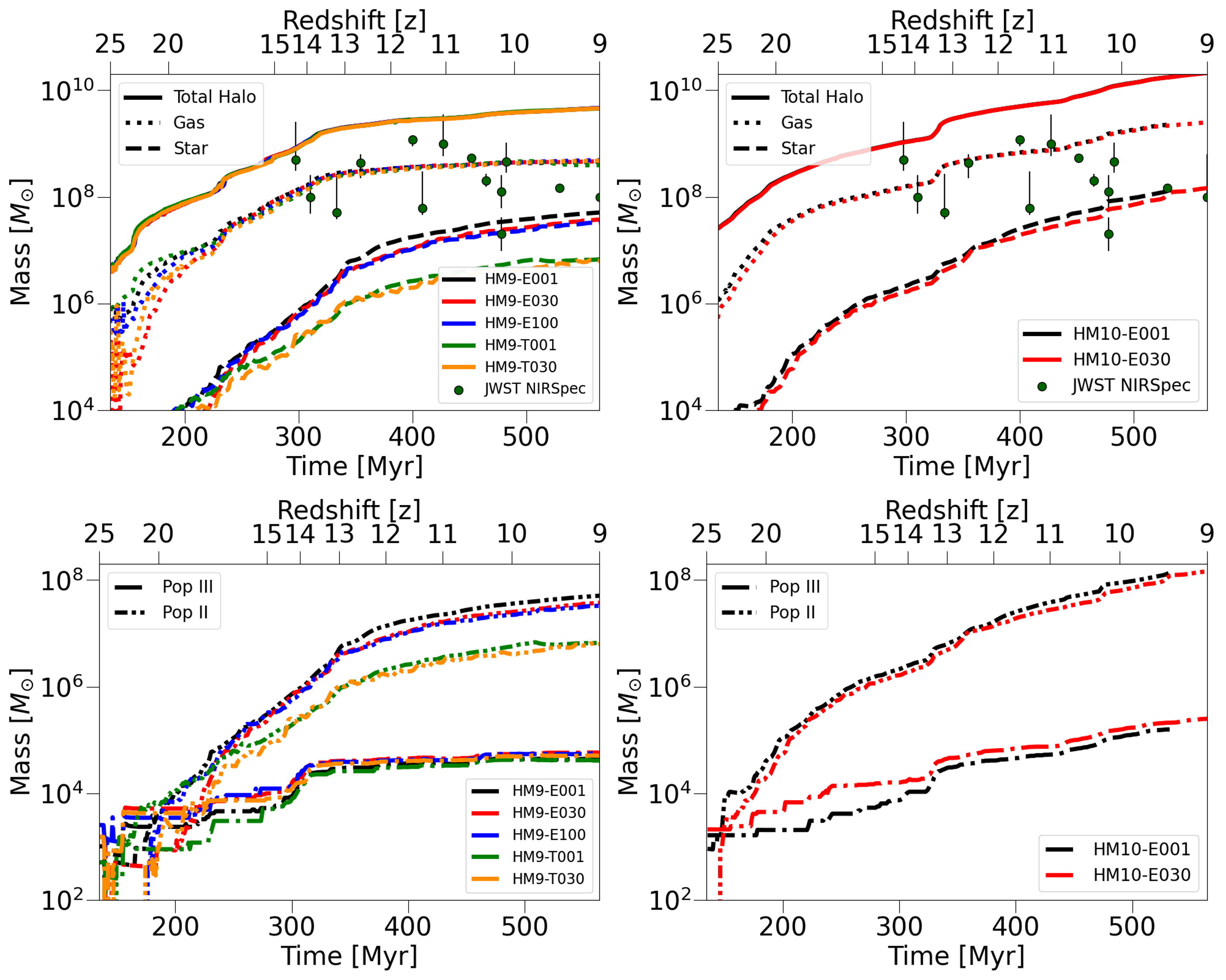}
    \caption{Mass evolution of our simulated sets. Each column depicts the mass evolution of simulated galaxies with different initial conditions, HM9 (left, $M_{\rm vir} \approx 10^9 \msun$ at $z =10$) and HM10 (right,  $M_{\rm vir} \approx 4.2\times 10^{10} \msun$ at $z =10$), respectively. The panels in the top row show the evolution of the total halo mass (solid), gas mass (dotted), and stellar mass (dashed) as a function of time, whereas the panels in the bottom row display the evolution of the stellar mass within $R_{\rm vir}$ of halos, separately for Pop~III stars (dash-dotted) and Pop~II stars (dash-double-dotted). The colored lines in both panels indicate different simulation sets: E001 (black, $\epsilon_{\rm ff} = 0.01$, Chabrier IMF for Pop II stars), E030 (red, $\epsilon_{\rm ff} = 0.3$, Chabrier IMF for Pop II stars), E100 (blue, $\epsilon_{\rm ff} = 1.0$, Chabrier IMF for Pop II stars), T001 (green, $\epsilon_{\rm ff} = 0.01$, Top-heavy IMF for Pop II stars), and T030 (orange, $\epsilon_{\rm ff} = 0.3$, Top-heavy IMF for Pop II stars), respectively. Green circles exhibit the stellar masses of observed high-$z$ galaxies spectroscopically confirmed by JWST NIRSpec.}
    \label{fig:mass evolution}
\end{figure*}
\par

\section{Results}
In this section, we present the results of our simulations, emphasizing the impact of sub-grid physics variations on the evolution and observability of our simulated galaxies. Specifically, we compare two scenarios with different star formation efficiencies and varied IMFs for Pop II stars, originating from distinct initial conditions.
In Section \ref{simulation results}, we examine the general mass assembly process and the associated star formation histories of our simulated galaxies. Section \ref{post-processing results} discusses the derived observabilities of our simulated galaxies from post-processing and explores the signatures of Pop III stars within these galaxies.

\subsection{Simulation results}
\label{simulation results}
\subsubsection{Mass evolution}
\label{mass evolution}

\par
In Figure \ref{fig:mass evolution}, we present the mass evolution as a function of cosmic time from our simulation sets, illustrating the composite evolution histories for each simulated galaxy. The left panel corresponds to the HM9 runs, which have a virial mass $M_{\rm vir}\approx4.2 \times 10^9 \msun$ at $z$ = 10. The right panel represents the HM10 sets, with $M_{\rm vir}\approx1.3 \times 10^{10} \msun$ at $z$ = 10, each with different initial conditions. For both panels in the top row, different line styles represent various components: solid lines indicate virial mass, dotted lines show gas mass, and dashed lines represent stellar mass. For each stellar population, we also illustrate the evolution of stellar mass by distinguishing between Pop III (dash and dot) and Pop II stars (dash and two dots) in the bottom-row panels. The colors of the lines in the figure denote different simulation sets with varied sub-grid physics, E001 ($\epsilon_{\rm ff} = 0.01$, Chabrier IMF for Pop II), E030 ($\epsilon_{\rm ff} = 0.3$, Chabrier IMF for Pop II), E100 ($\epsilon_{\rm ff} = 1.0$, Chabrier IMF for Pop II), T001 ($\epsilon_{\rm ff} = 0.01$, top-heavy IMF for Pop II), and T030 ($\epsilon_{\rm ff} = 0.3$, top-heavy IMF for Pop II). Additionally, the stellar masses of observed high-redshift galaxies, spectroscopically confirmed by JWST surveys (e.g., \citealp{Harikane2024, Hsiao2023, Curtis-Lake2023, Bunker2023, Carniani2024, Curti2024}), are compared using green circle symbols.

\par
By starting with the fiducial set, HM9-E001, we observe that when the virial mass of the halo reaches $M_{\rm vir} \approx  10^6 \msun$, the first star formation commences within the halo, forming Pop III clusters at $z \approx 25.7$.
Due to the photoionization heating, followed by SN explosions from Pop III stars, nearly half of the gas within the halo is evacuated, consequently suppressing star formation for a few $\sim 10$ Myr (e.g., \citealp{Ritter2012}). This period of suppressed star formation tends to exhibit similar durations across all HM9 sets. After this brief suppression period, Pop II stars are born out of the contaminated gas clouds that have been enriched by the SNe of Pop III star clusters. The transition from Pop III to Pop II stars is achieved rapidly, within $\rm \sim 50$ Myr after the suppression, making the Pop~II stars the predominant population within the galaxies. For instance, at $z\approx20$ the total mass of Pop~II stars is higher than that of Pop~III stars by a factor of 2.

It is noteworthy that the onset of Pop III stars may be postponed if the influence of a Lyman-Werner (LW) background—produced by early-forming Pop III stars outside the halo—is taken into account (e.g., \citealp{Fialkov2013, Hirano2015, Schauer2021, Kulkarni2021, Prole2023}). This study does not consider the LW background, which can hinder the collapse of star-forming gas by dissociating molecular hydrogen. For example, by varying the LW~background strength, $J_{\rm 21} = 0.01$ and $J_{\rm 21} = 0.1$ (in units of $10^{-21} \rm erg \, s^{-1} \, cm^{-2} \, Hz^{-1} \, sr^{-1}$), \citet{Prole2023} suggested an increased halo mass requirement for the birth of Pop~III stars by a factor of 1.3 at a minimum and 2.5 at maximum. Nevertheless, as previously mentioned, the era of Pop~III stars is brief due to the rapid transition to Pop~II stars, and the exclusion of the LW background thus does not significantly alter the global characteristics of the simulated galaxies.


\par
\begin{figure*}
    \centering
    \includegraphics[width = 185mm]{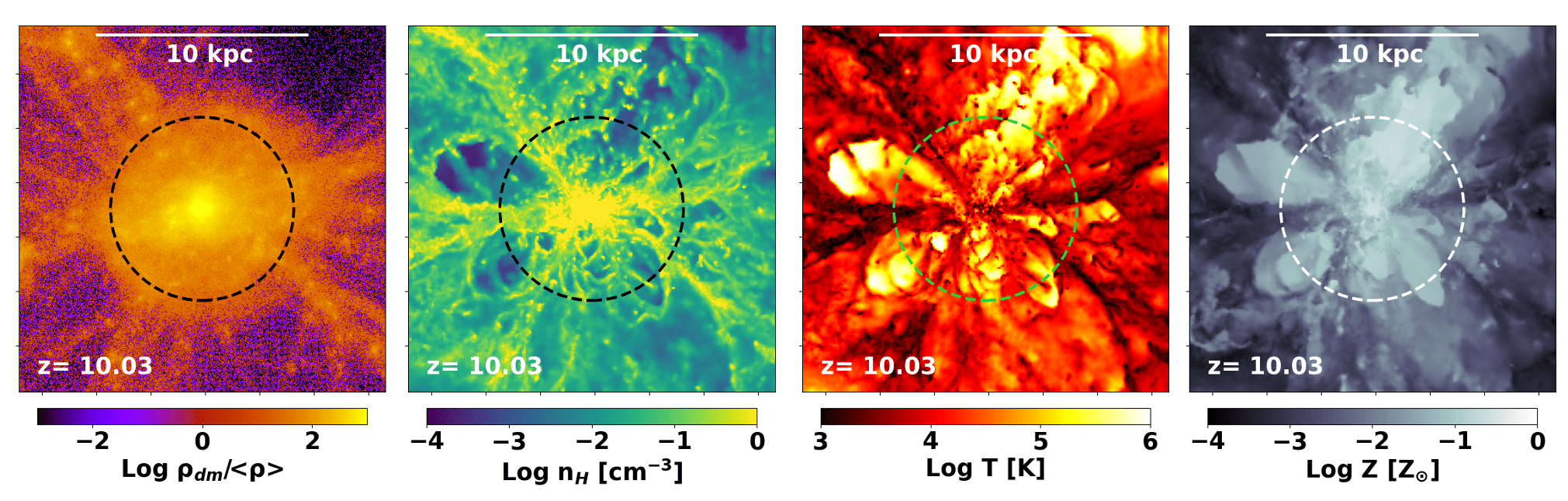}
    \caption{The morphology of simulated galaxy from the HM9-E001 set at $z\approx10$. From left to right, each panel illustrates the projected DM density, hydrogen number density, gas temperature, and gas metallicity along the line of sight within $4R_{\rm vir}$ from the center of the galaxy. The dashed circles in each panel indicate the virial radius of the galaxy.}
    \label{fig:projection plot}
\end{figure*}
\par

\par
After the transition of the stellar populations, the virial mass of the halo increases and surpasses the size of the atomic cooling halo ($M_{\rm vir} \approx 5 \times 10^7 \msun, z \approx 20.5$), which is defined as the minimum mass scale required for sufficient atomic cooling of hydrogen, the deepened potential well of the halo can retain gas mass despite the stellar feedback from Pop II stars. Thus, stellar feedback from Pop II clusters, primarily due to SNe, expels gas only from the center of the halo ($r \lesssim 0.2 R_{\rm vir}$) and the evacuated gas within the halo falls back on timescales of a few Myr. 
To analyze the mass of outflows and inflows as the halo evolves, we track gas particles across consecutive snapshots. Gas particles moving from inside to outside the virial radius, $R_{\rm vir}$, are classified as outflow, whereas those transitioning from outside to inside are considered inflow. Star particles formed between snapshots that meet these criteria are also included in their respective categories. We define $M_{\rm outflow}$ and $M_{\rm inflow}$ as the total mass of these outflowing and inflowing particles. As the halo grows, stellar feedback from newly formed Pop II stars intensifies, expelling a small fraction of gas ($M_{\rm outflow}/M_{\rm gas} < 10^{-2}$) within $R_{\rm vir}$. Nevertheless, inflow mass remains predominant over outflow mass. This indicates that the increased stellar feedback is insufficient to prevent cold, dense gas from cosmic filaments from accreting into the halo, leading to the continuous Pop II star formation from these gas clouds at the center of the halo. Consequently, both the gas mass and stellar mass within the halo increase proportionally with the virial mass of the halo. Finally, when the simulation completes ($z \sim 9$), the total stellar mass within the halo reaches $M_{\star} \approx 5.0 \times 10^7 \msun$, which overlaps with the lower end of the stellar mass range observed in high-$z$ galaxies that have been spectroscopically confirmed by JWST surveys.

\par
Figure \ref{fig:projection plot} displays the morphology of the simulated galaxy from the HM9-E001 set at $z \approx 10$. From left to right, the panels show the projected morphology of DM density, hydrogen number density, gas temperature and gas metallicity of the simulated galaxy, respectively. The dashed circles in each panel indicate the virial radius of the galaxy ($R_{\rm vir}\approx 4.31$ kpc at $z=10$). The impact of stellar feedback, especially SN feedback from Pop~II stars, is clearly evident as an inhomogeneous distribution, with stellar feedback preferentially acting along directions that avoid the cosmic filaments. On the other hand, cold dense gas from cosmic filaments can withstand the feedback and accrete into the center of the halo, triggering continuous star formation at the center of the galaxies.

\par
When we increase the star formation efficiency, $\epsilon_{\rm ff}$, we obtain results that deviate from our expectations, showing a slightly decreased stellar mass with the increased-$\epsilon_{\rm ff}$ values compared to the fiducial set, HM9-E001. The first stars in the $\epsilon_{\rm ff}$-increased sets (HM9-E030, HM9-E100) form at similar epochs as in the HM9-E001 set, but the transition from Pop III to Pop II occurs relatively later ($z \approx 17.6$ for HM9-E030, $z \approx 19.6$ for HM9-E100). This delay is because the more effective formation of Pop III stars with increased-$\epsilon_{\rm ff}$ values, accompanied by stronger stellar feedback, results in a longer suppression period, which compensates for the initially high star formation with subsequent suppressed star formation. Consequently, even with the boosted $\epsilon_{\rm ff}$, until the halo in both sets reaches a virial mass of $M_{\rm vir} \approx 10^8 \msun$, there are no significant differences in the total masses of gas and stars in the halos compared to the HM9-E001 set.

As the virial mass of the halos exceeds the threshold for efficient atomic cooling of hydrogen ($z \approx 20.17$ for HM9-E030 and $z \approx 20.41$ for HM9-E100), the halos can withstand the feedback and sustain star formation (see also \citealt{Bromm2011}). Star formation with the boosted $\epsilon_{\rm ff}$ values tends to exhibit more bursty behavior, which will be discussed in detail in Section \ref{SFR}, compared to the HM9-E001 set. The strong feedback associated with these starbursts can suppress star formation for certain periods. Consequently, we find that when the virial mass of halos in the increased $\epsilon_{\rm ff}$ sets exceeds $M_{\rm vir}\approx10^9 \msun$ at $z \approx 14.1$ for both sets, the growth trend of stellar mass within the halos shows a slight decrease compared to the HM9-E001 set. Eventually, the increased $\epsilon_{\rm ff}$ sets are likely to display more gradual star-forming histories and lower stellar masses, unable to surpass the stellar mass of the HM9-E001 set at $z\lesssim14$. Thus, contrary to the aforementioned expectation that enhancing $\epsilon_{\rm ff}$ could boost their stellar mass, increasing $\epsilon_{\rm ff}$ does not guarantee an increased stellar mass in high-redshift galaxies. Instead, such boost could slightly reduce their stellar mass to about 60\%-70\% of the value in HM9-E001 with $M_{\rm vir} \approx 4.2 \times 10^9 \msun$. These findings are consistent with simulations conducted on smaller scales that focus on detailed star formation histories within giant molecular clouds (GMCs) (e.g., \citealp{Gaudic2019}).


\par
The star formation histories of the top-heavy IMF adopted sets (HM9-T001, HM9-T030) differ significantly from those of the Chabrier IMF adopted sets (HM9-E001, HM9-E030, and HM9-E100). When the virial mass of halos reaches $M_{\rm vir}\approx10^8 \msun$, the stellar mass within these halos shows slower growth, and this trend becomes more pronounced after $z \lesssim 17$, resulting in a total stellar mass that is lower by a factor of $\sim 6-7$ compared to the Chabrier IMF sets. This reduced growth is attributed to the significant fraction of massive OB stars in the top-heavy IMF, which increases the frequency of SN explosions. These explosions, accompanied by significantly strong feedback effects on the surrounding medium, reduce subsequent star formation activities. Consequently, our findings show that adopting a top-heavy IMF, which is typically considered a scenario to boost UV luminosity while maintaining similar or reduced stellar masses in high-redshift galaxies, actually leads to a decrease in stellar mass. 

\par
\begin{figure}
    \centering
    \includegraphics[width = 85mm]{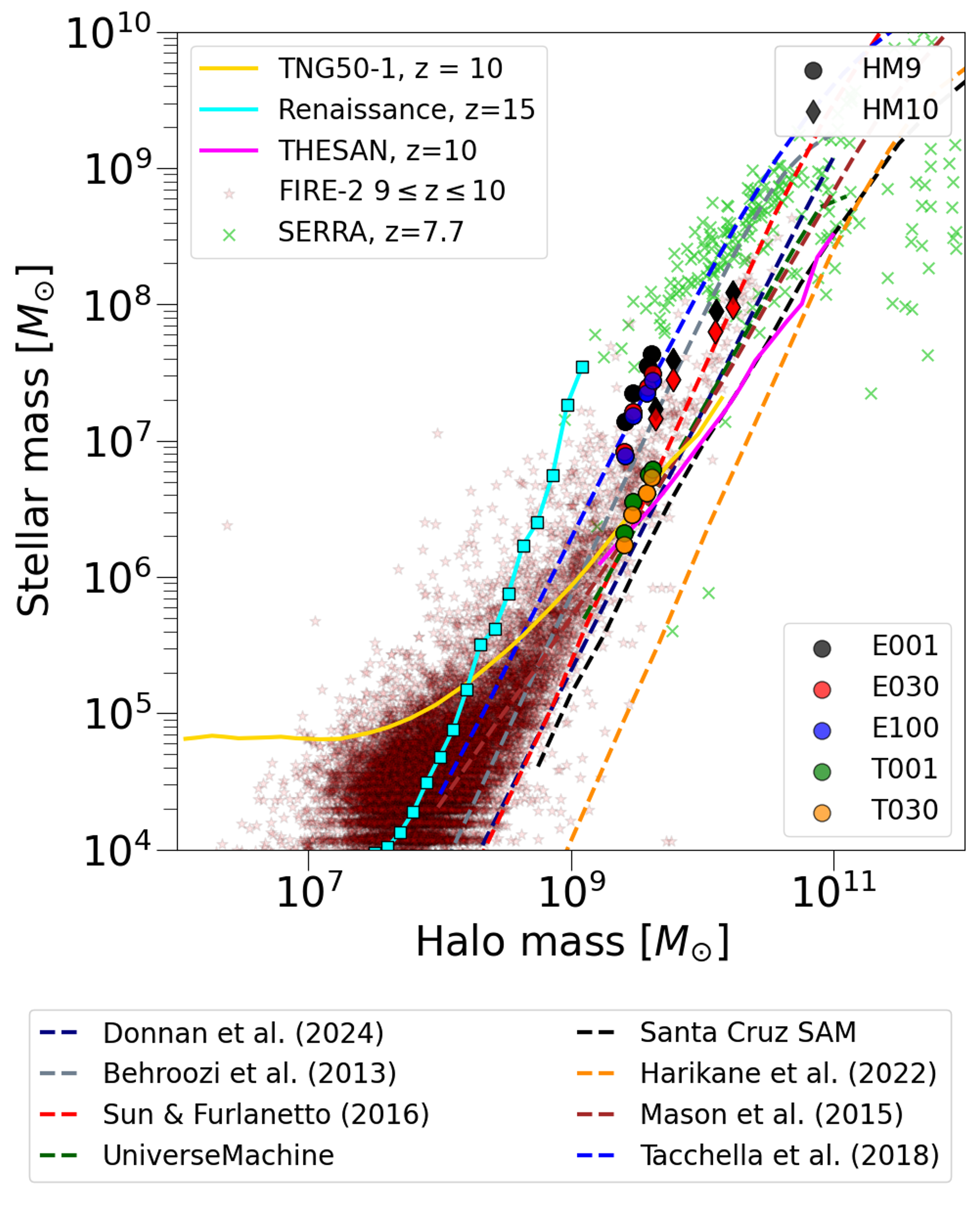}
    \caption{HMSM relation of our simulated galaxies, compared with high-$z$ results from other simulation works. The circle and diamond symbols represent the HM9 and HM10 sets at selected epochs, $z$ = 9.5, 10, 11, and 12, respectively. To compare our HMSM relation, we display high-$z$ median values from other simulations, TNG50 (yellow solid line, $z$ = 10), the Renaissance simulation, specifically for the rare peak (cyan solid line, $z$ = 15), and THESAN (red solid line, $z$ = 10). Although our results exhibit a slightly higher stellar mass at a given halo mass compared to the median values from TNG50 and THESAN, they still show a close match with other simulations such as FIRE-2 and SERRA. We also show HMSM results from analytic models using median values of \textsc{UniverseMachine} (\citealp{Behroozi2020}, green) and Santa Cruz SAM (\citealp{Yung2019}, black). Additionally, we include an analytic formula from \citet{Donnan2024} (navy) and formulae based on low-$z$ observations by \citet{Behroozi2013} (gray, extrapolated to $z=10$), \citet{Sun&Furlanetto2016} (red), \citet{Harikane2022} (orange), \citet{Mason2015} (brown), and \citet{Tacchella2018} (blue), all represented with dashed lines.}
    \label{fig:HMSM}
\end{figure}
\par

For the more massive halos in the HM10 sets (HM10-E001 and HM10-E030), where we solely test the impact of increased $\epsilon_{\rm ff}$, both sets exhibit a similar relationship to the HM9 sets. Since the halos in the HM10 sets are more massive than those in the HM9 sets, they exceed their virial mass threshold for efficient atomic cooling before $z \approx 25$. We find that the $\epsilon_{\rm ff}$-increased set (HM10-E030) tends to have a similar or slightly lower stellar mass compared to the HM10-E001 set, though the factor of difference is insignificant ($\lesssim$30 \%) at $z \leq 18.9$. By the end of the simulations, both sets reach a total stellar mass within the galaxy of $M_{\star}\approx 10^8 \msun$, with a virial mass of around $M_{\rm vir}\approx 1.7 \times 10^{10} \msun$.

\par
We find that the strength of stellar feedback is reflected in the distribution of stellar mass in our simulated galaxies. Due to bursty star formation histories and the associated feedback, sets with increased-$\epsilon_{\rm ff}$ are likely to exhibit a slightly extended radial distribution of their stellar mass. This trend is pronounced in the HM9-E030 set, which shows the most extended radial stellar distribution among our simulation sets. Nonetheless, most of the stellar mass in our simulated galaxies  ($M_{\star, 0.2 R_{\rm vir}}/M_{\star} \geq 0.7$) is contained within $R \leq 0.2 R_{\rm vir}$, for host virial masses exceeding $M_{\rm vir} \sim 10^9 \msun$. Further details on the radial distribution and size evolution of our simulated galaxies will be discussed in Section \ref{effective radius}.
\par
\begin{figure*}
    \centering
    \includegraphics[width = 170mm]{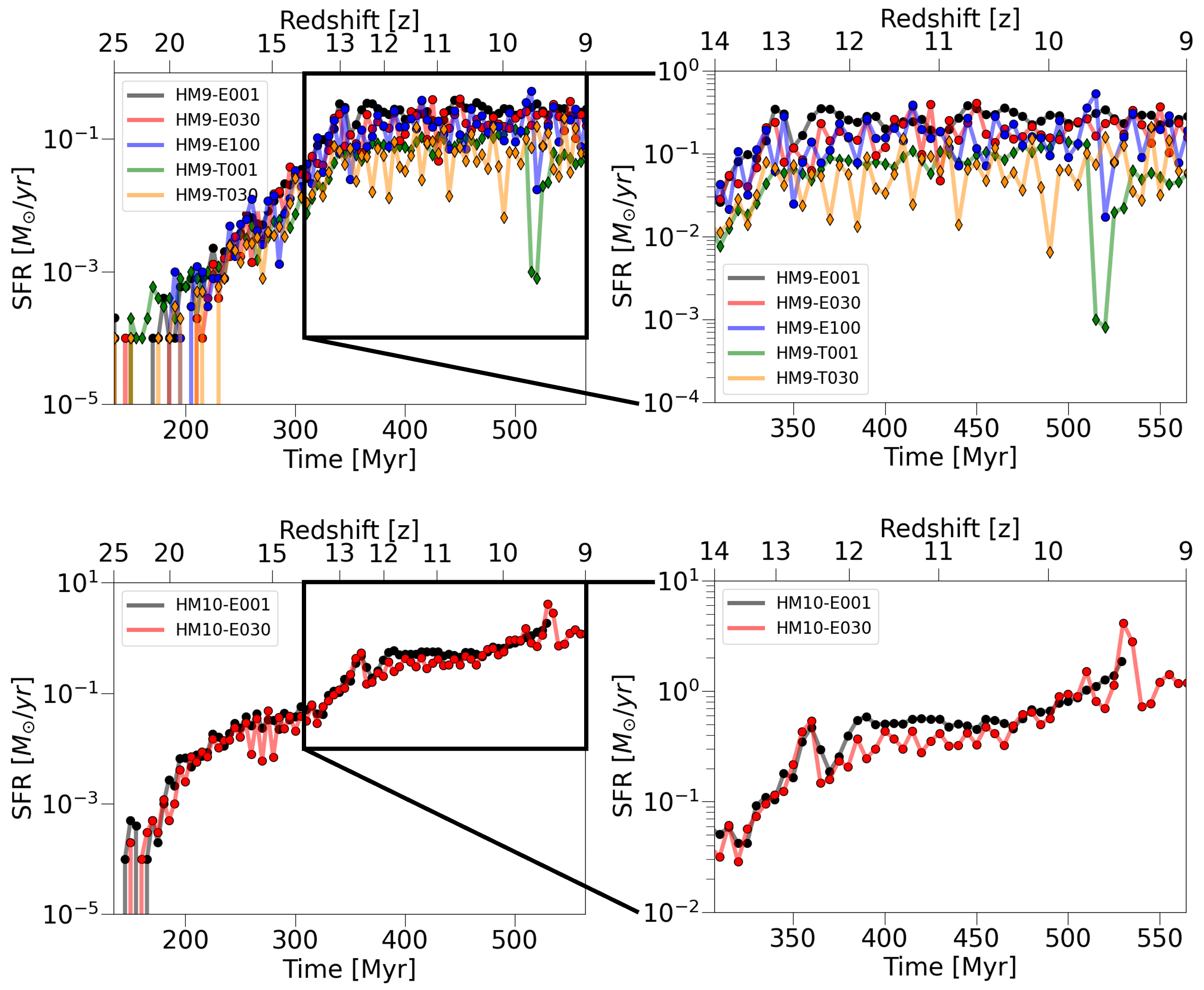}
    \caption{The SFR of our composite sets of simulated galaxies, separately displaying the total SFR history in the left panels and subset of SFR, in particular, within the redshift range $14 \leq z \leq 9$ in the right panels. The top panels indicate the results from HM9 sets, while the bottom panels show the results from HM10 sets. The colors of the lines and markers correspond to those used in Figure \ref{fig:mass evolution}. For better visibility, however, we vary the shapes for markers as circles and diamonds for Chabrier IMF adopted sets and top heavy adopted sets, respectively. The SFRs are calculated based on a time bin of $\Delta t = 5 \rm Myr$, which is comparable to the lifetime of massive stars in our simulated galaxies.}
    \label{fig:SFR}
\end{figure*}
\par
To validate our mass evolution results, we compare the halo mass-stellar mass (HMSM) relation of high-z galaxies with results from other simulation works and analytic studies. These include large-scale cosmological simulations such as TNG50 (\citealp{Nelson2019}) and THESAN (\citealp{Kannan2022}), which do not employ the zoom-in technique, as well as zoom-in simulations such as the Renaissance simulation (\citealp{Chen2014}), SERRA (\citealp{Pallottini2022}), and FIRE-2 (\citealp{Ma2018a, Ma2019, Ma2020}). We also consider median values from analytic models, such as \textsc{UniverseMachine} ($z=10$) (\citealp{Behroozi2020}) and Santa Cruz SAM (\citealp{Yung2019}), as well as results from analytic formulae based on low-$z$ observations (\citealp{Donnan2024, Harikane2022, Behroozi2013, Sun&Furlanetto2016, Mason2015, Tacchella2018}). In Figure \ref{fig:HMSM}, we present the HMSM results compared with the aforementioned ones. Our simulated results are marked with circle and diamond symbols, each indicating HM9 and HM10 sets, respectively, at selected epochs, z = 9.5, 10, 11, and 12. The median values for TNG50 (z = 10), the Renaissance simulation (rare peak, z = 15), and THESAN (z = 10) are illustrated as yellow, cyan, and magenta solid lines, respectively. Additionally, results from galaxies in SERRA (z = 7.7) and FIRE-2 ($9 \leq z \leq 10$) are marked with lime greened x symbols and red stars, respectively. Findings from analytic models and formulae based on low-$z$ observations are also shown as dashed lines.

\par

Generally, we find that our simulated results align well with the overall HMSM trends suggested in other simulation projects and analytic results with the exception of those proposed by \citet{Harikane2022}. Although our results tend to exhibit higher stellar masses within a similar halo mass range compared to the median values from TNG50 and THESAN, they still show a strong match with the HMSM trends from zoom-in simulations such as SERRA and FIRE-2. However, when compared to another zoom-in simulation, the Renaissance simulation, our results show a smaller stellar mass within the same halo mass range. This difference arises because the median results of the Renaissance simulation are derived from a rare peak region, specifically selected to explore overdense areas.

\par
\par

\subsubsection{Star formation rate}
\label{SFR}
In this subsection, we discuss the star formation trends and histories along with the SFR of our simulation sets. Figure \ref{fig:SFR} presents the SFR of our simulated galaxies as a function of cosmic time, calculated within a time bin of $\Delta t = 5\, {\rm Myr}$. The top row displays the SFR for the HM9 sets, while the bottom one shows the SFR for the HM10 sets. The colored lines indicate the same sets as in Figure \ref{fig:mass evolution}. However, for better visibility, we vary the symbols in each time bin such that circles represent the Chabrier IMF adopted sets, and diamonds depict the top-heavy IMF adopted sets. Additionally, we separately display the same SFR of the simulation sets, showing the total SFR histories for the redshift range $9 \leq z \leq 25$ (left panels) and a zoomed-in view for the redshift range $9 \leq z \leq 14$ (right panels), during which the halos of the target galaxies exceed $M_{\rm vir}=10^9 \msun$.

\par
\begin{figure*}
    \centering
    \includegraphics[width = 170mm]{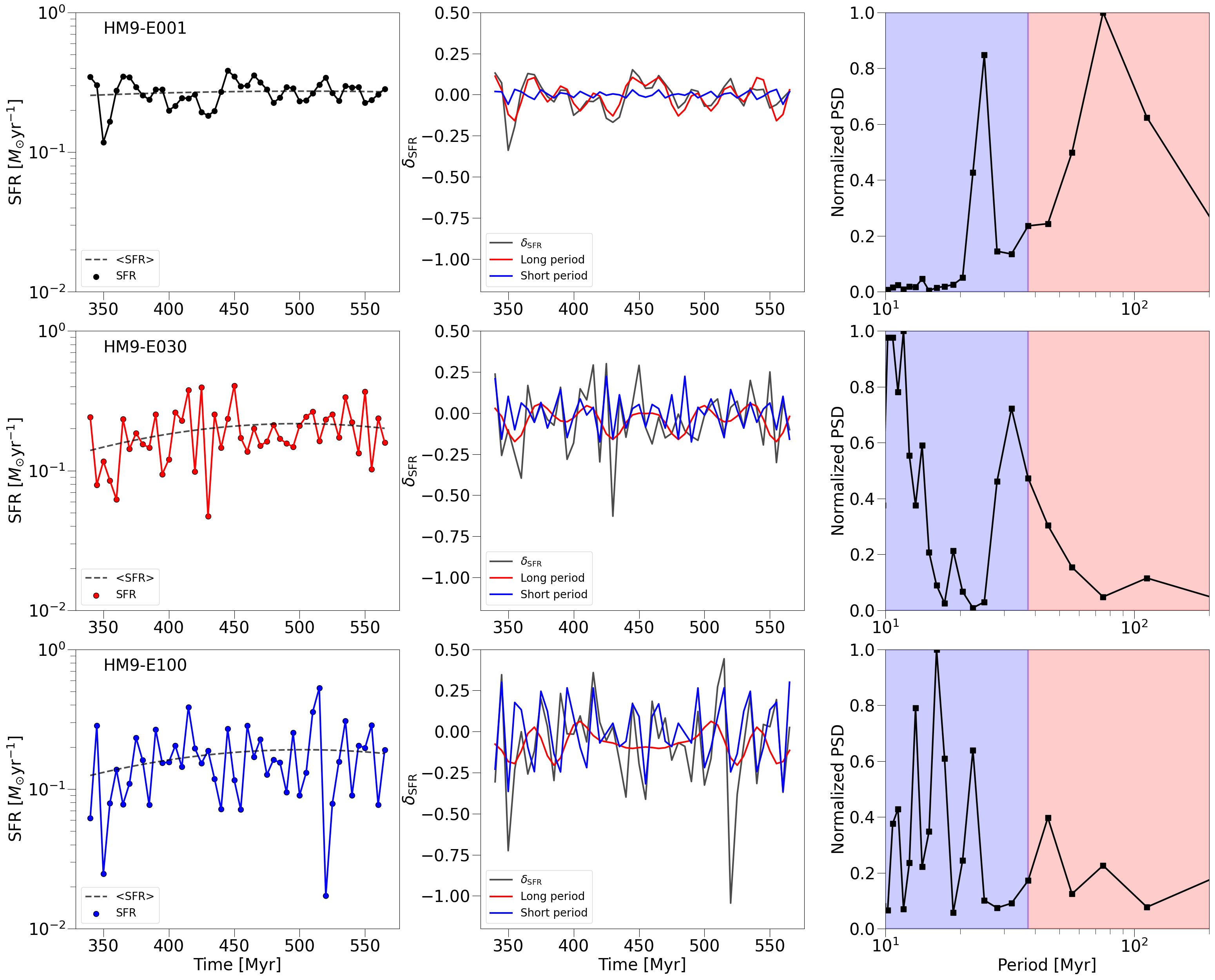}
    \caption{Periodicity of SFR within $9 \leq z \leq 13$ computed based on a time bin of $\Delta t=$ 5\,Myr. Rows show the results from each simulation, HM9-E001 (top), HM9-E030 (middle) and HM9-E100 (bottom), respectively. The panels in the left column display SFR, using the same colors as in Figure \ref{fig:SFR}, and the average SFR, <SFR>, denoted by the dashed line. The middle column panels show $\delta_{\rm SFR}$ (black solid), which can be separated into short-period (blue solid, $\tau_{\rm short} < 37.5$ Myr) and long-period (red solid, $\tau_{\rm long} > 37.5$ Myr) components. The right column panels present the normalized PSD as a function of the period, with short periods and long periods shaded in blue and red, respectively.}
    \label{fig:5myr periodicity}
\end{figure*}
\par
As shown in Figure \ref{fig:SFR}, the simulated galaxy in the HM9-E001 set experiences fluctuations in its SFR until $z \geq 13$ due to its susceptibility to stellar feedback, which is more pronounced in a relatively shallow potential well. As the halo mass increases, the central gas clouds become more resilient to stellar feedback, allowing for sustained star formation. This results in continuous star formation with small fluctuations in SFR, ranging from $\rm 0.2 \msun yr^{-1}$ to $\rm 0.4 \msun  yr^{-1}$. This period of continuous star formation persists until $z \approx 9$, at which point the simulation ends. However, the $\epsilon_{\rm ff}$-increased sets exhibit distinct star formation patterns compared to the HM9-E001 set. Owing to their highly efficient star formation over short timescales, they show bursty star formation behavior. During these bursts, the intense stellar feedback generated leads to longer quenching periods of star formation than observed in the HM9-E001 set. These cycles of bursty star formation and subsequent quenching repeat after $z \leq 13$, characterized as episodic star formation with specific periodicities.

\par
To investigate the periodicity of episodic trends in SFRs, we adopt the procedure suggested by \citet{Pallottini2023}. In this approach, the SFRs are normalized by the average SFR, <SFR>, as follows,
\begin{equation}
\label{eq5}
    \delta_{\rm SFR} \equiv \log_{10} \frac{\rm SFR}{<\rm SFR>},
\end{equation}
where the average SFR is expressed as a polynomial fit in log space,
\begin{equation}
\label{eq4}
    \log_{10} <\rm SFR/ \msun yr^{-1}> \equiv \sum_{n=0}^{2} \it p_n \Big( \frac{t}{\rm Myr} \Big)^{\it n},
\end{equation}
with $p_{\rm n}$ representing the coefficients for the $n$th-order polynomial terms. This method allows us to avoid the obscuration of small periodic patterns by the overall increasing trend in the total SFR evolution. To determine the periodicity of SFR across the redshift range of $9 \leq z \leq 13$, we compute the power spectral density (PSD) of $\delta_{\rm SFR}$ using Welch's method (\citealp{Welch1967}), implemented in the \texttt{scipy.signal.welch} module.
\par

Figure \ref{fig:5myr periodicity} displays the SFRs and $\delta_{\rm SFR}$ over cosmic time, along with the normalized PSD of $\delta_{\rm SFR}$ for each simulation set. Specifically, from left to right, the columns show the SFR and <SFR> (left), the original value of $\delta_{\rm SFR}$ separated into short-period (red solid line, $\tau_{\rm short} < 37.5 \rm Myr$) and long-period (blue solid line, $\tau_{\rm long} > 37.5 \rm Myr$) components (middle), and the derived PSD of $\delta_{\rm SFR}$ (right). For separating the short-period and long-period components of $\delta_{\rm SFR}$, we transform $\delta_{\rm SFR}$ into PSD form using the Fast Fourier Transform in  \texttt{scipy.fftpack}. Then we apply a cutoff window to the period components higher (lower) than $\tau_{\rm thr} = 37.5 \rm \,Myr$, and use the inverse transform module \texttt{scipy.fftpack} to extract the short (long)-period $\delta_{\rm SFR}$.

Each row corresponds to results from different sets, HM9-E001 (top), HM9-E030 (middle), and HM9-E100 (bottom), respectively. In the default HM9-E001 set, $\delta_{\rm SFR}$ calculated with short-period windows tend to have lower amplitude, compared with long-period values, also showing that the maximum PSD of the HM9-E001 set is located at $\tau_{\rm max} \approx $ 75 Myr. For the $\epsilon_{\rm ff}$-increased sets, periods with the highest PSD values fall within 10 Myr $\leq \tau \leq$ 20 Myr, with maximum PSD periods located at $\tau_{\rm max, E030} \approx$ 11.8 Myr and $\tau_{\rm max, E100} \approx$ 16.1 Myr, respectively. This implies that a more episodic star-forming trend generates shorter periodic fluctuations. This is because bursty star formation in the $\epsilon_{\rm ff}$-increased sets, accompanied by strong SN feedback, delays subsequent star formation as the gas takes time to recover, resulting in periodicity (e.g., \citealp{Martin-Alvarez2023, Dome2024}). In contrast, the HM9-E001 set, where bursty star formation is less pronounced, exhibits continuous star formation, leading to a lack of short periodicity.

\par
To quantify the degree of burstiness when comparing the two $\epsilon_{\rm ff}$-increased sets, HM9-E030 and HM9-E100, we introduce the burstiness parameter proposed by \citet{Applebaum2019} as follows,
\begin{equation}
\label{eq6}
    \rm B = \frac{\sigma/\mu-1}{\sigma/\mu+1},
\end{equation}
where $\sigma$ is the standard deviation of SFR and $\mu$ is the mean SFR, which had similarly been used for confirming the burstiness of SFR in \citet{Caplar2019} and \cite{Kang2024}.
Using the definition of the burstiness parameter, $\rm -1 \leq B \leq 1$, lower values for $\rm B$ indicate a more uniform distribution. The calculated burstiness parameters for each set in the redshift range $9 \leq z \leq 13$ are $\rm B_{E001} \approx -0.67$, $\rm B_{E030} \approx -0.39$ and $\rm B_{E100} \approx -0.27$. From the results of calculated $\rm B$, we can conclude that the HM9-E100 set tends to exhibit more burstiness in SFR than the HM9-E030 set but the variation in burstiness between these two sets is not as significant compared to the HM9-E001 set.

\par
Next, we focus on the top-heavy IMF-adopted sets, HM9-T001 and HM9-T030. As shown in Figure \ref{fig:SFR}, the SFRs in these sets are almost 1 dex lower than in the Chabrier IMF-adopted sets at $z \gtrsim 10.5$. These lower SFR trends are due to the more frequent occurrence of massive stars in each star particle when adopting a top-heavy IMF. These massive stars are more likely to end their lives in SN explosions, providing strong stellar feedback to nearby gas clouds and effectively blowing out the dense cold gas from the centers of galaxies.

\par
Although the stronger feedback from the top-heavy IMF initially results in a lower SFR, the galaxy's potential well becomes deeper, allowing it to withstand the stellar feedback. This leads to a slight increase in the SFR of the HM9-T001 set. Consequently, the difference in stellar masses between the top-heavy IMF and Chabrier IMF-adopted sets decreases to within a factor of 4-5. Interestingly, we observe a deeply star-forming quenched valley in the HM9-T001 set at $9.5 \leq z \leq 9.7$ (see, Figure \ref{fig:SFR}). Despite the reduced stellar mass due to strong feedback, star formation in the HM9-T001 set continues relatively steadily without significant fluctuations, except for the quenched valley. Intriguingly, bursty star formation behavior is most prominent in the increased $\epsilon_{\rm ff}$ set while adopting a top-heavy IMF (HM9-T030), illustrated as the orange lines in Figure \ref{fig:SFR}. The resultant SFR varies from a few 0.1 $\Msunyri$ to 0.01 $\Msunyri$, displaying a notable short periodicity on the order of $\lesssim$ 15 Myr.

\par
Finally, the galaxy in the HM10-E001 set tends to exhibit similarly continuous and almost constant star formation trends until $z \sim 10$, compared to the HM9-E001 set (see Figure \ref{fig:SFR}). After $z \sim 10$, the SFR increases and reaches values of $\rm SFR \geq 1 \msun yr^{-1}$, which is an order of magnitude higher than those in the HM9 sets. This increasing star formation trend continues until the end of the simulation. The difference between the SFRs of the HM10-E001 and HM10-E030 sets mirrors the pattern observed in the HM9-E001 and HM9-E030 sets, with more bursty star formation occurring in the $\epsilon_{\rm ff}$-increased sets. Despite the episodic star formation trends, the HM10-E030 set generally does not surpass the SFR of the HM10-E001 set in the range $10 \lesssim z \lesssim 12$. However, during the period of increased star formation ($z \lesssim 10$), the SFRs in the HM10-E030 set fluctuate, with a median value that coincides with the SFR of the HM10-E001 set.
\par

\subsection{Post-processing results}
\label{post-processing results}
\subsubsection{JWST observability}
\par
\begin{figure*}
    \centering
    \includegraphics[width = 170mm]{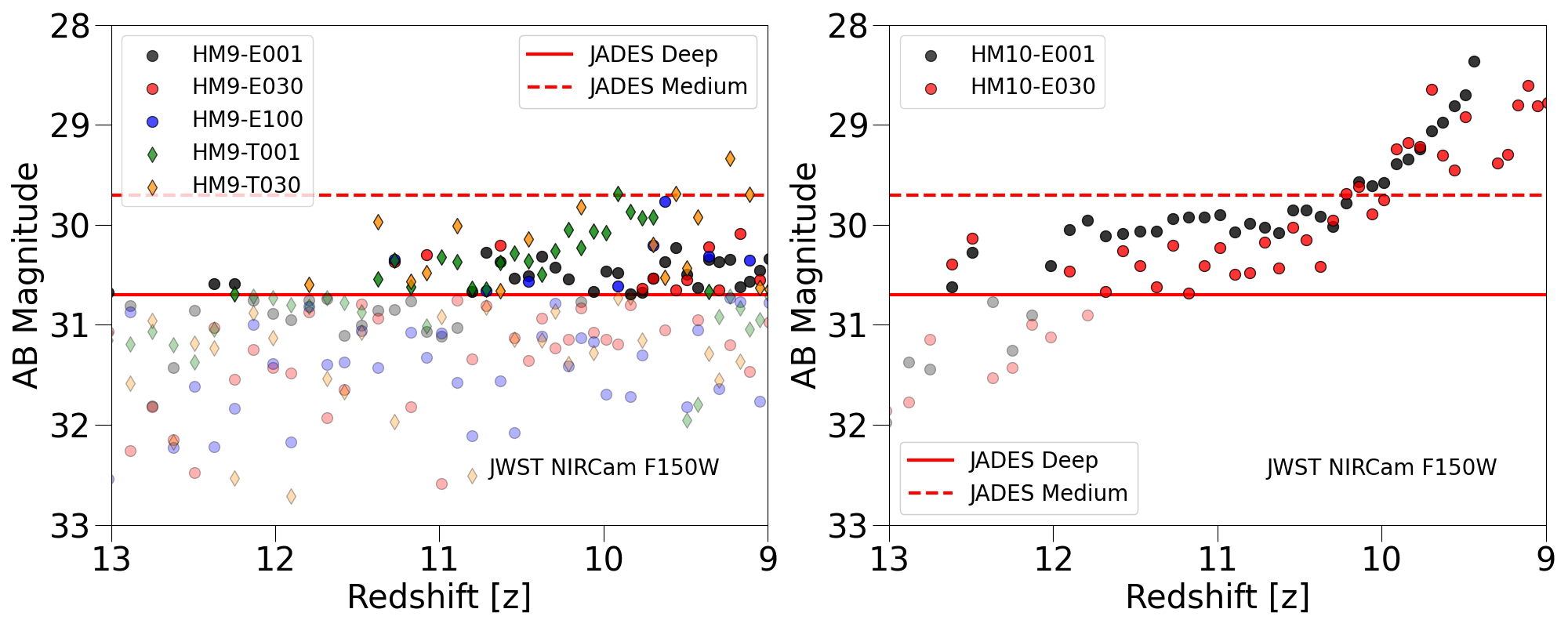}
    \caption{The expected AB magnitudes in the JWST NIRCam F150W filter as a function of redshift are derived using a straightforward relation between UV luminosity and SFR. The colors of the markers match those used in Figure \ref{fig:SFR}. For comparison, the red solid and dashed lines indicate the $5\sigma$ source limiting magnitudes for the JADES Deep and JADES Medium surveys in the JWST NIRCam F150W filter, respectively, as reported by \citet{Williams2018}. AB magnitudes that fall below the limiting magnitude of the JADES Deep survey are shown with reduced opacities.}
    \label{fig:mag by SFR}
\end{figure*}
\par
Before estimating the AB magnitude and observabilities of each set using our post-processing pipelines, we first calculate the expected AB magnitude based on the simple relation between UV magnitude and SFR. By modifying the expression from \citet{Inayoshi2022} (equ.~2), which converts SFR into UV luminosity, we transform our SFR from each time bin to UV flux at $\lambda_{\rm UV} = 1500 \Ang$, with units of $\rm erg \, s^{-1}\, Hz^{-1} \,cm^{-2}$, as follows,
\begin{equation}
\label{SFR to UV luminosity}
    f_{\rm UV} = \frac{1}{4 \pi d_{\rm L}^2}L_{\rm UV} = \frac{1}{4 \pi d_{\rm L}^2}\eta_{\rm UV} \rm SFR,
\end{equation}
and then convert the calculated UV flux into the AB magnitude. For a detailed description of parameters in Eq. (\ref{SFR to UV luminosity}), $d_{\rm L}$ is the luminosity distance which can be calculated from the redshift for a given simulation snapshot, $\eta_{\rm UV}$ is the conversion factor for SFR to UV luminosity, and $\rm SFR$ represents the SFRs from the simulation sets in each time bin, as illustrated in Figure~\ref{fig:SFR}.
\par

\par
For the conversion factors $\eta_{\rm UV}$, we adopt different values respectively for the Chabrier IMF and top-heavy IMF.
For the Chabrier IMF, we choose $\eta_{\rm UV} = 1 / K_{\rm UV}$, where $K_{\rm UV}$ has a value of $K_{\rm UV} = 10^{-28} \rm \msun yr^{-1}/(erg \, s^{-1} \, Hz^{-1})$ with a metallicity of $\log_{10} Z_{\star}/\Zsun =-1.0$ as determined by \citet{Madau2014}. This value was calculated using \fsps \nspace (\citealp{Conroy2010}), assuming a Salpeter IMF and constant SFR. For the top-heavy IMF, we utilize $\eta_{\rm UV} = 3.57 \times 10^{28} \rm erg \, s^{-1} \, Hz^{-1}/(\msun yr^{-1})$ derived from extreme top-heavy IMF values from \citet{Zackrisson2011}. We cautiously note that for these estimations, we use values only for $\lambda_{\rm UV} = 1500 \Ang$, and the functional IMFs of both values are not perfectly suited to our specific IMFs. As the observed wavelength depends on the redshift of the target galaxies, $\lambda_{\rm rest} = \lambda_{\rm obs}/(1+z)$, $f({1500 \Ang})$ may not accurately match the target wavelengths of the JWST NIRCam F150W filter at redshifts higher than $z \approx 10.12$. Also, dust attenuation is not included here, therefore these estimates represent the upper limit of observabilities for our simulated galaxies and we use these estimates solely for comparison with our post-processed results.

\par
Figure \ref{fig:mag by SFR} illustrates the computed AB magnitude as a function of redshift, derived from the straightforward relationship between SFR and UV luminosity. The left panel displays the results from the HM9 sets, while the right panel shows those from the HM10 sets. The colors of the symbols correspond to the same conventions as in Figure~\ref{fig:SFR}. The red solid and dashed lines represent the 5$\sigma$ point source limiting magnitudes for the JADES Deep ($m_{\rm AB}=30.7$) and Medium ($m_{\rm AB}=29.8$) surveys, respectively, as originally reported by \citet{Williams2018}. The estimated AB magnitudes that fall below the point source limiting magnitude of the JADES Deep survey are indicated with lower opacities.

\par
For the HM9 runs, at $z \lesssim 11$, all sets could surpass the limiting magnitude of the JADES Deep survey, despite detailed variations between different runs. The expected AB magnitudes of HM9-E001 at various redshifts remain nearly stable due to minor fluctuations in the SFR. In contrast, the AB magnitudes of the $\epsilon_{\rm ff}$-increased sets (HM9-E030 and HM9-E100) exhibit significant fluctuations, intermittently crossing the JADES Deep survey limiting magnitude, reflecting similar patterns observed in their SFR evolution. Interestingly, due to bursty star formation, where UV luminosity can be significantly enhanced during a starburst, the top-heavy IMF sets, HM9-T001 and HM9-T030, occasionally reach the limiting magnitude of the JADES Medium survey at $z \lesssim 10$, even with a low stellar mass ($M_{\star} \lesssim 10^7 \msun$). With higher SFR, the HM10 sets surpass the JADES Deep survey limiting magnitude at earlier epochs ($z \lesssim 12.5$). At $z \lesssim 10$, both sets experience increased star-forming periods, which proportionally boost UV luminosity and ultimately exceed the limiting magnitude of the JADES Medium survey.
\par
\par
\begin{figure*}
    \centering
    \includegraphics[width = 170mm]{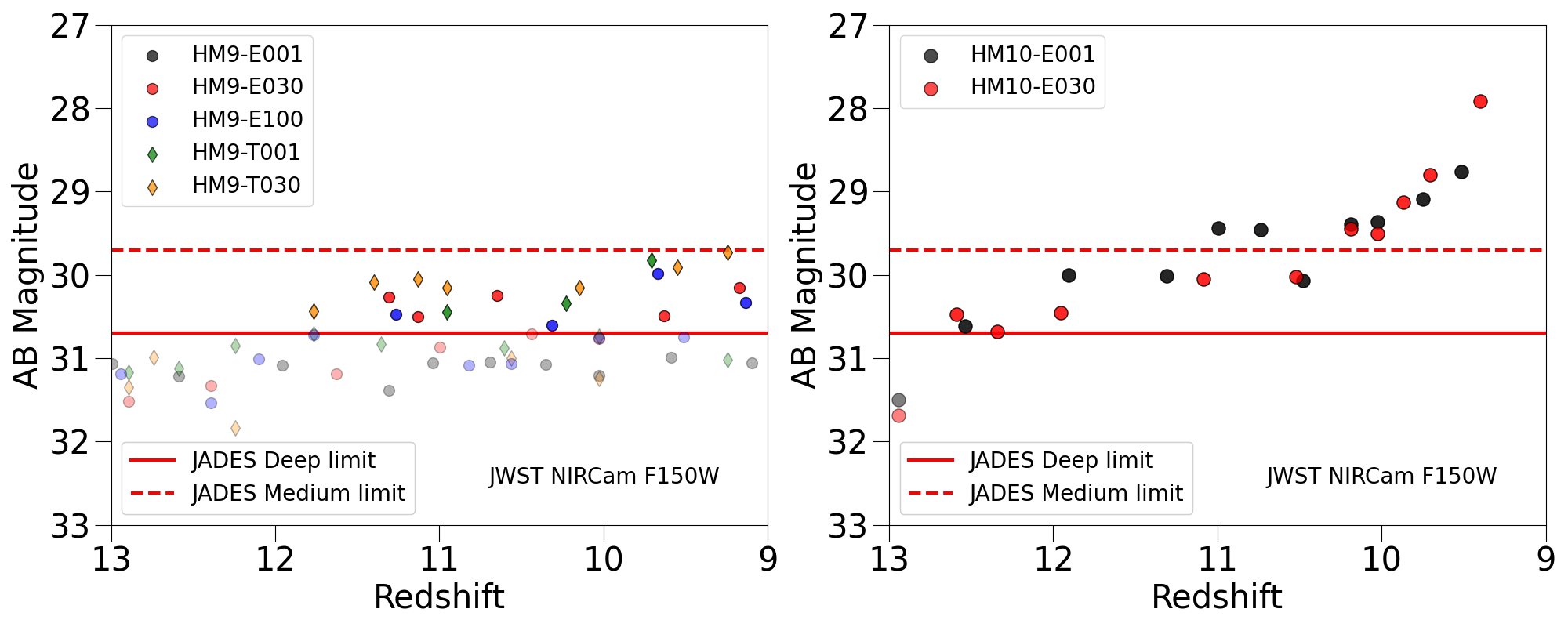}
    \caption{The derived AB magnitudes for the JWST NIRCam F150W filter as a function of redshift are shown. The symbols and lines correspond to those in Figure \ref{fig:mag by SFR}, but the results are calculated using post-processing pipelines. Similar trends are observed as in Figure \ref{fig:mag by SFR}, but with marginally lower magnitudes due to absorption and extinction from gas and dust particles. As same as Figure \ref{fig:mag by SFR}, AB magnitudes that fall below the limiting magnitude of the JADES Deep survey are shown with reduced opacities.}
    \label{fig:mag by post-processing}
\end{figure*}

\begin{figure}
    \centering
    \includegraphics[width = 85mm]{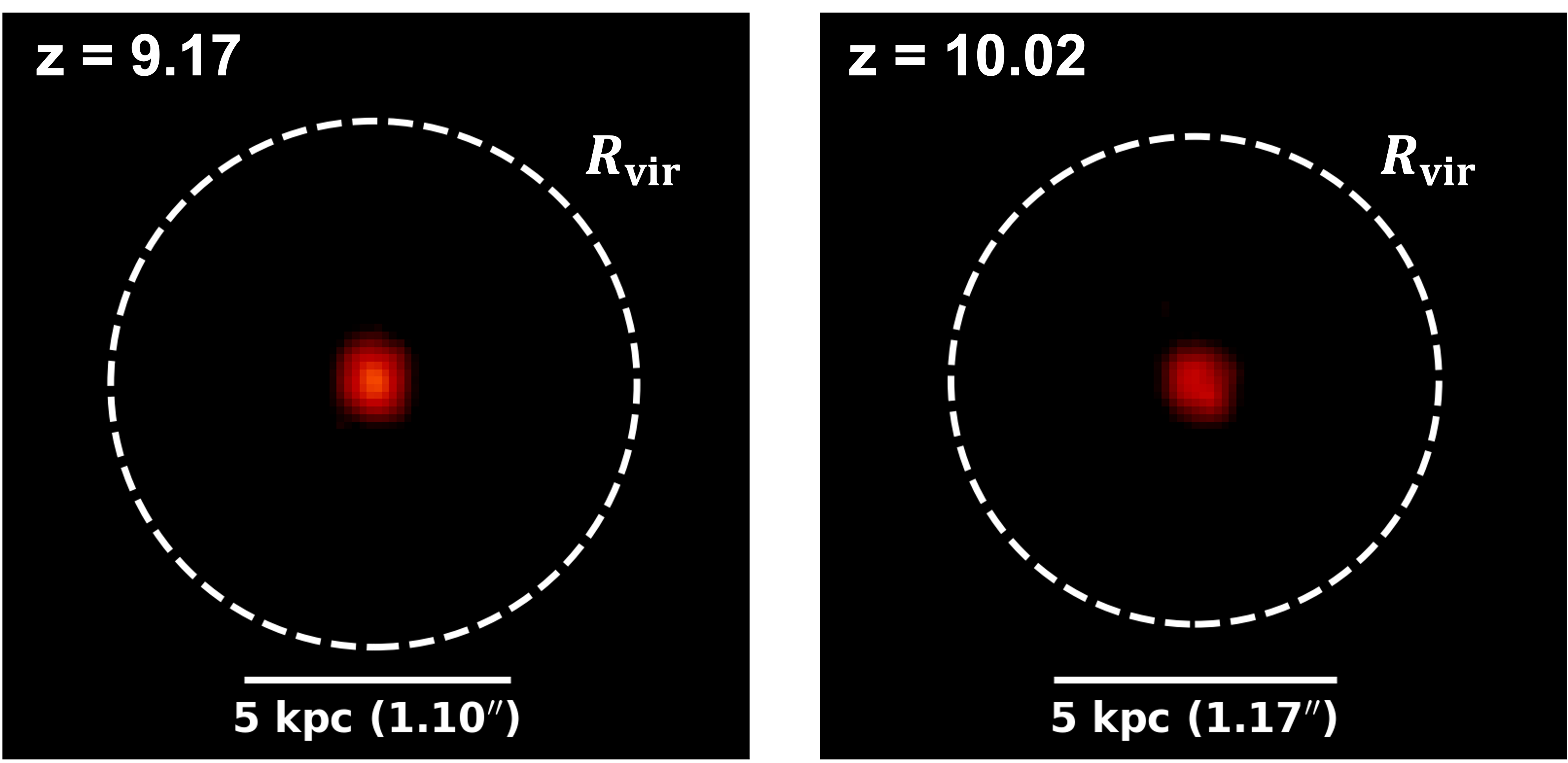}
    \caption{Noiseless mock images of a simulated galaxy from the HM9-E030 set in the JWST NIRCam F150W filter band. The resolution of the mock images is adjusted to match with the resolution of the JWST NIRCam F150W filter, with a pixel width of $W_{\rm pixel}=0.031^{\prime \prime}$. For better visibility, we illustrate a length unit of 5 physical kpc in arcsec. A Gaussian function with a Full Width at Half Maximum (FWHM) of 3 pixels is used for the point spread function. The left panel shows the galaxy when its AB magnitude exceeds the point source limiting magnitude of the JADES Deep surveys at $z \approx 9.17$, while the right panel depicts the galaxy at $z \approx 10$, where the AB magnitude is below the limiting magnitude. The dashed circles in both images indicate the virial radius of the halo.}
    \label{fig:Mock image}
\end{figure}
\par

On the other hand, AB magnitudes derived from our post-processing pipelines give similar trends but yield different values. Figure \ref{fig:mag by post-processing} presents the same plot as Figure \ref{fig:mag by SFR}, but with AB magnitude values estimated from our post-processing results. We note that due to the computational cost, we only post-process the snapshots which tend to have local minimum and maximum values for SFR. Also, in Figure \ref{fig:Mock image}, we illustrate example noiseless mock images from our simulation results for the HM9-E030 set under two scenarios: when the simulated galaxy is detectable (left, $z \approx 9.17$) and when it is undetectable (right, $z \approx 10$), compared to the limiting magnitude. The dashed circles in Figure \ref{fig:Mock image} indicate the virial radius of the halo at these redshifts.

\par
The overall trends of AB magnitude are similar to the results in Figure \ref{fig:mag by SFR}. However, unlike the results based on the simple relation between UV luminosity and SFR, the AB magnitudes of HM9-E001 do not exceed the limiting magnitude of the JADES Deep survey and instead exhibit a constant evolution until the end of the simulation. Nevertheless, the AB magnitudes from the $\epsilon_{\rm ff}$-increased sets can exceed the limiting magnitude at $z \lesssim 11$. These magnitudes tend to progress with a fluctuating trend, where the luminosity oscillates around the limiting magnitude of the JADES Deep survey. However, this magnitude fluctuation does not show a significant difference between HM9-E030 and HM9-E100, as their SFR evolution is not substantially different either.
\par

For the top-heavy IMF cases, the HM9-T030 set begins to surpass the limiting magnitude at $z \approx 11$. After $z \approx 11$, the magnitude evolution from post-processed results exhibits more significant fluctuations compared to the $\epsilon_{\rm ff}$-increased sets. At $z \lesssim 9.5$, we observe an extremely decreased value in magnitude evolution, which coincides with the deep star-forming quenched valley evident in the SFR evolution. Also, the AB magnitude evolution of HM9-T030 exceeds the limiting magnitude of the JADES Deep survey at $z \lesssim 11.7$. With fluctuations in its magnitude, the results from HM9-T030 tend to have the highest values. At $z \approx 9.25$, the AB magnitude of the simulated galaxy nearly reaches just below the limiting magnitude of the JADES Medium survey, making it the most luminous case among the composite HM9 sets.
\par

For the HM10 sets, the magnitudes of HM10-E001 exceed the limiting magnitude of the JADES Deep survey at $z \simeq 12.5$, which is earlier than the HM9 sets. Once the AB magnitude of the simulated galaxy in HM10-E001 surpasses the Deep limiting magnitude, it exhibits an increasing trend and eventually exceeds the Medium limiting magnitude between $11 \leq z \leq 11.3$. Following a brief fluctuation period in AB magnitudes from $10.08 \leq z \leq 10.73$, the UV luminosity of the simulated galaxy in the HM10-E001 set shows a rapidly increasing trend. Eventually, between $9.51 \leq z \leq 9.74$, the AB magnitude of the simulated galaxy surpasses $\sim$ 29 mag, which, we note, corresponds to the limiting magnitude of the CEERS survey (\citealp{Finkelstein2023}).
\begin{figure}
    \centering
    \includegraphics[width = 85mm]{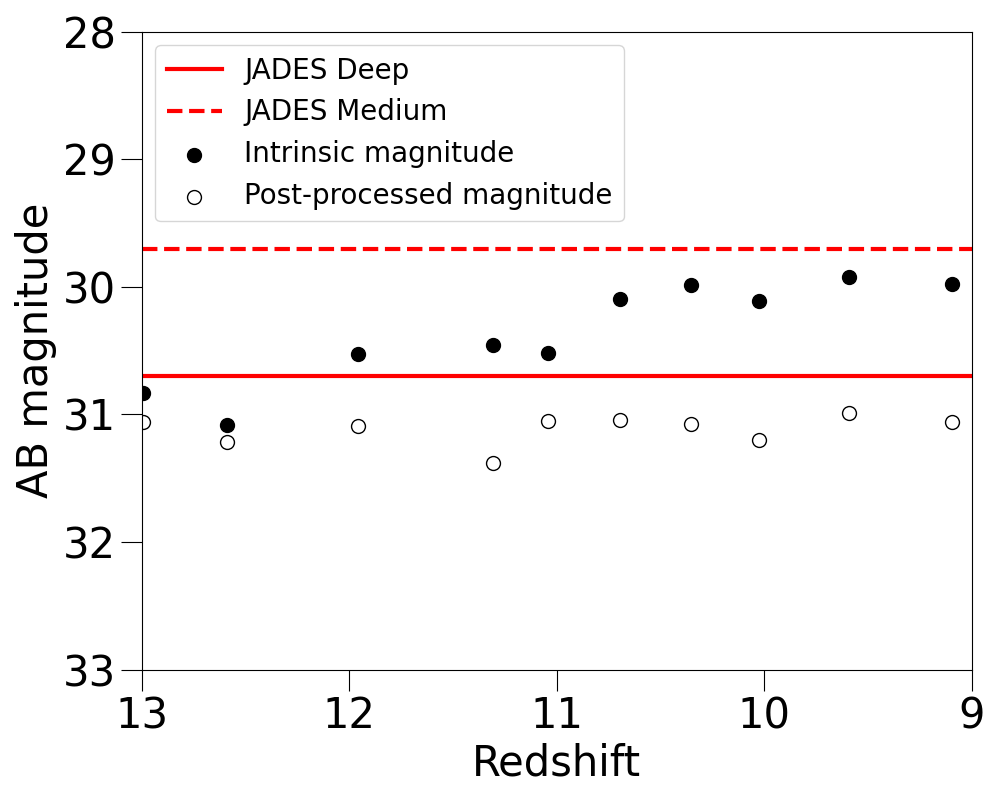}
    \caption{The dust attenuation effect is demonstrated by comparing the derived AB magnitudes of the HM9-E001 set for the JWST NIRCam F150W filter. The filled black circles represent the AB magnitude results from the intrinsic SED, while the open black circles indicate the AB magnitude from post-processed results. The red dashed and solid lines correspond to the source limiting magnitudes of the JADES Medium and Deep surveys, respectively. This comparison clearly shows that the AB magnitude can be significantly reduced by $\sim$ 1 magnitude due to the dust attenuation effect.}
    \label{fig:E001 intrinsic to final magnitude compare}
\end{figure}
\par

\par
The simulated galaxy in the HM10-E030 set exceeds the limiting magnitude of the JADES Deep survey at $z \approx 12.6$, which occurs slightly earlier compared to the HM10-E001 set due to more efficient star formation. However, the HM10-E030 set surpasses the limiting magnitude of the JADES Medium survey between $10.18 \leq z \leq 10.52$, which is later than the HM10-E001 set, owing to delayed star formation caused by previous bursty starbursts. During a period of intense star formation around $z \sim 9.74$, the AB magnitude of the HM10-E030 set surpasses the limiting magnitude of the CEERS survey earlier than the HM10-E001 set. Nevertheless, similar to the HM9 sets, both the HM10-E001 and HM10-E030 sets tend to show lower magnitudes than those predicted by the simple relation between SFR and UV luminosity.
\par

In summary, our simulations and post-processing results show that employing fiducial sub-grid physics, a standard IMF and moderate star formation efficiency reflective of the local Universe yields the lowest UV luminosity. However, when using alternative sub-grid physics, we find that these different scenarios can lead to significant variations in UV magnitudes, causing them to either increase dramatically or fluctuate, potentially surpassing the limiting magnitude of JWST surveys. It is important to note that while our simulations may represent the faintest high-redshift galaxies observed in JWST surveys, similar flickering or UV-boosting mechanisms could also occur in massive high-redshift galaxies, making them more detectable (e.g., \citealp{Ferrara2024}).
\par

\subsubsection{Dominant physical properties influencing the UV luminosity}
In this subsection, we explore the physical properties of galaxies that influence their observability. As detailed in Section \ref{SFR}, the HM9-E001 set shows almost constant SFRs with $0.2 \rm \msun yr^{-1} \leq SFR \leq 0.4 \msun \rm yr^{-1}$. This consistent SFR leads to the target galaxy in the HM9-E001 set becoming the most massive at redshifts $z \leq 13$ compared to other HM9 sets. However, our post-processing results, illustrated in Figure \ref{fig:mag by post-processing}, indicate that the HM9-E001 set is not detectable at redshifts $z \geq 9$. These contrasting findings from our simulations and post-processing raise the question of why the HM9-E001 set is unable to surpass the JWST survey's limiting magnitude despite having the highest stellar mass and average SFR, with young stars being the primary sources of UV luminosity.
\par
\begin{figure}
    \centering
    \includegraphics[width = 85mm]{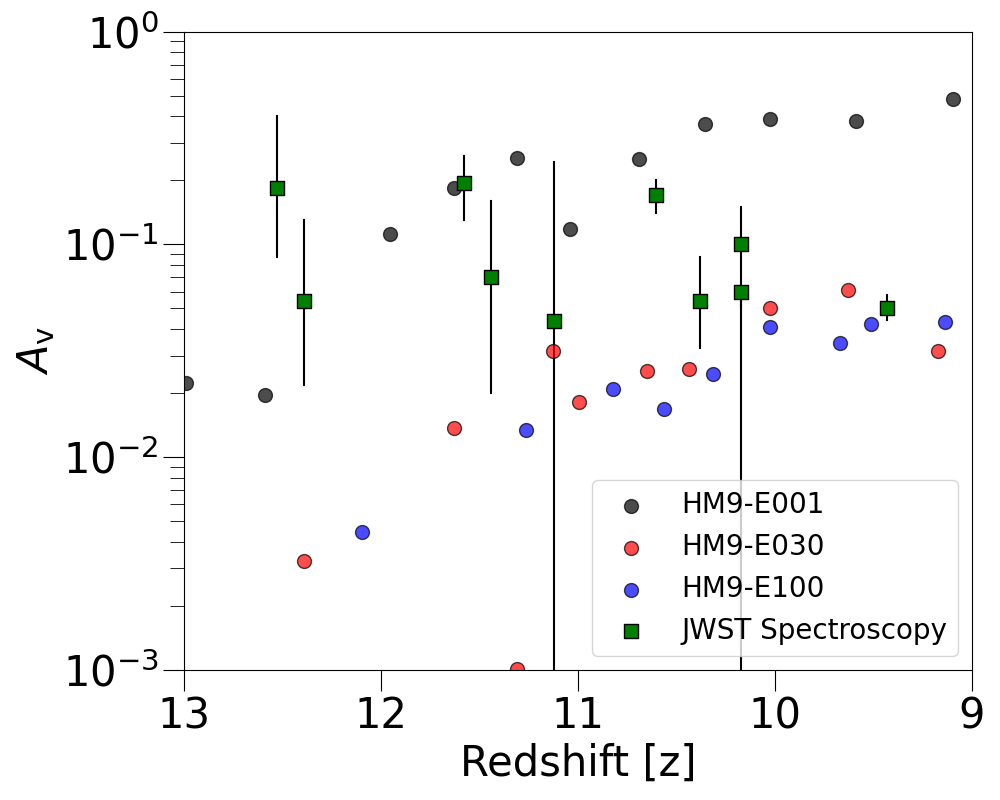}
    \caption{Evolution of dust attenuation ($A_{\rm v}$) as a function of redshift. The circled symbols represent $A_{\rm v}$ results from the HM9 set using the Chabrier IMF, while $A_{\rm v}$ results from JWST spectroscopic surveys (\citealp{Hainline2024, Bunker2023, Hsiao2023, Curtis-Lake2023, Curti2024, Wang2023}) are shown as green squares, with their $1\sigma$ errors indicated by black solid lines. Although the $A_{\rm v}$ values from the HM9-E001 run closely match the upper range of $A_{\rm v}$, as observed in the JWST surveys, the values for sets with increased $\epsilon_{\rm ff}$ differ by nearly an order of magnitude compared to the JWST data.}
    \label{fig:A_v}
\end{figure}
\par

\par
\begin{figure}
    \centering
    \includegraphics[width = 85mm]{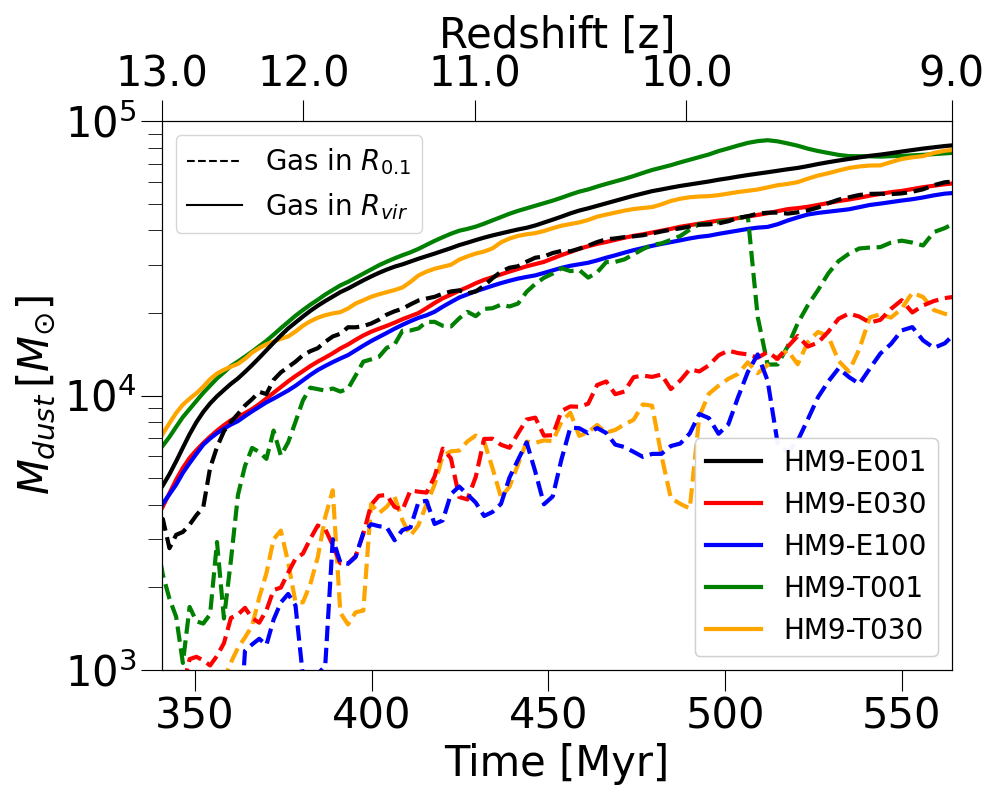}
    \caption{The evolution of dust mass as a function of cosmic time is depicted. The line colors correspond to those in Figure \ref{fig:mass evolution}. Solid lines represent the total dust mass within $R_{\rm vir}$, while dashed lines indicate the dust mass within $0.1 R_{\rm vir}$. The evolution of dust mass within $0.1 R_{\rm vir}$ shows fluctuations that tend to align with the periodicities of the SFRs.}
    \label{fig:dust mass evolution}
\end{figure}
\par
To figure out the mechanisms that could reduce UV luminosity, we examine the dust evolution and their impact on UV luminosity in our simulated galaxies. Figure \ref{fig:E001 intrinsic to final magnitude compare} presents the derived AB magnitude of the HM9-E001 set for the JWST NIRCam F150W filter as a function of redshift. The filled and open black circles represent the AB magnitude with and without the dust attenuation effect, respectively. As clearly shown in Figure \ref{fig:E001 intrinsic to final magnitude compare}, during $z \lesssim 11.5$, the AB magnitude results that account for dust attenuation are about 1 magnitude lower than those without the effect. This indicates that dust extinction reduces the AB magnitude of the simulated galaxies, pushing them below the limiting magnitude of JWST. We find that the impact of dust attenuation is most pronounced in the HM9-E001 set compared to other simulation sets. Thus, despite having the highest average SFRs, the galaxy in the HM9-E001 set experiences the strongest extinction, significantly reducing its AB magnitude.

\par
The degree of dust attenuation in the simulated galaxies tend to be higher than the values reported by \citet{Jaacks2018}, who found a maximum value of $A_{\rm V, max} \approx 0.3$. However, their study only considered baseline enrichment from Pop III star formation and excluded metal enrichment from Pop II~stars, thus setting a lower limit for dust attenuation. Another possible explanation for the discrepancy could be the absence of photoionization heating from Pop~II stars in our model. Otherwise, the feedback mechanisms might disperse the metals more effectively, reducing the dust extinction effect. Moreover, differences in dust distribution within our simulated galaxies and the assumptions in dust physics used during post-processing, which will be discussed in Section \ref{assumptions and limitations}, could also contribute to the variation in dust attenuation effects.

\par
To assess the effect of dust, we calculate the $A_{\rm v}$ values for the HM9 sets that use the Chabrier IMF (specifically, HM9-E001, HM9-E030, and HM9-E100) and compare those values with estimates from JWST spectroscopic surveys (\citealp{Hainline2024, Bunker2023, Hsiao2023, Curtis-Lake2023, Curti2024, Wang2023}), as depicted in Figure \ref{fig:A_v}. As shown in the figure, the $A_{\rm v}$ values for the HM9-E001 run are among the highest observed in the JWST surveys. On the other hand, the sets with increased $\epsilon_{\rm ff}$ exhibit $A_{\rm v}$ values nearly an order of magnitude lower than those of the HM9-E001 set, aligning more closely with the lower limits of the JWST survey results. The findings from Figures \ref{fig:E001 intrinsic to final magnitude compare} and \ref{fig:A_v} indicate that variations in dust evolution and physical properties across each set significantly influence their observability.

\par
\begin{figure*}
    \centering
    \includegraphics[width = 170mm]{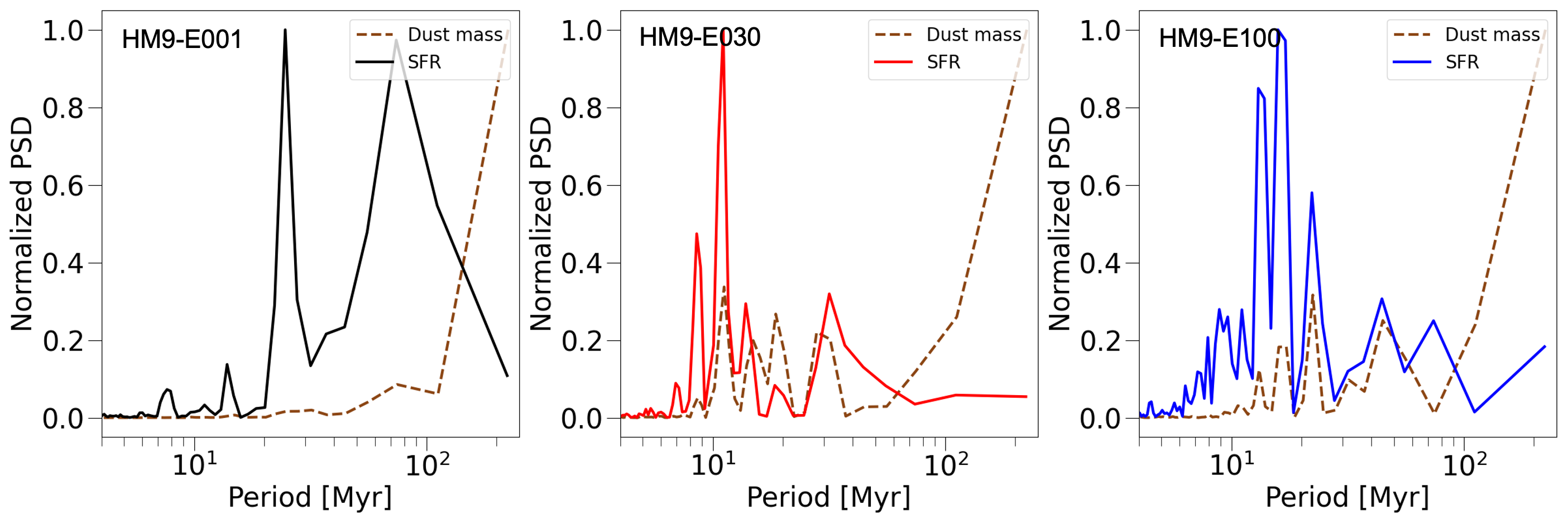}
    \caption{Periodicity comparison between dust mass and SFR evolution in different simulation sets, HM9-E001 (left), HM9-E030 (middle) and HM9-E100 (right) set, respectively. Each colored solid line represents the normalized PSD of $\delta_{\rm SFR}$ as a function of periods, while the dashed brown lines indicate the normalized PSD of $\delta_{\rm dust}$. We find that in the $\epsilon_{\rm ff}$-increased sets, the SFR and dust mass evolution appear to be correlated in the short period range ($\rm 10 Myr \leq \tau_{\rm dust} \leq 100 Myr$), as the peaks of the normalized PSD of $\delta_{\rm SFR}$ and $\delta_{\rm dust}$ show good agreement.}
    \label{fig:Periodicity dust and SFR}
\end{figure*}
\par
\begin{figure*}
    \centering
    \includegraphics[width = 170mm]{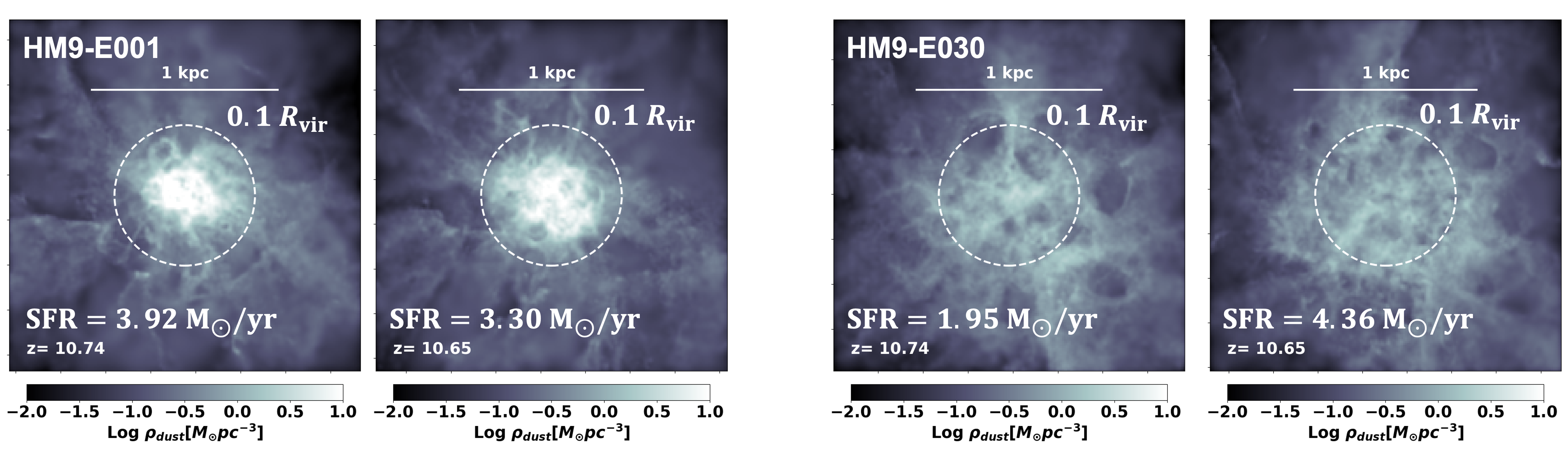}
    \caption{The central region dust density projection of a halo. The two left panels show the dust density for the HM9-E001 set at $z = 10.74$ (left) and $z = 10.65$ (right). The right panels display the dust density for the HM9-E030 set at the same redshifts. The dashed circles represent the radius of $0.1 R_{\rm vir}$ of the halos. In the HM9-E001 set, dust tends to accumulate in the central region despite stellar feedback from continuous star formation. In contrast, the HM9-E030 set exhibits decreased dust density due to ejection advected with the thermally-heated gas by intense stellar feedback from bursty star formation in the efficiency-enhanced sets.}
    \label{fig:Dust morphology}
\end{figure*}
\par
\begin{figure}
    \centering
    \includegraphics[width = 85mm]{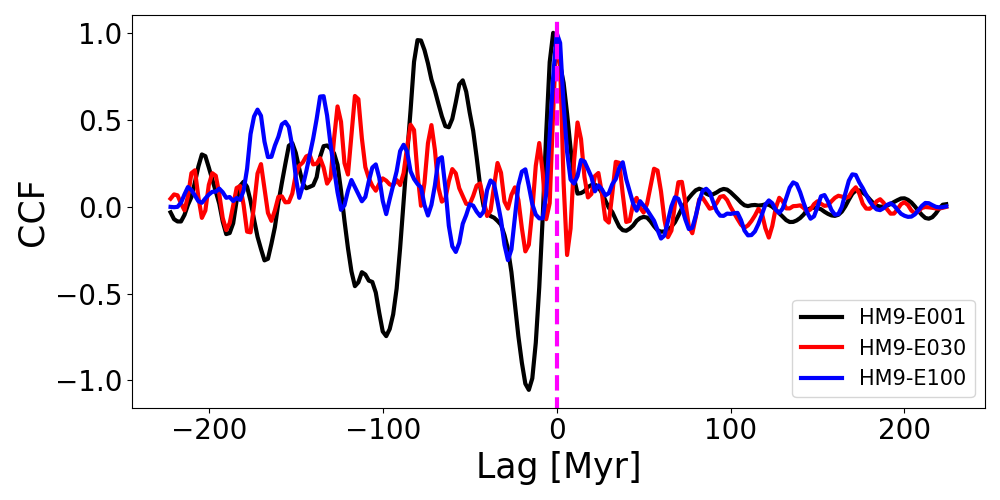}
    \caption{Cross-correlation function (CCF) results between $\delta_{\rm dust}$ and $\delta_{\rm SFR}$ for each set. The color of each line corresponds to the same simulation set as depicted in Fig.~\ref{fig:mass evolution}. Due to the insufficient masking of $\delta_{\rm dust}$, the CCF results exhibit fluctuations, with the HM9-E001 set showing the most significant fluctuations. Nevertheless, the maximum CCF values for each set occur at a lag of 0 (indicated by the dashed magenta line), implying that $\delta_{\rm dust}$ and $\delta_{\rm SFR}$ evolve synchronously in the same time series for each set.}
    \label{fig:delta cross-correlation}
\end{figure}

\par
To understand the detailed physical properties and evolution of dust in our composite sets, we first examine the evolution of dust mass in the simulated galaxies over cosmic time, as shown in Figure \ref{fig:dust mass evolution}. The dust mass is calculated within $R_{\rm vir}$ (solid line) and $0.1 R_{\rm vir}$ (dashed line), representing the total dust mass within the halos and near the star-forming regions, respectively. As depicted in Figure \ref{fig:dust mass evolution}, the HM9-E001 set among the Chabrier IMF sets exhibits the highest dust mass within $R_{\rm vir}$. These findings are consistent with the study by \citet{Tsuna2023}, which investigated dust evolution to explain high-redshift galaxies observed by the JWST using the semi-analytic code {\sc a-sloth} (\citealp{Hartwig2022}). They suggested that constant star-forming behavior likely leads to continuous dust accumulation inside galaxies due to moderate feedback effects. Otherwise, dust is easily ejected into the IGM. Moreover, the HM9-E001 set tends to have a significantly higher dust mass concentrated within the central region of the halo ($\leq 0.1 R_{\rm vir}$), larger by an order of magnitude compared to the $\epsilon_{\rm ff}$-increased sets. The $\epsilon_{\rm ff}$-increased sets exhibit lower dust mass in the central regions of halos, attributed to evacuation by bursty stellar feedback from episodic star-forming activities.

\par
Given that the evolution of dust mass within $0.1 R_{\rm vir}$ shows fluctuations that tend to align with the periodicities of the SFRs (see Figure \ref{fig:dust mass evolution}), we aim to confirm the interplay between these two quantities. To do this, we reconstruct the evolution of dust mass and SFRs with a time bin resolution of $\Delta t = 2$\,Myr for each set. We then calculate periodicity using the same methods described in Section \ref{SFR}, adopting the same parameters in Equations \ref{eq5} and \ref{eq4} for the dust mass, $<M_{\rm dust}>$ and $\delta_{\rm dust}$. This approach helps us avoid confusion with the global increasing trends seen in SFR and dust mass evolution. To confirm $\delta_{\rm dust}$, we also perform a KS-test between $M_{\rm dust}$ and $<M_{\rm dust}>$, yielding significant p-values of $p_{\rm E001} \approx 0.87$, $p_{\rm E030} \approx 0.87$, and $p_{\rm E100} \approx 0.21$. These results are consistent with the null hypothesis that both samples are drawn from the same distributions. While reconstructing the SFR and dust mass evolution, we focus on the interval from $9 \leq z \leq 13$, where we confirm that the simulated galaxies are detectable.

\par
Figure \ref{fig:Periodicity dust and SFR} presents the periodicity results of the fluctuations in SFR (solid lines) and dust mass (dashed lines) evolution, using the values of $\delta_{\rm SFR}$ and $\delta_{\rm dust}$. Each panel shows the resultant normalized PSD as a function of periods for HM9-E001 (left), HM9-E030 (middle), and HM9-E100 (right). The solid lines represent the PSD of $\delta_{\rm SFR}$, while the dashed brown lines represent the PSD of $\delta_{\rm dust}$. The colors of the solid lines correspond to the set information, consistent with those in Figure \ref{fig:mass evolution}. In all three sets, the periods with the maximum PSDs for $\delta_{\rm dust}$ are located in the long period region, corresponding to $\tau_{\rm dust} \geq 100\, \rm Myr$. This indicates that even using $\delta_{\rm dust}$, we are unable to perfectly mask the general increasing trend of dust mass. Considering the short period range (10 Myr $\leq \tau_{\rm dust}\leq$ 100 Myr), we can detect local peaks for $\tau_{\rm dust}$ in the $\epsilon_{\rm ff}$-increased sets that match with the peaks of the PSD of $\delta_{\rm SFR}$, suggesting that the SFR and dust mass evolution in the $\epsilon_{\rm ff}$-increased sets, in the short period range, appears to be correlated. In contrast, the PSD of $\delta_{\rm dust}$ in the HM9-E001 set shows shallow peaks even in the short period range, with normalized PSD values of these local peaks being very quiescent and negligible. This results from the continuous star-forming behavior observed in the HM9-E001 set, which lacks small fluctuations. Therefore, we suggest that fluctuations in dust mass evolution and SFR are related in the $\epsilon_{\rm ff}$-increased sets, likely due to dust outflows driven by short, intense stellar feedback resulting from bursty star-forming behavior.

\par
\begin{figure}
    \centering
    \includegraphics[width = 85mm]{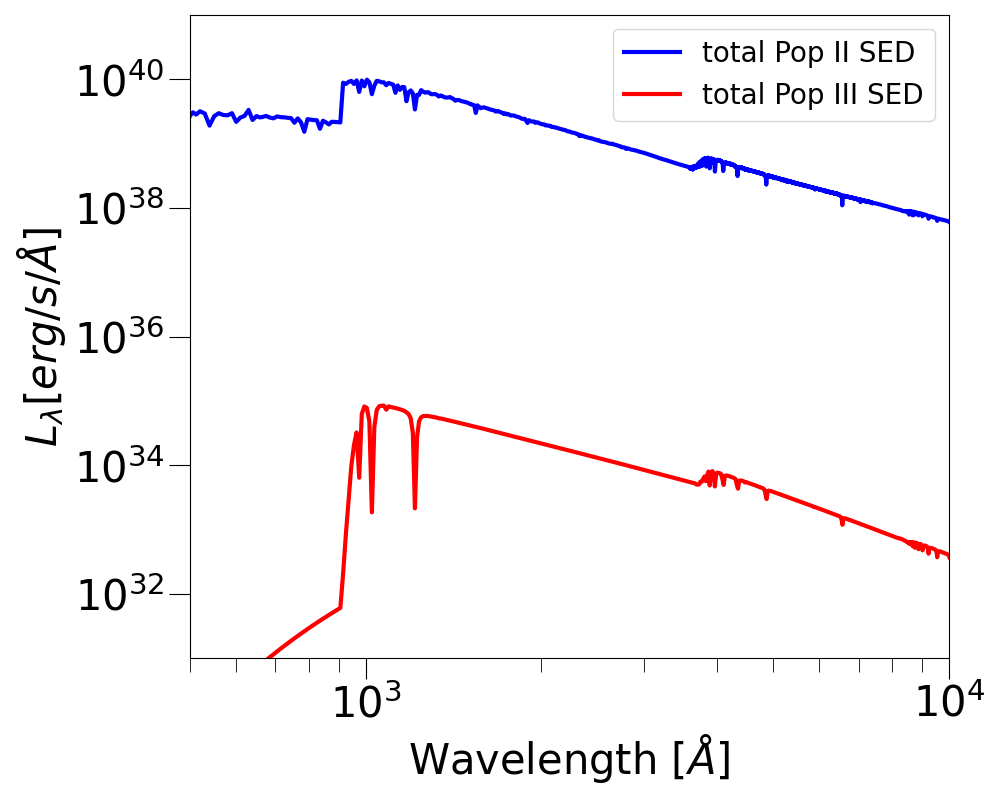}
    \caption{Intrinsic SED of total Pop~III and Pop~II stars, shown before post-processing, in the galaxy from the HM9-E030 set at $z = 9.63$. The red solid line represents the SED of total Pop III stars, while the blue line shows the SED of total Pop II stars in the galaxy. When the simulated galaxy surpasses the detection limits of JWST, we find that the luminosity of total Pop III stars is almost 5 orders of magnitude lower than that of the Pop II stars, contributing insignificantly to the total SED of the galaxy.}
    \label{fig:flux compare between pop2 stars and pop3 stars}
\end{figure}
\par
\par
To confirm that dust is driven out of the central region due to strong stellar feedback in the $\epsilon_{\rm ff}$-increased sets, Figure \ref{fig:Dust morphology} illustrates the projection plots of dust density in the center of halos. The two plots on the left side correspond to the HM9-E001 set at $z = 10.74$ (left) and $z = 10.65$ (right), while the two plots on the right side depict the dust density of the enhanced efficiency run, HM9-E030, at the same redshifts. Note that the dashed circles represent the radius of $0.1 R_{\rm vir}$ of the halos, and the SFRs are indicated at the corresponding redshifts. As seen in Figure \ref{fig:Dust morphology}, the HM9-E001 set shows dust accumulation in the center of the halo, while the $\epsilon_{\rm ff}$-increased sets exhibit reduced dust density due to outflows from bursty star formation.
\par

In addition, to examine the similarities between $\delta_{\rm dust}$ and $\delta_{\rm SFR}$, we calculate their cross-correlation using \texttt{scipy.signal.correlate}. Figure \ref{fig:delta cross-correlation} presents the resulting cross-correlation function (CCF) for the HM9-E001 (black), HM9-E030 (red), and HM9-E100 (blue) sets. As mentioned before, $\delta_{\rm dust}$ does not perfectly mask the general growth trend of $M_{\rm dust}$, which leads to large fluctuations in the CCF results. This unmasked trend particularly impacts the CCF for the HM9-E001 set, resulting in significant fluctuations for $\rm Lag_{E001} < 0$. However, all sets generally show maximum CCF values at $\rm Lag = 0$ (indicated by the dashed magenta line), suggesting that $\delta_{\rm dust}$ and $\delta_{\rm SFR}$ evolve synchronously within each set. It is important to note that the observed simultaneous relationship between $\delta_{\rm dust}$ and $\delta_{\rm SFR}$ could be due to the modeling of SN explosions for Pop II stars, which are assumed to occur immediately after their birth. Adjusting the SN delay time, therefore, could shift the maximum cross-correlation values, reflecting the specified SN delay distribution.

\par
We also observe similar trends in the top-heavy IMF adopted sets regarding the relationship between the evolution of SFRs and dust mass. Notably, the HM9-T001 set experiences significantly intense stellar feedback at $z \geq 9.7$, causing the dust mass in the central region of the halo within $R_{\rm vir}$ to decrease by nearly an order of magnitude (see Fig.~\ref{fig:dust mass evolution}). This intense outflow clears the dusty clouds near the young stars, which are the primary sources of UV luminosity. Unlike the HM9-E001 set, this strong dust outflow in the HM9-T001 set allows it to exceed the limiting magnitude at $z \leq 9.5$. As a result, the dust attenuation effect on the intrinsic SED of the galaxy is reduced, making it detectable.

\par
\par

\subsubsection{Impact of Pop III stars on UV luminosity}
\label{Impact of Pop III stars on UV luminosity}
\par
\begin{figure}
    \centering
    \includegraphics[width = 85mm]{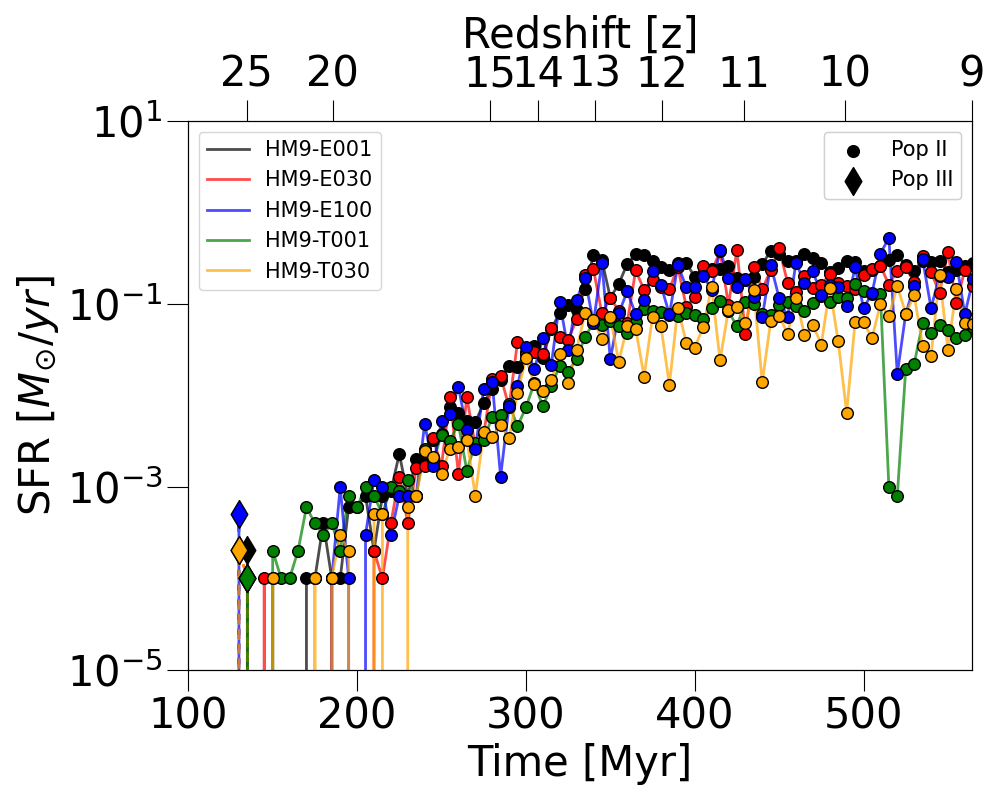}
    \caption{The SFRs of Pop III and Pop II stars in the HM9 sets,  with line colors and symbols matching those in Figure \ref{fig:mass evolution}. Diamond symbols represent the SFR of Pop III stars, while circle symbols denote the SFR of Pop II stars. The transition from Pop III to Pop II stars occurs around $z \gtrsim 20$. Once galaxies reach a stellar mass of $M_{\star} \geq 10^6 \msun$, Pop II stars become the dominant population.}
    \label{fig:SFR p2 p3}
\end{figure}
\par
In this subsection, we investigate the signatures of Pop III stars in our simulated galaxies. Figure \ref{fig:flux compare between pop2 stars and pop3 stars} shows the intrinsic stellar continuum in the wavelength range $500 \Ang \leq \lambda \leq 10000 \Ang$ for the total Pop III stars (red solid) and Pop II stars (blue solid) in the HM9-E030 set at $z = 9.63$, a point at which this simulated galaxy could be observable by JADES surveys. We emphasize that these SEDs are intrinsic stellar continuums, highlighting the differences in SEDs depending on each stellar population in the simulated galaxy. Despite the boosted UV luminosity of massive Pop III stars due to their top-heavy IMF, the total luminosity of Pop III stars in the galaxy is almost 5 dex lower than that of the total Pop II stars. This can be further explained by the SFR evolution of our sets, as shown in Figure \ref{fig:SFR p2 p3}, which illustrates the SFR of Pop III and Pop II stars separately in our HM9 sets. The colors of the lines and symbols match those in Figure \ref{fig:SFR}, with different symbols used to distinguish the stellar populations, diamonds for Pop III stars, and circles for Pop II stars.

\par
The general trend is as follows, the transition from Pop III to Pop II stars occurs at $z \gtrsim 20$ within a short timescale due to the low critical metallicity threshold, $Z_{\rm thr}=10^{-5.5}\zsun$, which can be easily achieved by a single Pop III SN event. This rapid transition from Pop III to Pop II stars has been predicted by other simulation studies (e.g., \citealp{Jeon_2017, Lee2024, Katz2023}) as well. After this transition, Pop II stars become the predominant population, effectively ending the era of Pop III star formation.
However, we also expect that Pop III stars could originate from other progenitor halos that eventually merge into the main progenitor halo. Nonetheless, the fraction of stars from this external origin is insignificant in contributing Pop III stars. Therefore, given that the high-redshift galaxies observed by JWST have already evolved significantly enough to be observable, we expect that the UV continuum of these galaxies is more likely to originate from Pop II stars, with the contribution to the UV luminosity from Pop III stars being negligible.
\par

Although our simulations favor an insignificant contribution from Pop III stars, the properties of Pop III stars remain uncertain. Therefore, there is still a possibility of detecting the signatures of Pop III stars in high-$z$ galaxies (see also \citealt{Venditti2024_PISN}). For instance, using simulated results from FOREVER22 (\citealp{Yajima2022}), \citet{Yajima2023} indicated that despite Pop II stars being the dominant population, Pop III stars might significantly impact Lyman-continuum fluxes in high-$z$ galaxies. They found a higher contribution to the UV luminosity by Pop III stars ($7.5 \times 10^{-2}$ at $\lambda = 1500 \Ang$ at $z = 12$) compared to our results, suggesting a notable contribution from Pop III stars. Additionally, \citet{Trussler2023} explored theoretical model spectra of Pop III-only galaxies and suggested AB magnitude depths required to achieve $5\sigma$ continuum detection with JWST NIRCam and MIRI. Therefore, future observational surveys with deeper limiting magnitudes using JWST might reveal and capture the signatures of primordial galaxies, predominantly composed of Pop III stars.

\section{Discussion}
In this section, we outline the observable properties and their features and discuss the limitations of our work. In Section \ref{observable properties}, we present observable properties such as effective radius, the UV slope of the derived spectra, and emission line ratios when our simulated galaxies are detectable. In Section \ref{Caveat}, we address the caveats and limitations, focusing on uncertainties and variations related to adopting a top-heavy IMF and assumptions about dust properties during post-processing.

\subsection{Observable properties}
\label{observable properties}
\subsubsection{Effective radius}
\label{effective radius}
\par
\begin{figure}
    \centering
    \includegraphics[width = 85mm]{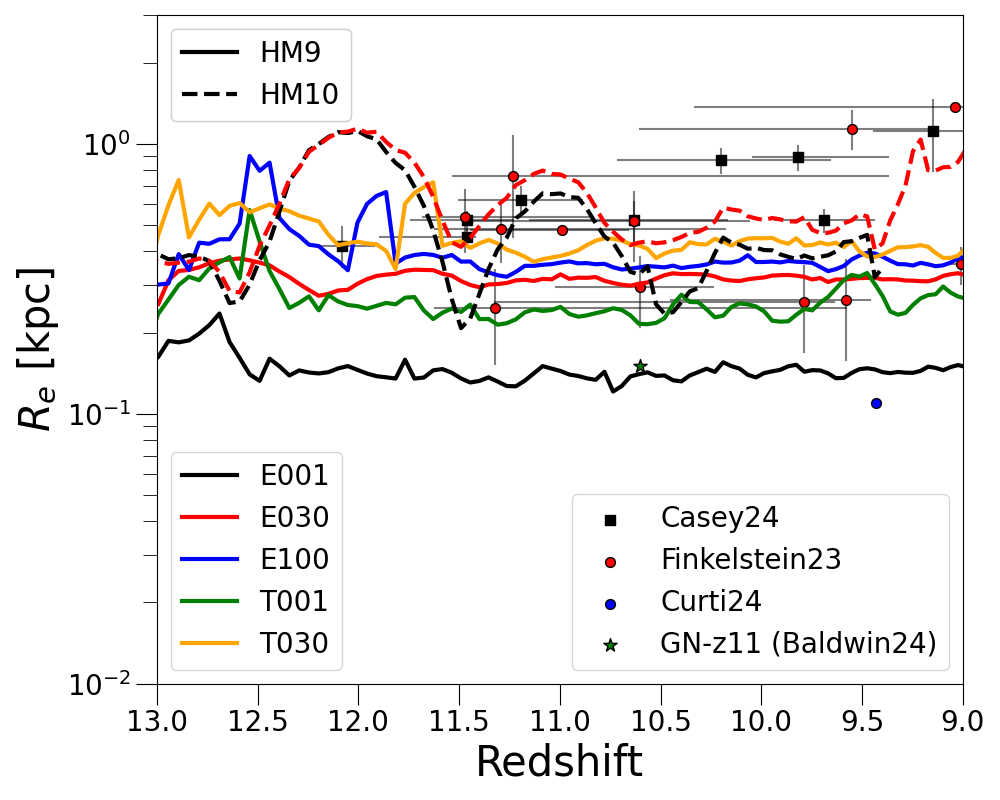}
    \caption{The effective radius of our simulated galaxies as a function of redshift is depicted with solid lines representing the HM9 sets and dashed lines representing the HM10 sets. Observational results from JWST are indicated by colored symbols, with horizontal and vertical lines denoting the 1$\sigma$ estimations from \citet{Finkelstein2023}, \citet{Casey2024}, \citet{Curti2024}, and \citet{Baldwin2024}.}
    \label{fig:effective radius}
\end{figure}
\par
Most observational studies determine the effective radius of galaxies by fitting their luminosities at each filter to the Sérsic profile (\citealp{Sersic1963}). However, this method is computationally intensive. To avoid the need for post-processing our simulated galaxies to track the continuous evolution of the effective radius, we instead adopt the half-mass radius as a proxy. The half-mass radius is defined as the radius within which half of the stellar mass of the simulated galaxies is enclosed, measured from the center of the halos. Figure \ref{fig:effective radius} illustrates the evolution of the effective radius of our simulated galaxies as a function of redshift, focusing only on the range discussed in Section \ref{post-processing results} regarding the observability of our simulated galaxies. The solid and dashed lines represent the HM9 and HM10 sets, respectively. For comparison, we include observed results from JWST surveys, denoted by colored symbols, black squares (\citealp{Casey2024}), red circles (\citealp{Finkelstein2023}), blue circles (\citealp{Curti2024}), and green starred symbols (\citealp{Baldwin2024}).
\par

For the HM9 sets, during $z \gtrsim 11.5$, the effective radius, $R_{\rm e}$, of simulated galaxies exhibits significant fluctuations, with the largest value reaching $\sim$0.9 kpc in the HM9-E100 set. Since $z \sim 11.5$, these fluctuations diminish, and the effective radius evolves more consistently within the range of 0.1 kpc $\leq R_{\rm e} \leq$ 0.5 kpc, which corresponds to $0.022^{\prime \prime} \lesssim R_{\rm e} \lesssim 0.133^{\prime \prime}$. On the other hand, the episodic nature of starbursts and the associated stellar feedback are reflected in the derived effective radius for the $\epsilon_{\rm ff}$-increased sets. In these sets, more intense feedback causes stars to become more diffuse, leading to greater variations and higher $R_{\rm e}$ values compared to the HM9-T001 set. This trend is also observed in sets with a top-heavy IMF due to increased stellar feedback from massive stars, these sets tend to have larger $R_{\rm e}$ values compared to those with a Chabrier IMF, even when using the same $\epsilon_{\rm ff}$.

\par
Focusing on the HM10 sets, a slight positive correlation between $\epsilon_{\rm ff}$ and $R_{\rm e}$ is found as well. Additionally, it is important to note that the HM10 sets exhibit two significant peaks in $R_{\rm e}$ during the intervals (11.5 $\lesssim z \lesssim$ 11.5) and (10.5 $\lesssim z \lesssim$ 11.5), with the maximum $R_{\rm e}$ reaching 1.14 kpc ($0.303^{\prime \prime}$) in the HM10-E030 set. These two peaks in the effective radius evolution are attributed to merging events. Before the first peak (12.5 $\lesssim z \lesssim$ 11.5), a small halo with a stellar mass of $M_{\star} \approx 5.3 \times 10^5 \msun$ and a virial mass of $M_{\rm vir} \approx 7.12 \times 10^8 \msun$ merges with the main progenitor. This merging event is completed at $z \sim 12.34$, around the time when $R_{\rm e}$ reaches the first peak. During the second peak, another small halo merges with the main progenitor, having a stellar mass of $M_{\star} \approx 6.0 \times 10^5 \msun$ and a virial mass of $M_{\rm vir} \approx 6.79 \times 10^8 \msun$. This merging event concludes at $z \sim 10.2$, corresponding to the second peak in $R_{\rm e}$. Our simulated galaxies remain among the faintest observed high-redshift galaxies by JWST, displaying effective radius $R_{\rm e}$ values comparable to those reported by \citet{Finkelstein2023} and \citet{Casey2024}, even when considering the increased impact of merging events, especially in the HM10 sets. However, our $R_{\rm e}$ values are higher than those found by \citet{Baldwin2024} and \citet{Curti2024}, who noted potential AGN features in several studies (e.g., \citealp{Maiolino2024, Curti2024}).
Lastly, in comparison with Little Red Dots (LRDs) recently observed by JWST in the redshift range $4 \leq z \leq 9$ (e.g., \citealp{Kokorev2024a, Kokorev2024b, Matthee2024ApJ}), we find that although our simulated galaxies show similar or lower stellar mass than LRDs, these LRDs tend to have a similar effective radius. This is likely because LRDs are expected to host massive AGNs in their center, which contribute to their compact features. The effective radius of LRDs is mostly in the range of $R_{\rm e} \lesssim 500 \rm pc\,(0.0713^{\prime \prime})$ (e.g., \citealp{Kokorev2024a, Kokorev2024b, Matthee2024ApJ}), which is comparable to the effective radius range of our simulated galaxies.

\subsubsection{UV slope}
\par
\begin{figure}
    \centering
    \includegraphics[width = 85mm]{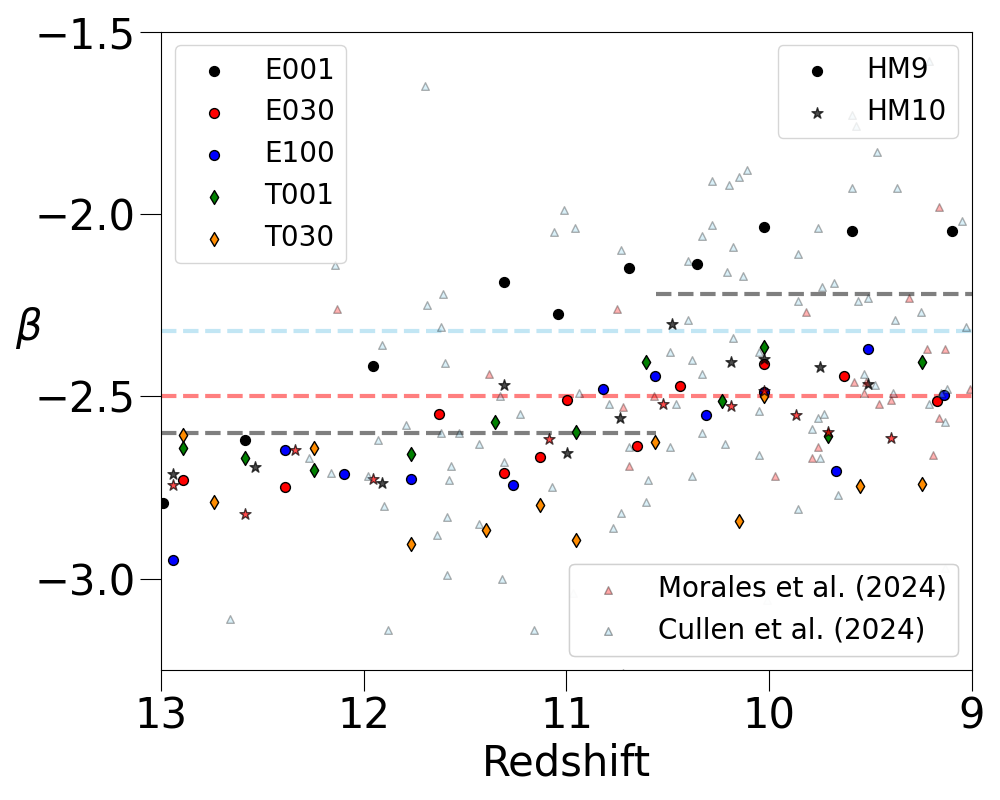}
    \caption{The estimated UV slope, $\beta$, from our simulated galaxies depicted with circle symbols representing the HM9 sets and star-shaped symbols representing the HM10 sets. The shaded triangles show the observational results from \citet{Morales2024} (red) and \citet{Cullen2024} (sky blue). The median values for \citet{Morales2024} and the inverse-variance averaged value for \citet{Cullen2024} are illustrated as dashed lines in their respective colors. Additionally, we show the step function model for the UV slope at $z \simeq 10.56$ from \citet{Cullen2024} as gray dashed lines.}
    \label{fig:UV slope}
\end{figure}
\par

In this subsection, we discuss the UV slope of our simulated galaxies using post-processed results. Various studies have suggested different ranges for the minimum wavelength of the UV slope, typically between 1216 $\text{\AA} \leq \lambda_{\rm min} \leq$ 1500 $\text{\AA}$ (e.g., \citealp{Austin2023}, \citealp{Cullen2023B}, \citealp{Cullen2024}, \citealp{Morales2024}). Given these varying suggestions, we adopt the definition and wavelength range for the UV slope from \citet{Morales2024}, which provides a specific functional form and wavelength range as follows,
\begin{equation}
    \begin{split}
        \log_{10}(f_{\lambda}) = \beta \log_{10}(\lambda) + \log_{10}(y_{\rm int}), \, \\ 1500 \Ang \leq \lambda_{\rm rest} \leq 3000 \Ang.
    \end{split}
\end{equation}
Using the continuum-only SED from our post-processed results, we determine the best fit for the UV slope within a specific UV wavelength range by employing the linear regression module in \texttt{scipy.stats.linregress}. 

\par
Figure \ref{fig:UV slope} presents our estimated UV slope results in comparison with the UV slopes of observed high-redshift galaxies, plotted as a function of redshift. The circle symbols represent the UV slope results from the HM9 sets, while the star symbols represent those from the HM10 sets. The shaded triangles indicate observational results from \citet{Morales2024} (red, hereafter M24) and \citet{Cullen2024} (sky blue, hereafter C24). Our simulated results tend to show a slight positive correlation with decreasing redshifts, particularly at $z \leq 12$. We emphasize that the top-heavy IMF adopted set, HM9-T030, tends to exhibit the steepest $\beta$ values, while the HM9-E001 set shows a shallower $\beta$ value, approaching $\beta \approx -2.0$ after $z \approx 10$. Except for the HM9-E001 set, our simulated results closely match the median value in M24 ($\beta_{\rm M24} = -2.5$) and are lower than the inverse-variance average value in C24 ($<\beta_{\rm C24}> = -2.32$).

Moreover, when comparing our results with the step function model from C24, which has a functional form with $<\beta_{\rm C24, \it z \geq 10.56}> \simeq -2.60, \, <\beta_{\rm C24, \it z \leq 10.56}> \simeq -2.22$ (see the gray dashed lines in Figure \ref{fig:UV slope}), we find that our results show an excellent match with $<\beta_{\rm C24, \it z \geq 10.56}>\approx-2.60$ and a slight discrepancy with <$\beta_{\rm C24, \it z \leq 10.56}$>$\approx-2.22$. We attribute the slight discrepancy at $z \leq 10.5$ to the population bias between our results and the analyzed samples in C24. The analyzed samples in C24 show the average absolute UV magnitude of $<M_{\rm UV}> = -19.3$. However, using $f(1500\Ang)$ from our post-processed results, most of our simulated galaxies tend to be less bright, with $M_{\rm UV} \lesssim -19.0$, except for HM10-E030 at $z \sim 9.40$, which has $M_{\rm UV} \approx -19.55$, which are in the range of faint samples in C24. Due to the diversity of samples in $M_{\rm UV}$, about 69\% of the samples at $z \leq 10.5$ from C24 fall within a higher UV magnitude range compared to our results. This leads to flattened $\beta$ values and a slight positive correlation between $\beta$ and $M_{\rm UV}$.

\par
Finally, unlike observational results from C24, our simulated results do not exhibit an extremely blue UV slope ($\beta \leq -3$), which would indicate the presence of extremely metal-poor galaxies with
$Z \lesssim 0.01\Zsun$. As described in Sections \ref{mass evolution} and \ref{Impact of Pop III stars on UV luminosity}, the transition from Pop III to Pop II stars occurs rapidly at $z \approx20$, and our simulated galaxies become sufficiently enriched to levels above $Z \approx 0.01\Zsun$ before $z = 13$. This metal enrichment leads to a proportional increase in dust mass, which in turn makes it challenging to produce an extremely blue UV slope in our simulations, even when adopting a top-heavy IMF.

\subsubsection{Emission line properties}
\begin{figure}
    \centering
    \includegraphics[width = 85mm]{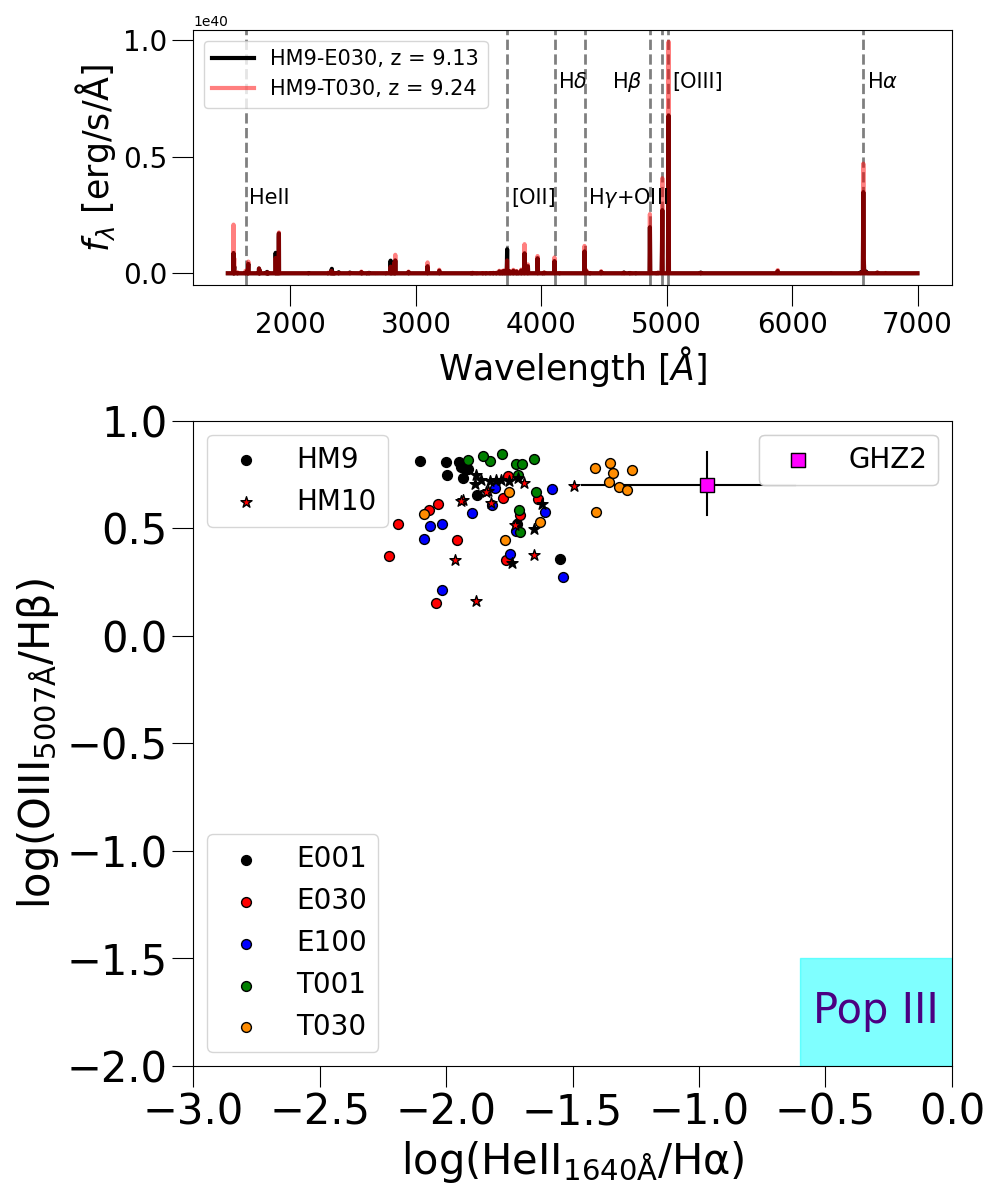}
    \caption{The top panel shows the emission line results in the rest frame from our simulated galaxies. To represent each IMF result, we present emission line data from the HM9-E030 and HM9-T030 sets at redshifts $z = 9.13$ and $z = 9.24$ when they are observable. For comparison, the specific locations of the emission lines are marked with dashed lines. The bottom panel illustrates the ratio $\log_{10}(\rm HeII1640\Ang / H\alpha)$ versus $\log_{10}(\rm [OIII]5007\Ang / H\beta)$. Circle and star-shaped symbols denote the ratios from the HM9 and HM10 sets, respectively. The cyan-shaded region in the bottom right indicates the expected location of Pop III-dominated galaxies, as proposed by \citet{Katz2023}. The magenta-colored square marker represents the ratio results combining emission line flux data from NIRSpec (\citealp{Castellano2024}) and MIRI (\citealp{Zavala2024}).}
    \label{fig:Emission lines}
\end{figure}
\par

Lastly, we present the emission line properties derived from post-processing results of our simulated galaxies at redshifts $9 \leq z \leq 13$. The top panel of Figure \ref{fig:Emission lines} shows the emission lines derived from the HM9-E030 and HM9-T030 datasets at redshifts $z = 9.13$ and $z = 9.24$ when they are observable. Both datasets exhibit significant emission features, particularly for oxygen and the Balmer lines of hydrogen. Notably, the doubly-ionized oxygen $\rm [OIII]$ doublet (4959$\Ang$, 5007$\Ang$) and the $\rm H\alpha$ and $\rm H\beta$ lines are prominent. It is important to note that the significant $\rm [OIII]$ doublet results are comparable to those observed in green pea galaxies in the local Universe (e.g., \citealp{Rhoads2023}).

\par
Also, utilizing our emission line data, we estimate the relationship between the emission line ratios $\log_{10}(\rm HeII1640\Ang / H\alpha)$ versus $\log_{10}(\rm [OIII]5007\Ang / H\beta)$, as shown in the bottom panel of Figure \ref{fig:Emission lines}. Provided that the $\rm HeII1640\Ang$ lines are generated by highly ionizing sources such as Pop III stars or AGN, validating this relationship could demonstrate that our top-heavy IMF models replicate the properties of Pop III stars. In the bottom panel of Figure \ref{fig:Emission lines}, we present the emission line ratio results from our simulated galaxies, with circle markers representing HM9 and star-shaped markers representing HM10. The cyan-shaded region in the bottom right denotes the expected location of galaxies dominated by Pop III stars ($\log_{10}(\rm HeII1640\Ang / H\alpha) > -0.6$ and $\log_{10}(\rm [OIII]5007\Ang / H\beta) < -1.5$), as suggested by \citet{Katz2023}. For comparison with JWST data, we also plot the line ratio of an example galaxy, GHZ2, using combined data from NIRSpec (\citealp{Castellano2024}) and MIRI (\citealp{Zavala2024}). GHZ2 tends to have extremely compact feature ($R_{\rm e} = 105 \pm 9 \rm pc$) with hard ionizing sources as indicated by the detection of emission lines such as N IV, C IV, He II and O III (\citealp{Castellano2024}).

\par
Our results indicate a high flux ratio for $\log_{10}(\rm [OIII]5007\Ang / H\beta)$, with values $\log_{10}(\rm [OIII]5007\Ang / H\beta) \geq 0$. Moreover, when comparing the sets with top-heavy IMF to those with Chabrier IMF, the top-heavy IMF sets exhibit higher ratios for both $\log_{10}(\rm [OIII]5007\Ang / H\beta)$ and $\log_{10}(\rm HeII1640\Ang / H\alpha)$. This increase is likely due to a greater population of massive stars in the top-heavy IMF sets, which produce harder spectra and emit more high-energy ionizing photons compared to the Chabrier IMF sets. In comparing our simulated results with the ratios from GHZ2, we find that the $\log_{10}(\rm [OIII]5007\Ang / H\beta)$ values are similar. However, there is a slight discrepancy in the $\log_{10}(\rm HeII1640\Ang / H\alpha)$ values between our simulated galaxies and GHZ2.

\par
It is important to note that, although we adopt a top-heavy IMF for Pop II stars in the HM9-T001 and HM9-T030 sets, their emission line properties are distinct from those of Pop III star-dominated galaxies. This discrepancy can be attributed to two main factors.
Firstly, the functional form and mass range of our top-heavy IMF for Pop II stars differ significantly from those of the Pop III IMF, particularly within the mass range considered in this study. As described in Section \ref{subsubsec:p2}, the mass range for the top-heavy IMF for Pop II stars is $1\msun \leq m \leq 100 \msun$, which is narrower than the typical mass range for Pop III stars, extending up to $\sim 500 \msun$. This difference could lead to a decreased flux of $\rm HeII1640 \Ang$. Secondly, by the time our simulated galaxies become observable, the gas is already enriched. Moreover, adopting a top-heavy IMF for Pop II stars results in an overproduction of metals compared to the same stellar mass in Chabrier IMFs, which could also elevate the $\log_{10}(\rm [OIII]5007\Ang / H\beta)$ values. Further details on the metal enrichment histories of the simulated galaxies will be discussed in Section \ref{MZR}.


\subsection{Caveats and limitations}
\label{Caveat}
\subsubsection{Uncertainties in the shape of top-heavy IMF}
\label{MZR}
\par
\begin{figure}
    \centering
    \includegraphics[width = 85mm]{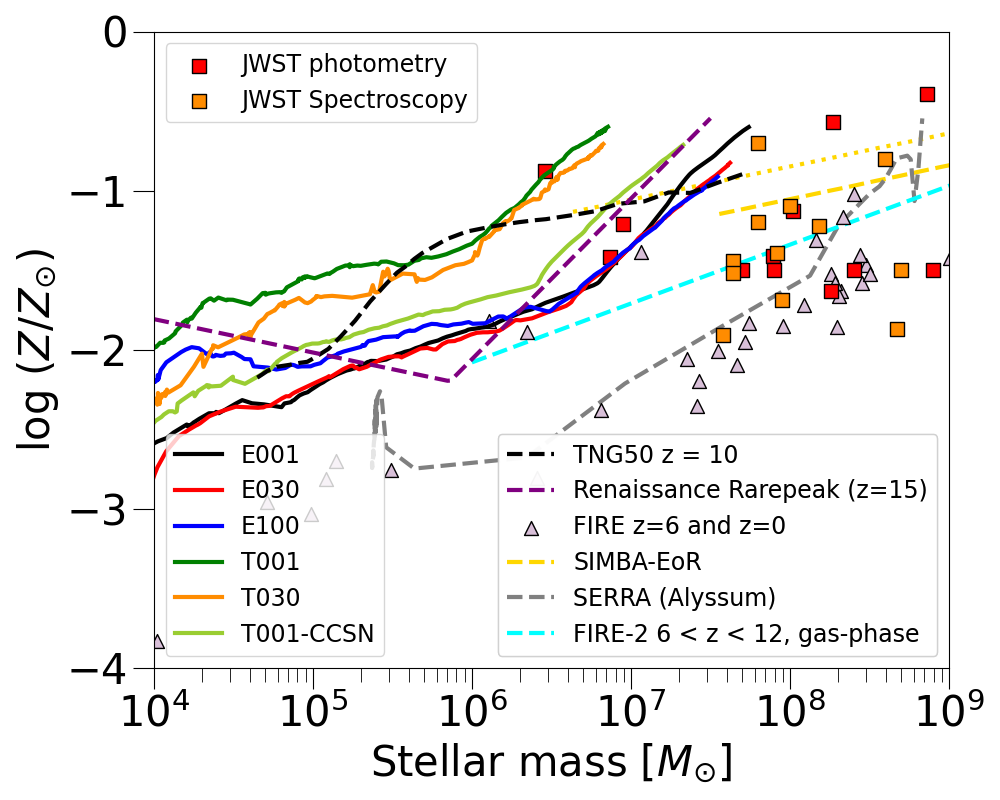}
    \caption{The stellar mass-metallicity relation, MZR, of our HM9 sets is represented by solid lines. The additional lime-green line indicates a top-heavy IMF adopted set, with the upper mass limit for Pop II CCSNe reduced to $m_{\star} = 40\msun$. The dashed and dotted lines represent high-z median results from other simulations, TNG50 (black), Renaissance simulation (purple), SIMBA-EoR (gold), and FIRE-2 (cyan, gas-phase within $0.2R_{\rm vir}$). The gray dashed line illustrates one example galaxy from the SERRA simulation. Triangle symbols denote results from the FIRE simulations at $z = 6$ and $z = 0$. Lastly, square symbols indicate observational results from the JWST, with red squares for photometry-only data (\citealp{Furtak2023, Robertson2023}) and orange squares for spectroscopically confirmed data (\citealp{Curtis-Lake2023, Carniani2024, Curti2024, Tacchella2023a, Wang2023}).}
    \label{fig:MZR}
\end{figure}
\par
In this subsection, we examine the uncertainty surrounding the functional form of a top-heavy IMF and its implications for simulated outcomes. As noted in Section \ref{subsubsec:p2}, we adopt a top-heavy IMF for Pop II stars, yet the IMF for high-$z$ galaxies remains uncertain, with various proposed shapes. For instance, using a SAM, \citet{Trinca2023} proposed a non-universal IMF that varies with $z$, $Z_{\star}$, and $M_{\star}$ over cosmic time to explain high-z observational data. Also, \citet{Inayoshi2022} suggested that adopting a top-heavy IMF similar to that of Pop III stars could account for the enhanced UV luminosity observed during the cosmic dawn. While this is a plausible hypothesis, the precise functional form of a top-heavy IMF at high-$z$ is still under debate (\citealp{Lazar2022, Liu2024}). 

\par
Also, we adopt a mass range of $m \geq 8\msun$ for Pop II SNeII across all simulation sets. However, the upper mass limit of SNeII is particularly significant for a top-heavy IMF, as variations in this mass range can influence the associated SN feedback and the amount of produced metals, determining the relationship between the stellar mass and metallicity of simulated galaxies, known as the mass-metallicity relation (MZR). Figure \ref{fig:MZR} illustrates the MZR results for our simulated galaxies. The solid lines represent the HM9 sets. To investigate the impact of varying the mass range for Pop II SNe when adopting a top-heavy IMF, we conduct an additional simulation for the HM9-T001 set, altering the SNII mass range of Pop II stars to $8\msun \leq m \leq 40\msun$. This result is depicted as a lime green line in Figure \ref{fig:MZR}. For comparison, high-$z$ galaxies spectroscopically confirmed by \citet{Curtis-Lake2023}, \citet{Carniani2024}, \citet{Curti2024}, \citet{Tacchella2023a}, and \citet{Wang2023} are marked with orange squares, while those identified only through photometry by \citet{Furtak2023} and \citet{Robertson2023} are indicated with red squares.

\par
The median values from other simulation projects are depicted as dashed lines, TNG50 (black, $z = 10$, \citealp{Nelson2019}), Renaissance simulation for the rare peak region (purple, $z = 15$, \citealp{Chen2014}), SIMBA-EoR (m50n1024 (dashed), m25n1024 (dotted), gold, \citealp{Jones2024}), and FIRE-2 simulation (cyan, $7 \leq z \leq 12$, \citealp{Marszewski2024}). Note that the results from the FIRE-2 simulations are based on gas-phase metallicity and are calculated within $0.2R_{\rm vir}$ of halos. The gray dashed line represents the MZR evolution of a sample galaxy from the SERRA simulation (Alyssum, \citealp{Pallottini2022}), which evolves into a galaxy with $M_{\star} \approx 5 \times 10^8 \msun$ at $z = 7.7$. Finally, the results from the FIRE simulations (\citealp{Ma2016}) are indicated by gray triangle symbols, denoting the MZR relation at $z=6$ and $z=0$.

\par
Our results demonstrate the expected trend of increasing metallicity with rising stellar mass. However, it is evident that sets adopting a top-heavy IMF tend to have higher $Z_{\star}$ compared to those adopting a Chabrier IMF, with a median difference of 0.9 dex at the same stellar mass range. The elevated MZR trends in the top-heavy IMF scenarios are anticipated, as metal enrichment is driven by a higher population of massive stars that end their lives as SNe. In top-heavy scenarios, the number of SNe per unit mass is $n_{\rm SN, TH} \approx 0.027 \msun^{-1}$, compared to $n_{\rm SN, Chabrier} \approx 0.012 \msun^{-1}$ in Chabrier IMF scenarios. Due to this significant difference, the top-heavy IMF sets exceed $\log_{10} Z_{\star} \sim -1.0$ at $M_{\star} \simeq 2.3 - 3 \times 10^6 \msun$, whereas the Chabrier IMF sets reach the same level at $M_{\star} \simeq 2.5 - 3 \times 10^7 \msun$.


\par
However, if we rearrange the mass range of Pop II CCSNe (hereafter T001-CCSN), represented by the lime green line in Figure \ref{fig:MZR}, the metallicity gap between T001-CCSN and HM9-E001 decreases, with a median value of 0.32 dex at the same stellar mass range. The simulated galaxy in the T001-CCSN set surpasses $\log_{10} Z_{\star} \gtrsim -1.0$ at $M_{\star} \simeq 9.2 \times 10^6 \msun$, positioning it between the top-heavy IMF adopted sets and the Chabrier IMF adopted sets. Also, due to the reduced SN feedback resulting from the lower SN number per unit mass ($n_{\rm SN, CCSN} \approx 0.02 \msun^{-1}$), the simulated galaxy in the T001-CCSN set reaches a stellar mass of $M_{\star} \approx 1.48 \times 10^7 \msun$, which is $\sim$ 2.6 times more massive than the HM9-T001 set at the same redshift. When compared with results from the JWST, the Chabrier IMF adopted sets align with the highest regions of the photometry and spectroscopic data. On the other hand, the top-heavy IMF adopted sets, while showing similar results to certain outliers in the photometry data, generally exhibit higher metallicity than most of the JWST measurements. Even with the decreased MZR in the T001-CCSN set, the frequent metal enrichment by massive stars increases the metallicity gap when compared to observed high-$z$ galaxies. The frequency of high-$z$ SNe, and thus the dependence on the uncertain IMF mass cut-offs, may be directly testable with future absorption spectroscopy of the early IGM, using gamma-ray bursts as bright background sources (\citealp{Wang2012}).


\par
Next, we compare our results with those from other simulation projects. While the top-heavy IMF adopted sets tend to exhibit the highest metallicity values among all simulations, the Chabrier IMF adopted sets align closely with the median values from the TNG50, Renaissance, and SIMBA-EoR projects within a similar mass range. Although the FIRE-2 results, based on gas-phase metallicity within $0.2 R_{\rm vir}$, show slightly lower metallicity than our simulations, our results are nearly 1 dex higher than those from the SERRA simulations and the FIRE simulations. This MZR gap between various simulation projects and our results arises from differences in assumptions and numerical recipes used in each simulation. We will briefly discuss the potential factors contributing to this discrepancy.

\par
The first factor could be the variation in sub-grid recipes for SN feedback modeling, which determines how energy is ejected into the ISM and how nearby gas clouds are enriched. This effect is discussed in \citet{Ibrahim2024}, where they compared the resultant MZR by adopting four different sub-grid recipes for SN feedback models such as thermal, kinetic, stochastic, and mechanical. Although their work is based on low-$z$ results ($z \lesssim 3$), they showed that thermal and stochastic models are likely to overproduce metals compared to kinetic and mechanical models. This discrepancy could be because thermal and stochastic feedback scenarios tend to have higher SFR than kinetic and mechanical scenarios. Also, the kick velocities from SN feedback in thermal and stochastic models are weaker than those in kinetic feedback models, making them insufficient to blow the loaded metals out of the galaxies into the IGM. Since our feedback mechanism is similar to the thermal and stochastic feedback cases in \citet{Ibrahim2024}, the reduced ejection of metals into the IGM could result in higher metallicities within the simulated galaxies.

\par
The second factor could be variations in the functional form of the IMF and the mass range for SNeII from Pop~II stars. While we adopt the Chabrier IMF for our fiducial IMF of Pop II stars, considering a Salpeter IMF with the same SNII mass range results in a lower number of SNe per unit mass, $n_{\rm SN, Salpeter} \approx 0.0074$, which is 63\% of the value used in our simulations. Additionally, if we apply the same mass range as T001-CCSN for each IMF, the number of SNe per unit mass for SNeII in each IMF decreases to $n_{\rm CCSN, Chabrier} \approx 0.010$ and $n_{\rm CCSN, Salpeter} \approx 0.0068$. These variations in the functional form of the IMF and the mass range for Pop II SNeII suggest that adopting different IMF parameters for Pop II stars could yield different outcomes, potentially aligning our results more closely with those from the FIRE simulations or SERRA. Lastly, as described in \citet{AGORA2024}, even with the same initial conditions, different MZR and radial metallicity gradients can be estimated across various simulation codes, often showing differences of about 1 dex. This discrepancy is attributed to the different numerical methodologies employed in each simulation.

\par
\par

\subsubsection{Several assumptions and limitations in our work}
\label{assumptions and limitations}
In this subsection, we discuss the assumptions and limitations related to our post-processing methodologies. The evolution and growth of dust particles are highly sensitive to diverse physical processes and parameters such as efficient dust destruction in SN reverse shocks (e.g., \citealp{Jones2011, Ji2014, Salim2020}). However, for simplicity, we have only considered dust evolution in terms of gas-phase metallicity, expressed as a simple metal-to-dust ratio, adopting a value of 0.07. This value is consistent with \citet{Barrow2017}, but the exact metal-to-dust ratio of high-$z$ galaxies remains uncertain \citep[e.g.,][]{DeRossi2023}. Several simulation studies have used values that can vary up to 0.4 (\citealp{Ma2018b, Jeon_2019}). Also, while running the post-processing, we employed the MW dust model ($R_v = 3.1$) from \citet{Draine2003}. However, the dust properties of high-$z$ galaxies are not well-known, introducing uncertainties in the dust models for these galaxies. Using different dust physics models, such as those from \citet{Calzetti1994}, which discuss dust profiles in star-forming galaxies from the local Universe, or models for low-metallicity environments like the SMC and LMC (\citealp{LMCSMC}), could yield different outcomes for the observability of our simulated galaxies.

\par
Second, although we adopted \yggdrasil \nspace and \fsps \nspace to generate synthetic spectra for Pop III and Pop II stars, there are various other stellar population and spectral synthesis codes available. For instance, other simulations and theoretical works have used the \bpass \nspace model (\citealp{BPASS}) and the \st \nspace package (\citealp{Starburst99}) to perform post-processing and calculate spectral features (\citealp{Sun2023a, Dekel2023}). As discussed in \citet{Ma2018b}, even within \bpass, binary and single-star models can yield different SED results despite having the same star formation histories and adopted IMF. Therefore, variations and uncertainties between different stellar population and spectral synthesis models, even with identical simulation results, could produce different post-processed outcomes and observability. Finally, due to computational costs, our work targeted only one isolated high-$z$ galaxy per simulation using the zoom-in technique to maintain high resolution. Consequently, the limited number of samples in our simulations prevents us from providing statistically robust values and parameters, such as the UVLF and cosmic UV luminosity density, $\rho_{\rm UV}$.

\par
Finally, as mentioned in Section~\ref{RT}, we neglect the radiation effects from Pop II stars on the simulated galaxies due to high computational costs. However, radiation effects are likely important for understanding the visibility of high-redshift galaxies. Recently, \citet{Ferrara2024a} proposed that if high-$z$ galaxies become super-Eddington, a strong radiation-driven outflow might clear away most of the dust, making them observable during a post-starburst phase. In future research, we aim to include radiation effects in our models to test this scenario and better explain UV-bright galaxies.


\section{Comparable works}
In this section, we discuss other theoretical work that focuses on the effects of varying star formation efficiency or adopting a top-heavy IMF to elucidate the high-redshift galaxies observed by the JWST, as detailed in Section \ref{simulations and theoretical works}. We also compare the physical characteristics of our simulated galaxies with observational data, particularly focusing on stellar mass, $M_{\star}$, and SFR in Section \ref{observational works}.

\subsection{Comparison with theoretical work}
\label{simulations and theoretical works}
\subsubsection{Stochasticity and time variability in the star formation}

Our simulations indicate that stochasticity and time variability in star formation may be driven by episodic star formation, achieved either by increasing $\epsilon_{\rm ff}$ or by adopting a top-heavy IMF for Pop~II stars. The significance of stochasticity in star formation, particularly in explaining the overabundance of bright high-redshift galaxies, is also emphasized in other theoretical studies (e.g., \citealp{Mirocha2023, Pallottini2023, Shen2023, Gelli2024}). For instance, \citet{Mirocha2023} utilized a semi-empirical model to suggest that a high scatter in the mass accretion rate onto halos during their assembly process could introduce stochasticity in star formation and variability in the SFR-halo mass relationship, thereby increasing the prevalence of bright blue galaxies. Furthermore, \citet{Shen2023} employed an empirical model assuming that observed UV luminosity follows a log-normal distribution with a fixed median value. Based on this assumption, they proposed that a large UV variability of $\sigma_{\rm UV}\approx2.0$, especially at high-$z$ ($z\approx12$), to be aligned with the UVLF constructed from observations of high-$z$ galaxies.

\par
Using hydrodynamic simulations, \citet{Pallottini2023} found stochastic star formation behavior in the simulated galaxies of SERRA (\citealp{Pallottini2022}). By employing the $\delta_{\rm SFR}$ values mentioned in Section \ref{SFR}, they calculated the periodicity of the SFR and suggested that the zero-mean Gaussian distribution of $\delta_{\rm SFR}$ has a standard deviation of $\sigma_{\rm \delta} = 0.24$. From these $\sigma_{\rm \delta}$ values, they estimated a UV variability of $\sigma_{\rm UV} \simeq 0.61$ based on the relationship between SFR and UV luminosity. However, they pointed out that $\sigma_{\rm UV}$ needs to be three times higher than their estimations to explain the UVLF constructed from the observed high-z galaxies by the JWST. Utilizing the FIRE-2 simulations (\citealp{Ma2018a}), \citet{Sun2023a, Sun2023b} investigated the time variabilities in star formation in high-redshift simulated galaxies. They considered two scenarios, "bursty" and "smoothed". In the bursty scenario, star particles are represented by a SSP model, whereas in the smoothed scenario, each star particle in the simulated galaxies forms at a constant rate within a specific time bin. Without any fine-tuning, they demonstrated that the UVLF at high redshift is well explained by the bursty scenario, while the smoothed case underpredicted the observed UVLF.
\par

\subsubsection{Variations in the IMF functional form in high-z galaxies}

The results of adopting a top-heavy IMF show differences between works that primarily focus on semi-analytic models and our own simulations. For instance, using the Santa Cruz Semi-Analytic Model (SCSAM), \citet{Yung_2023} suggested that adopting a top-heavy IMF in high-redshift galaxies, in terms of idealized boosting factors, could help match the observed UVLF. Similarly, utilizing the Cosmic Archaeology Tool (CAT), \citet{Trinca2023} proposed a transitional IMF, which combines the Kroupa IMF (\citealp{Kroupa2001}) with the log-flat IMF obtained by \citet{Chon2022}, resembling a top-heavy IMF. By weighting these two IMFs based on the metallicities and redshifts of galaxies, they adopted a new conversion factor for the transitional IMF when constructing the UVLF. Their findings suggested that a transitional IMF could reproduce recent observational results. However, it is important to note that both \citet{Yung_2023} and \citet{Trinca2023} did not consider the effects of stellar feedback from top-heavy and transitional IMFs in their models.

Our simulations support the straightforward expectation that adopting a top-heavy IMF would enhance the UV luminosity of high-redshift galaxies due to the high light-to-mass ratio of massive stars. Indeed, we find a lower stellar mass but increased luminosity with a top-heavy IMF compared to a conventional one. This result is due to stronger feedback from the more frequent massive stars produced by the top-heavy IMF, which suppresses star-forming activities. A similar trend was reported in the work by \citet{Cueto2024}, who used the semi-numerical simulation \astraeus. They studied the effects of a Salpeter IMF as a fiducial model, comparing it to an evolving IMF from \citet{Chon2022}, which combines the Salpeter IMF with a log-flat shape that varies with redshift and metallicity. By employing star formation efficiencies as free parameters to match the observed UVLF, they found that the star formation efficiency for the evolving IMF needed to be about 2.5 times lower than for the Salpeter IMF due to stronger feedback from the higher abundance of massive stars. They, however, concluded that an evolving IMF alone could not fully explain the bright blue galaxies at high-$z$. Additionally, they reported a higher MZR trend with the evolving IMF compared to the Salpeter IMF, which aligns with our findings when comparing the top-heavy IMF adopted set to the Chabrier IMF.

\subsubsection{Feedback-free starbursts}
\par
\begin{figure}
    \centering
    \includegraphics[width = 85mm]{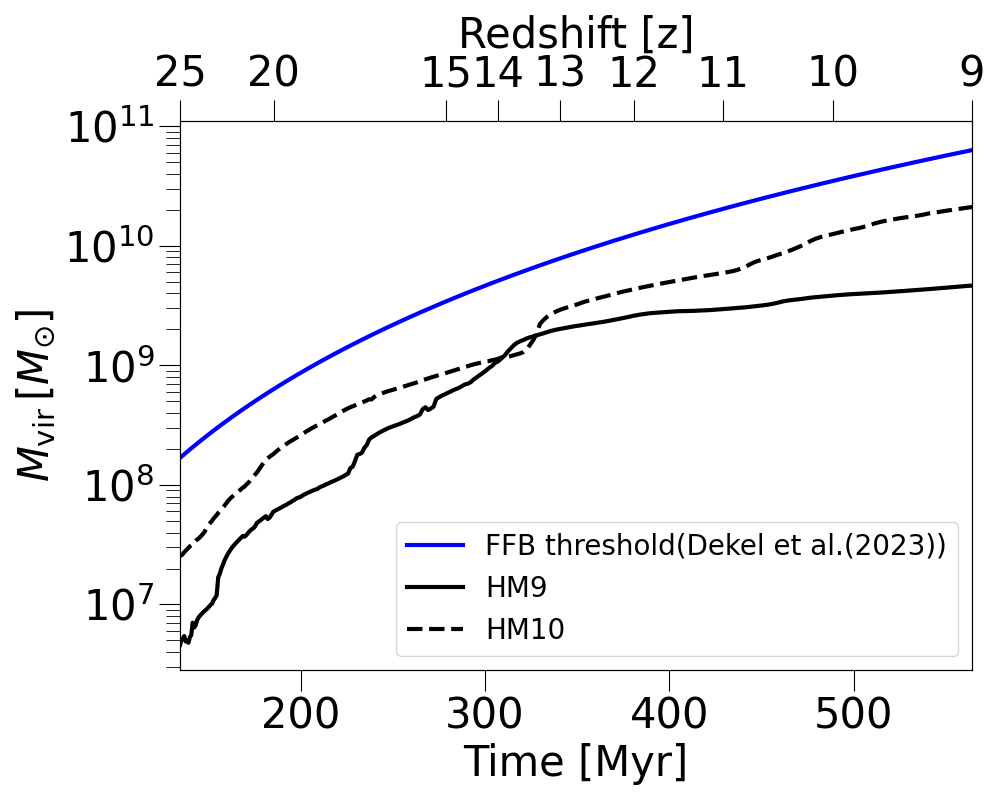}
    \caption{The virial mass comparison with the minimum threshold virial mass of the FFB model from \citet{Dekel2023}. The blue line represents the FFB threshold mass, while the black dashed and solid lines show the virial mass of the HM10 and HM9 sets, respectively. Although the virial mass of our simulated galaxies remains below the threshold of the FFB model, our simulated galaxies can occasionally be observable via bursty star forming activities.}
    \label{fig:FFB}
\end{figure}
\par
Using an analytical approach, \citet{Dekel2023} and \citet{Li2023} demonstrated the possibility for extremely efficient star formation in high-redshift galaxies when the freefall time of star-forming clouds is $\sim$ 1 Myr, shorter than the delay time of stellar feedback from nearby newly born stars. This scenario allows for continuous star formation without interruption from feedback, known as a feedback-free starburst (FFB). To investigate this possibility in our simulations, we compare the minimum threshold of virial mass for the FFB model from \cite{Dekel2023} with the virial mass evolution of our simulated galaxies in Figure \ref{fig:FFB}. The black solid and dashed lines represent the virial mass of the HM9 and HM10 sets, respectively, while the blue dashed line indicates the minimum threshold of virial mass for the FFB model ($M_{\rm FFB}$). As evident in Figure \ref{fig:FFB}, our simulated galaxies do not exceed $M_{\rm FFB}$ by the end of the simulations, showing lower virial masses by an order of 0.5 to 1, and even 1.5 for the HM9 sets at $z \gtrsim 22.8$, and more than 1 dex for the HM9 sets at $z \lesssim 9.7$.

\par
\begin{figure}
    \centering
    \includegraphics[width = 85mm]{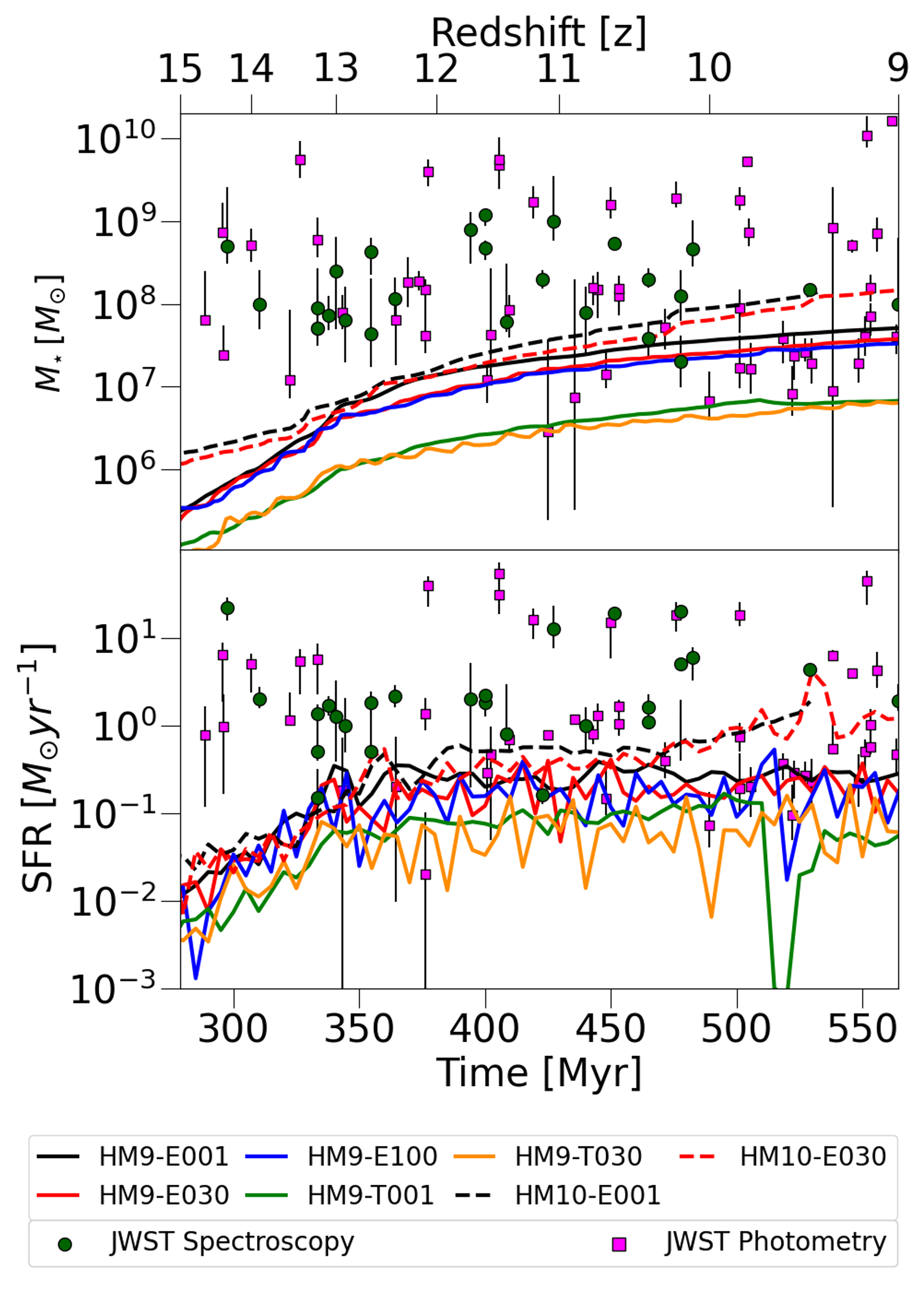}
    \caption{The estimated stellar mass and SFR of our simulations are compared with observational results from JWST. The stellar masses and SFR of our simulated galaxies are depicted by solid and dashed lines for the HM9 and HM10 sets, respectively. For the JWST results, magenta squares and green circles represent photometric results (\citealp{Furtak2023, Casey2024, Robertson2024, Morales2024}) and spectroscopic results (\citealp{Harikane2024, Hsiao2023, Bunker2023, Carniani2024, Robertson2023, Curti2024, Curtis-Lake2023, Wang2023, Hainline2024}), respectively. We also show the $1\sigma$ range estimation for observational results as black lines.}
    \label{fig:Stellar mass comparison}
\end{figure}

\par
For halo masses below $M_{\rm FFB}$, adopting an extremely high star formation efficiency, $\epsilon_{\rm ff} = 1.0$, as suggested by the FFB model in \citet{Dekel2023}, does not initiate the FFB process. Our simulations, specifically the HM9-E100 set, demonstrate that within this mass range, strong feedback from bursty star formation is more effective, resulting in prolonged quenching periods. It is important to note that even in the absence of the FFB process, the increased $\epsilon_{\rm ff}$ and a top-heavy IMF enable these smaller galaxies to occasionally exceed the limiting magnitude of the JADES survey. Consequently, we propose that while the FFB process is one mechanism for enhancing the UV luminosity of massive galaxies at high-$z$, other mechanisms are also possible.

\subsection{Comparison with observational data}
\label{observational works}

Lastly, we compare our results with recent observational findings from the JWST, focusing on two key physical properties, the stellar mass of galaxies and SFR, as illustrated in Figure \ref{fig:Stellar mass comparison}. The top panel presents the stellar mass comparison, while the bottom panel shows the SFR comparison with the observed data. In both panels, the solid and dashed lines represent our simulation results from the HM9 and HM10 sets, respectively. The line colors correspond to the same simulation sets as in the previous figures. Magenta squares and green circles denote JWST photometric results (\citealp{Furtak2023, Casey2024, Robertson2024, Morales2024}) and spectroscopic results (\citealp{Harikane2024, Hsiao2023, Bunker2023, Carniani2024, Robertson2023, Curti2024, Curtis-Lake2023, Wang2023, Hainline2024}). Also, we indicate the $1\sigma$ range of the estimated stellar mass and SFR from observational results using black lines.

\par
As shown in the top panel of Figure \ref{fig:Stellar mass comparison}, for $z\gtrsim 12$, the total stellar mass of our simulated galaxies is below the lower mass boundary of observational results. However, after $z=12$, when most of our simulated galaxies surpass the limiting magnitude of the JADES surveys, they exceed the lowest stellar mass range for both photometric and spectroscopic results. Notably, the sets with a top-heavy IMF tend to have total stellar masses just below the lowest stellar mass confirmed by spectroscopy, but still within the boundary of the photometry results.
In the bottom panel of Figure \ref{fig:Stellar mass comparison}, which depicts the SFR results, most observed galaxies exhibit higher SFRs compared to our simulations. Nonetheless, our simulated results align well with the lower SFR range of the observed galaxies. This is true even for the sets with a top-heavy IMF, which show the lowest SFR trends in our simulations, yet still match the lower region of the observed SFR data.

\par
To summarize, our simulated results typically fall within the lowest stellar mass range, corresponding to the faintest region of observed results from JWST. Varying sub-grid physics parameters, such as $\epsilon_{\rm ff}$ and the IMF, could enhance the UV luminosity of our simulated galaxies, potentially allowing them to exceed the JADES survey's limiting magnitude. Additionally, in terms of SFR, our simulated galaxies show a good match with the SFR derived from observed galaxies. Furthermore, considering the gravitational lensing effect, our simulated galaxies could easily surpass the limiting magnitude of the JWST, even for the faintest results (the HM9-E001 set), which otherwise do not exceed the limiting magnitude (e.g., \citealp{Atek2024}).

\section{Summary and Conclusions}
In this study, we have investigated the characteristics and observability of high-redshift galaxies observed by JWST, with a focus on the early galaxy problem where these galaxies appear unexpectedly bright compared to theoretical predictions. To address this issue, we conducted cosmological hydrodynamic simulations of high-$z$ galaxy formation, specifically for galaxies with virial masses $M_{\rm vir} \sim 4.2 \times 10^9 \msun$ and $M_{\rm vir} \sim 1.3 \times 10^{10} \msun$ at $z = 10$. To reconcile the discrepancies between observations and theoretical models, we adopted two specific scenarios related to star formation physics and methodologies. We adjusted the star formation efficiency, $\epsilon_{\rm ff}$, and employed a top-heavy IMF for the dominant stellar populations in the simulated galaxies. Specifically, we increased the star formation efficiency from a fiducial value of 1\% to 100\%, representing cases of extremely efficient star formation. For the top-heavy IMF of the dominant Pop II stars, we used a simple power-law IMF with a shallower slope than the Salpeter value, which increases the fraction of massive stars in the stellar population. Our findings indicate that the adopted scenarios can significantly enhance the UV luminosity of high-$z$ galaxies, allowing the simulated galaxies to surpass the limiting magnitude of JWST surveys. Furthermore, by analyzing our simulated and post-processed results, we explored various physical properties of high-$z$ galaxies and identified which factors critically influence the observability of our simulated galaxies.

\par
Our main finding can be summarized as follows:

\par
\begin{enumerate}
    \item[$\bullet$] The galaxies from our simulations exhibited total stellar masses ranging from $M_{\star}=10^6 \msun$ to $M_{\star}=2 \times 10^8 \msun$ at redshifts $9 \leq z \leq 11$, which corresponds to the lower end of the high-$z$ galaxies observed by JWST. Despite increasing the star formation efficiency in our simulations, this variation did not directly result in a higher total stellar mass within the galaxies, contrary to prevailing expectations. Also, adopting a top-heavy IMF for Pop II stars led to a higher frequency of massive stars. This, in turn, triggered strong stellar feedback that regulated star formation, ultimately reducing the total stellar mass compared to cases using the Chabrier IMF.

    \item[$\bullet$] In the efficiency-increased simulation sets, we found episodic star formation behavior, where stars are more likely to form in bursts followed by brief quenching periods, characterized by periodicity within the range of 10 to 20 Myr. In contrast, the simulation set with the fiducial star formation efficiency (HM9-E001) exhibited continuous star formation. This episodic nature is more pronounced in the set adopting a top-heavy IMF, due to the enhanced feedback from massive stars, resulting in star formation rates that are almost 1 dex lower compared to the sets using the Chabrier IMF.

    \item[$\bullet$] Our post-processing results revealed that increasing star formation efficiency or adopting a top-heavy IMF could enhance the UV luminosity of our simulated galaxies, surpassing the limiting magnitude of the JADES deep survey. This enhancement occurs because, in the efficiency-increased set, intense feedback accompanied by bursty star formation effectively clears out dust within the star-forming regions, which would otherwise obscure and suppress UV luminosity. By examining the periodic relationship between star formation trends and dust mass fluctuations within the star-forming regions, we confirmed that these two quantities are correlated. The boosted UV luminosity in the top-heavy IMF-adopted set, even with less stellar mass, is attributed to the frequent presence of massive stars.
    
    \item[$\bullet$] We found that Pop III stars contributed insignificantly to our simulated galaxies once they exceeded the limiting magnitude of surveys. This is because the transition from Pop III to Pop II stars occurs early at $z \approx 20$ in our simulations, resulting in only a brief period of Pop III star formation. This finding was further validated by the emission line ratios of the simulated galaxies derived from post-processing.

    \item[$\bullet$] We also derived several observable quantities, including effective radius, UV slope, and emission line ratios from our simulations. We identified a trend where more bursty star formation leads to significant variations in the effective radius due to intense feedback effects. Moreover, merging events can result in puffed-up effective radii. The estimated UV slopes align with observed values, showing the expected trend of less-steep $\beta$ values with decreasing redshift. Note that our results do not exhibit the very steep value of $\beta \sim -3$ predicted by some observations. 
    
    \item[$\bullet$] Adopting a top-heavy IMF for Pop II stars requires careful consideration. The frequent presence of massive stars significantly enhances metal enrichment, resulting in an increase in galaxy metallicity by $\sim$1 dex compared to using the Chabrier IMF. This leads to a higher stellar mass-metallicity relation than observed. This issue can be addressed by varying the mass range of stars eligible for supernova explosions or by employing different supernova feedback recipes to reduce metallicity within galaxies by ejecting metals into the IGM.    

    \item[$\bullet$] Compared to other studies, particularly those employing semi-analytic methods to account for increased efficiency or adopting a top-heavy IMF effect, our hydrodynamic simulations indicate that feedback mechanisms play a crucial role in regulating star formation, resulting in decreased stellar mass contrary to expectations.   
\end{enumerate}

\par
Even though our simulated galaxies are among the smallest in high-redshift samples, our findings suggest that starbursts with effective star formation or a higher frequency of massive stars in high-redshift galaxies might be key to narrowing the discrepancy between recent observational results and theoretical models. Also, our post-processed results could serve as valuable theoretical templates for forthcoming spectroscopic data from JWST's NIRSpec and MIRI instruments. However, as highlighted in previous sections, our study is limited by having performed zoom-in simulations for only two high-redshift galaxies, thus not providing statistical properties of these galaxies. Future work will involve exploring expanded simulation suites, including galaxies massive enough to test the feedback-free scenario. Furthermore, larger spectroscopically confirmed samples from follow-up JWST observations will help elucidate the detailed physical properties of high-redshift galaxies. These combined theoretical and observational efforts will help resolve the current discrepancies observed in high-redshift galaxies and enhance our understanding of the early Universe.

\section*{acknowledgements}
We are grateful to Volker Springel, Joop Schaye, and Claudio Dalla
Vecchia for permission to use their versions of \textsc{gadget}.
T.~B.~J. is supported by the National Research Foundation (NRF) grants funded by the Korean government (MSIT) (No.2021R1A2C109491713, No. 2022M3K3A1093827).


\begin{thebibliography}{}
\expandafter\ifx\csname natexlab\endcsname\relax\def\natexlab#1{#1}\fi
\providecommand{\url}[1]{\href{#1}{#1}}
\providecommand{\dodoi}[1]{doi:~\href{http://doi.org/#1}{\nolinkurl{#1}}}
\providecommand{\doeprint}[1]{\href{http://ascl.net/#1}{\nolinkurl{http://ascl.net/#1}}}
\providecommand{\doarXiv}[1]{\href{https://arxiv.org/abs/#1}{\nolinkurl{https://arxiv.org/abs/#1}}}

\bibitem[{{Adams} {et~al.}(2023){Adams}, {Conselice}, {Ferreira}, {Austin}, {Trussler}, {Juod{\v{z}}balis}, {Wilkins}, {Caruana}, {Dayal}, {Verma}, \& {Vijayan}}]{Adams2023}
{Adams}, N.~J., {Conselice}, C.~J., {Ferreira}, L., {et~al.} 2023, \mnras, 518, 4755, \dodoi{10.1093/mnras/stac3347}

\bibitem[{Applebaum {et~al.}(2019)Applebaum, Brooks, Quinn, \& Christensen}]{Applebaum2019}
Applebaum, E., Brooks, A.~M., Quinn, T.~R., \& Christensen, C.~R. 2019, Monthly Notices of the Royal Astronomical Society, 492, 8–21, \dodoi{10.1093/mnras/stz3331}

\bibitem[{{Arrabal Haro} {et~al.}(2023){Arrabal Haro}, {Dickinson}, {Finkelstein}, {Kartaltepe}, {Donnan}, {Burgarella}, {Carnall}, {Cullen}, {Dunlop}, {Fern{\'a}ndez}, {Fujimoto}, {Jung}, {Krips}, {Larson}, {Papovich}, {P{\'e}rez-Gonz{\'a}lez}, {Amor{\'\i}n}, {Bagley}, {Buat}, {Casey}, {Chworowsky}, {Cohen}, {Ferguson}, {Giavalisco}, {Huertas-Company}, {Hutchison}, {Kocevski}, {Koekemoer}, {Lucas}, {McLeod}, {McLure}, {Pirzkal}, {Seill{\'e}}, {Trump}, {Weiner}, {Wilkins}, \& {Zavala}}]{ArrabalHaro2023}
{Arrabal Haro}, P., {Dickinson}, M., {Finkelstein}, S.~L., {et~al.} 2023, \nat, 622, 707, \dodoi{10.1038/s41586-023-06521-7}

\bibitem[{{Atek} {et~al.}(2024){Atek}, {Labb{\'e}}, {Furtak}, {Chemerynska}, {Fujimoto}, {Setton}, {Miller}, {Oesch}, {Bezanson}, {Price}, {Dayal}, {Zitrin}, {Kokorev}, {Weaver}, {Brammer}, {Dokkum}, {Williams}, {Cutler}, {Feldmann}, {Fudamoto}, {Greene}, {Leja}, {Maseda}, {Muzzin}, {Pan}, {Papovich}, {Nelson}, {Nanayakkara}, {Stark}, {Stefanon}, {Suess}, {Wang}, \& {Whitaker}}]{Atek2024}
{Atek}, H., {Labb{\'e}}, I., {Furtak}, L.~J., {et~al.} 2024, \nat, 626, 975, \dodoi{10.1038/s41586-024-07043-6}

\bibitem[{{Austin} {et~al.}(2023){Austin}, {Adams}, {Conselice}, {Harvey}, {Ormerod}, {Trussler}, {Li}, {Ferreira}, {Dayal}, \& {Juod{\v{z}}balis}}]{Austin2023}
{Austin}, D., {Adams}, N., {Conselice}, C.~J., {et~al.} 2023, \apjl, 952, L7, \dodoi{10.3847/2041-8213/ace18d}

\bibitem[{{Bakx} {et~al.}(2023){Bakx}, {Zavala}, {Mitsuhashi}, {Treu}, {Fontana}, {Tadaki}, {Casey}, {Castellano}, {Glazebrook}, {Hagimoto}, {Ikeda}, {Jones}, {Leethochawalit}, {Mason}, {Morishita}, {Nanayakkara}, {Pentericci}, {Roberts-Borsani}, {Santini}, {Serjeant}, {Tamura}, {Trenti}, \& {Vanzella}}]{Bakx2023}
{Bakx}, T. J.~L.~C., {Zavala}, J.~A., {Mitsuhashi}, I., {et~al.} 2023, \mnras, 519, 5076, \dodoi{10.1093/mnras/stac3723}

\bibitem[{{Baldwin} {et~al.}(2024){Baldwin}, {Nelson}, {Johnson}, {Oesch}, {Tacchella}, {Illingworth}, {Gibson}, \& {Hartley}}]{Baldwin2024}
{Baldwin}, J.~O., {Nelson}, E., {Johnson}, B.~D., {et~al.} 2024, Research Notes of the American Astronomical Society, 8, 29, \dodoi{10.3847/2515-5172/ad220a}

\bibitem[{{Barrow} {et~al.}(2017){Barrow}, {Wise}, {Norman}, {O'Shea}, \& {Xu}}]{Barrow2017}
{Barrow}, K. S.~S., {Wise}, J.~H., {Norman}, M.~L., {O'Shea}, B.~W., \& {Xu}, H. 2017, \mnras, 469, 4863, \dodoi{10.1093/mnras/stx1181}

\bibitem[{{Behroozi} {et~al.}(2020){Behroozi}, {Conroy}, {Wechsler}, {Hearin}, {Williams}, {Moster}, {Yung}, {Somerville}, {Gottl{\"o}ber}, {Yepes}, \& {Endsley}}]{Behroozi2020}
{Behroozi}, P., {Conroy}, C., {Wechsler}, R.~H., {et~al.} 2020, \mnras, 499, 5702, \dodoi{10.1093/mnras/staa3164}

\bibitem[{{Behroozi} {et~al.}(2013{\natexlab{a}}){Behroozi}, {Wechsler}, \& {Conroy}}]{Behroozi2013}
{Behroozi}, P.~S., {Wechsler}, R.~H., \& {Conroy}, C. 2013{\natexlab{a}}, \apj, 770, 57, \dodoi{10.1088/0004-637X/770/1/57}

\bibitem[{{Behroozi} {et~al.}(2013{\natexlab{b}}){Behroozi}, {Wechsler}, \& {Wu}}]{Behroozi_2013}
{Behroozi}, P.~S., {Wechsler}, R.~H., \& {Wu}, H.-Y. 2013{\natexlab{b}}, \apj, 762, 109, \dodoi{10.1088/0004-637X/762/2/109}

\bibitem[{{Bouwens} {et~al.}(2023{\natexlab{a}}){Bouwens}, {Illingworth}, {Oesch}, {Stefanon}, {Naidu}, {van Leeuwen}, \& {Magee}}]{Bouwens2023a}
{Bouwens}, R., {Illingworth}, G., {Oesch}, P., {et~al.} 2023{\natexlab{a}}, \mnras, 523, 1009, \dodoi{10.1093/mnras/stad1014}

\bibitem[{{Bouwens} {et~al.}(2023{\natexlab{b}}){Bouwens}, {Stefanon}, {Brammer}, {Oesch}, {Herard-Demanche}, {Illingworth}, {Matthee}, {Naidu}, {van Dokkum}, \& {van Leeuwen}}]{Bouwens2023b}
{Bouwens}, R.~J., {Stefanon}, M., {Brammer}, G., {et~al.} 2023{\natexlab{b}}, \mnras, 523, 1036, \dodoi{10.1093/mnras/stad1145}

\bibitem[{{Boylan-Kolchin}(2023)}]{Boylan-Kolchin2023}
{Boylan-Kolchin}, M. 2023, Nature Astronomy, 7, 731, \dodoi{10.1038/s41550-023-01937-7}

\bibitem[{{Bromm}(2013)}]{Bromm2013}
{Bromm}, V. 2013, Reports on Progress in Physics, 76, 112901, \dodoi{10.1088/0034-4885/76/11/112901}

\bibitem[{{Bromm} {et~al.}(2001){Bromm}, {Kudritzki}, \& {Loeb}}]{Bromm2001}
{Bromm}, V., {Kudritzki}, R.~P., \& {Loeb}, A. 2001, \apj, 552, 464, \dodoi{10.1086/320549}

\bibitem[{{Bromm} \& {Yoshida}(2011)}]{Bromm2011}
{Bromm}, V., \& {Yoshida}, N. 2011, \araa, 49, 373, \dodoi{10.1146/annurev-astro-081710-102608}

\bibitem[{{Bunker} {et~al.}(2023){Bunker}, {Saxena}, {Cameron}, {Willott}, {Curtis-Lake}, {Jakobsen}, {Carniani}, {Smit}, {Maiolino}, {Witstok}, {Curti}, {D'Eugenio}, {Jones}, {Ferruit}, {Arribas}, {Charlot}, {Chevallard}, {Giardino}, {de Graaff}, {Looser}, {L{\"u}tzgendorf}, {Maseda}, {Rawle}, {Rix}, {Del Pino}, {Alberts}, {Egami}, {Eisenstein}, {Endsley}, {Hainline}, {Hausen}, {Johnson}, {Rieke}, {Rieke}, {Robertson}, {Shivaei}, {Stark}, {Sun}, {Tacchella}, {Tang}, {Williams}, {Willmer}, {Baker}, {Baum}, {Bhatawdekar}, {Bowler}, {Boyett}, {Chen}, {Circosta}, {Helton}, {Ji}, {Kumari}, {Lyu}, {Nelson}, {Parlanti}, {Perna}, {Sandles}, {Scholtz}, {Suess}, {Topping}, {{\"U}bler}, {Wallace}, \& {Whitler}}]{Bunker2023}
{Bunker}, A.~J., {Saxena}, A., {Cameron}, A.~J., {et~al.} 2023, \aap, 677, A88, \dodoi{10.1051/0004-6361/202346159}

\bibitem[{{Calzetti} {et~al.}(1994){Calzetti}, {Kinney}, \& {Storchi-Bergmann}}]{Calzetti1994}
{Calzetti}, D., {Kinney}, A.~L., \& {Storchi-Bergmann}, T. 1994, \apj, 429, 582, \dodoi{10.1086/174346}

\bibitem[{{Cameron} {et~al.}(2024){Cameron}, {Katz}, {Witten}, {Saxena}, {Laporte}, \& {Bunker}}]{Cameron2024}
{Cameron}, A.~J., {Katz}, H., {Witten}, C., {et~al.} 2024, \mnras, 534, 523, \dodoi{10.1093/mnras/stae1547}

\bibitem[{{Caplar} \& {Tacchella}(2019)}]{Caplar2019}
{Caplar}, N., \& {Tacchella}, S. 2019, \mnras, 487, 3845, \dodoi{10.1093/mnras/stz1449}

\bibitem[{{Carniani} {et~al.}(2024){Carniani}, {Hainline}, {D'Eugenio}, {Eisenstein}, {Jakobsen}, {Witstok}, {Johnson}, {Chevallard}, {Maiolino}, {Helton}, {Willott}, {Robertson}, {Alberts}, {Arribas}, {Baker}, {Bhatawdekar}, {Boyett}, {Bunker}, {Cameron}, {Cargile}, {Charlot}, {Curti}, {Curtis-Lake}, {Egami}, {Giardino}, {Isaak}, {Ji}, {Jones}, {Maseda}, {Parlanti}, {Rawle}, {Rieke}, {Rieke}, {Rodr{\'\i}guez Del Pino}, {Saxena}, {Scholtz}, {Smit}, {Sun}, {Tacchella}, {{\"U}bler}, {Venturi}, {Williams}, \& {Willmer}}]{Carniani2024}
{Carniani}, S., {Hainline}, K., {D'Eugenio}, F., {et~al.} 2024, arXiv e-prints, arXiv:2405.18485, \dodoi{10.48550/arXiv.2405.18485}

\bibitem[{{Casey} {et~al.}(2024){Casey}, {Akins}, {Shuntov}, {Ilbert}, {Paquereau}, {Franco}, {Hayward}, {Finkelstein}, {Boylan-Kolchin}, {Robertson}, {Allen}, {Brinch}, {Cooper}, {Ding}, {Drakos}, {Faisst}, {Fujimoto}, {Gillman}, {Harish}, {Hirschmann}, {Jin}, {Kartaltepe}, {Koekemoer}, {Kokorev}, {Liu}, {Long}, {Magdis}, {Maraston}, {Martin}, {McCracken}, {McKinney}, {Mobasher}, {Rhodes}, {Rich}, {Sanders}, {Silverman}, {Toft}, {Vijayan}, {Weaver}, {Wilkins}, {Yang}, \& {Zavala}}]{Casey2024}
{Casey}, C.~M., {Akins}, H.~B., {Shuntov}, M., {et~al.} 2024, \apj, 965, 98, \dodoi{10.3847/1538-4357/ad2075}

\bibitem[{{Castellano} {et~al.}(2024){Castellano}, {Napolitano}, {Fontana}, {Roberts-Borsani}, {Treu}, {Vanzella}, {Zavala}, {Arrabal Haro}, {Calabr{\`o}}, {Llerena}, {Mascia}, {Merlin}, {Paris}, {Pentericci}, {Santini}, {Bakx}, {Bergamini}, {Cupani}, {Dickinson}, {Filippenko}, {Glazebrook}, {Grillo}, {Kelly}, {Malkan}, {Mason}, {Morishita}, {Nanayakkara}, {Rosati}, {Sani}, {Wang}, \& {Yoon}}]{Castellano2024}
{Castellano}, M., {Napolitano}, L., {Fontana}, A., {et~al.} 2024, arXiv e-prints, arXiv:2403.10238, \dodoi{10.48550/arXiv.2403.10238}

\bibitem[{{Chabrier}(2003)}]{Chabrier2003}
{Chabrier}, G. 2003, \pasp, 115, 763, \dodoi{10.1086/376392}

\bibitem[{{Chatzikos} {et~al.}(2018){Chatzikos}, {Ferland}, {Guzman}, {van Hoof}, \& {Williams}}]{Chatzikos2018}
{Chatzikos}, M., {Ferland}, G., {Guzman}, F., {van Hoof}, P., \& {Williams}, R. 2018, in Walking the Line 2018, 13, \dodoi{10.5281/zenodo.1209973}

\bibitem[{{Chen} {et~al.}(2014){Chen}, {Wise}, {Norman}, {Xu}, \& {O'Shea}}]{Chen2014}
{Chen}, P., {Wise}, J.~H., {Norman}, M.~L., {Xu}, H., \& {O'Shea}, B.~W. 2014, \apj, 795, 144, \dodoi{10.1088/0004-637X/795/2/144}

\bibitem[{{Chon} {et~al.}(2022){Chon}, {Ono}, {Omukai}, \& {Schneider}}]{Chon2022}
{Chon}, S., {Ono}, H., {Omukai}, K., \& {Schneider}, R. 2022, \mnras, 514, 4639, \dodoi{10.1093/mnras/stac1549}

\bibitem[{{Clark} {et~al.}(2011){Clark}, {Glover}, {Smith}, {Greif}, {Klessen}, \& {Bromm}}]{Clark2011}
{Clark}, P.~C., {Glover}, S. C.~O., {Smith}, R.~J., {et~al.} 2011, Science, 331, 1040, \dodoi{10.1126/science.1198027}

\bibitem[{{Conroy} \& {Gunn}(2010)}]{Conroy2010}
{Conroy}, C., \& {Gunn}, J.~E. 2010, {FSPS: Flexible Stellar Population Synthesis}, Astrophysics Source Code Library, record ascl:1010.043.
\newblock \doeprint{1010.043}

\bibitem[{{Cueto} {et~al.}(2024){Cueto}, {Hutter}, {Dayal}, {Gottl{\"o}ber}, {Heintz}, {Mason}, {Trebitsch}, \& {Yepes}}]{Cueto2024}
{Cueto}, E.~R., {Hutter}, A., {Dayal}, P., {et~al.} 2024, \aap, 686, A138, \dodoi{10.1051/0004-6361/202349017}

\bibitem[{{Cullen} {et~al.}(2023){Cullen}, {McLure}, {McLeod}, {Dunlop}, {Donnan}, {Carnall}, {Bowler}, {Begley}, {Hamadouche}, \& {Stanton}}]{Cullen2023B}
{Cullen}, F., {McLure}, R.~J., {McLeod}, D.~J., {et~al.} 2023, \mnras, 520, 14, \dodoi{10.1093/mnras/stad073}

\bibitem[{{Cullen} {et~al.}(2024){Cullen}, {McLeod}, {McLure}, {Dunlop}, {Donnan}, {Carnall}, {Keating}, {Magee}, {Arellano-Cordova}, {Bowler}, {Begley}, {Flury}, {Hamadouche}, \& {Stanton}}]{Cullen2024}
{Cullen}, F., {McLeod}, D.~J., {McLure}, R.~J., {et~al.} 2024, \mnras, 531, 997, \dodoi{10.1093/mnras/stae1211}

\bibitem[{{Curti} {et~al.}(2024){Curti}, {Witstok}, {Jakobsen}, {Kobayashi}, {Curtis-Lake}, {Hainline}, {Ji}, {D'Eugenio}, {Chevallard}, {Maiolino}, {Scholtz}, {Carniani}, {Arribas}, {Baker}, {Bhatawdekar}, {Boyett}, {Bunker}, {Cameron}, {Cargile}, {Charlot}, {Eisenstein}, {Ji}, {Johnson}, {Kumari}, {Maseda}, {Robertson}, {Silcock}, {Tacchella}, {Ubler}, {Venturi}, {Williams}, {Willmer}, \& {Willott}}]{Curti2024}
{Curti}, M., {Witstok}, J., {Jakobsen}, P., {et~al.} 2024, arXiv e-prints, arXiv:2407.02575, \dodoi{10.48550/arXiv.2407.02575}

\bibitem[{{Curtis-Lake} {et~al.}(2023){Curtis-Lake}, {Carniani}, {Cameron}, {Charlot}, {Jakobsen}, {Maiolino}, {Bunker}, {Witstok}, {Smit}, {Chevallard}, {Willott}, {Ferruit}, {Arribas}, {Bonaventura}, {Curti}, {D'Eugenio}, {Franx}, {Giardino}, {Looser}, {L{\"u}tzgendorf}, {Maseda}, {Rawle}, {Rix}, {Rodr{\'\i}guez del Pino}, {{\"U}bler}, {Sirianni}, {Dressler}, {Egami}, {Eisenstein}, {Endsley}, {Hainline}, {Hausen}, {Johnson}, {Rieke}, {Robertson}, {Shivaei}, {Stark}, {Tacchella}, {Williams}, {Willmer}, {Bhatawdekar}, {Bowler}, {Boyett}, {Chen}, {de Graaff}, {Helton}, {Hviding}, {Jones}, {Kumari}, {Lyu}, {Nelson}, {Perna}, {Sandles}, {Saxena}, {Suess}, {Sun}, {Topping}, {Wallace}, \& {Whitler}}]{Curtis-Lake2023}
{Curtis-Lake}, E., {Carniani}, S., {Cameron}, A., {et~al.} 2023, Nature Astronomy, 7, 622, \dodoi{10.1038/s41550-023-01918-w}

\bibitem[{{Dalla Vecchia} \& {Schaye}(2012)}]{DallaVecchia2012}
{Dalla Vecchia}, C., \& {Schaye}, J. 2012, \mnras, 426, 140, \dodoi{10.1111/j.1365-2966.2012.21704.x}

\bibitem[{{De Rossi} \& {Bromm}(2023)}]{DeRossi2023}
{De Rossi}, M.~E., \& {Bromm}, V. 2023, \apjl, 946, L20, \dodoi{10.3847/2041-8213/acc32e}

\bibitem[{{Dekel} {et~al.}(2023){Dekel}, {Sarkar}, {Birnboim}, {Mandelker}, \& {Li}}]{Dekel2023}
{Dekel}, A., {Sarkar}, K.~C., {Birnboim}, Y., {Mandelker}, N., \& {Li}, Z. 2023, \mnras, 523, 3201, \dodoi{10.1093/mnras/stad1557}

\bibitem[{{Dome} {et~al.}(2024){Dome}, {Martin-Alvarez}, {Tacchella}, {Yuan}, \& {Sijacki}}]{Dome2024}
{Dome}, T., {Martin-Alvarez}, S., {Tacchella}, S., {Yuan}, Y., \& {Sijacki}, D. 2024, arXiv e-prints, arXiv:2410.00113, \dodoi{10.48550/arXiv.2410.00113}

\bibitem[{{Donnan} {et~al.}(2023{\natexlab{a}}){Donnan}, {McLeod}, {McLure}, {Dunlop}, {Carnall}, {Cullen}, \& {Magee}}]{Donnan2023b}
{Donnan}, C.~T., {McLeod}, D.~J., {McLure}, R.~J., {et~al.} 2023{\natexlab{a}}, \mnras, 520, 4554, \dodoi{10.1093/mnras/stad471}

\bibitem[{{Donnan} {et~al.}(2023{\natexlab{b}}){Donnan}, {McLeod}, {Dunlop}, {McLure}, {Carnall}, {Begley}, {Cullen}, {Hamadouche}, {Bowler}, {Magee}, {McCracken}, {Milvang-Jensen}, {Moneti}, \& {Targett}}]{Donnan2023a}
{Donnan}, C.~T., {McLeod}, D.~J., {Dunlop}, J.~S., {et~al.} 2023{\natexlab{b}}, \mnras, 518, 6011, \dodoi{10.1093/mnras/stac3472}

\bibitem[{{Donnan} {et~al.}(2024){Donnan}, {McLure}, {Dunlop}, {McLeod}, {Magee}, {Arellano-C{\'o}rdova}, {Barrufet}, {Begley}, {Bowler}, {Carnall}, {Cullen}, {Ellis}, {Fontana}, {Illingworth}, {Grogin}, {Hamadouche}, {Koekemoer}, {Liu}, {Mason}, {Santini}, \& {Stanton}}]{Donnan2024}
{Donnan}, C.~T., {McLure}, R.~J., {Dunlop}, J.~S., {et~al.} 2024, arXiv e-prints, arXiv:2403.03171, \dodoi{10.48550/arXiv.2403.03171}

\bibitem[{{Draine}(2003)}]{Draine2003}
{Draine}, B.~T. 2003, \araa, 41, 241, \dodoi{10.1146/annurev.astro.41.011802.094840}

\bibitem[{{Eldridge} \& {Stanway}(2009)}]{BPASS}
{Eldridge}, J.~J., \& {Stanway}, E.~R. 2009, \mnras, 400, 1019, \dodoi{10.1111/j.1365-2966.2009.15514.x}

\bibitem[{{Ellis} {et~al.}(2013){Ellis}, {McLure}, {Dunlop}, {Robertson}, {Ono}, {Schenker}, {Koekemoer}, {Bowler}, {Ouchi}, {Rogers}, {Curtis-Lake}, {Schneider}, {Charlot}, {Stark}, {Furlanetto}, \& {Cirasuolo}}]{Ellis2013}
{Ellis}, R.~S., {McLure}, R.~J., {Dunlop}, J.~S., {et~al.} 2013, \apjl, 763, L7, \dodoi{10.1088/2041-8205/763/1/L7}

\bibitem[{{Faucher-Gigu{\`e}re}(2018)}]{Faucher-Giguere2018}
{Faucher-Gigu{\`e}re}, C.-A. 2018, \mnras, 473, 3717, \dodoi{10.1093/mnras/stx2595}

\bibitem[{{Ferland} {et~al.}(1998){Ferland}, {Korista}, {Verner}, {Ferguson}, {Kingdon}, \& {Verner}}]{1998PASP..110..761F}
{Ferland}, G.~J., {Korista}, K.~T., {Verner}, D.~A., {et~al.} 1998, \pasp, 110, 761, \dodoi{10.1086/316190}

\bibitem[{{Ferrara}(2024{\natexlab{a}})}]{Ferrara2024}
{Ferrara}, A. 2024{\natexlab{a}}, \aap, 684, A207, \dodoi{10.1051/0004-6361/202348321}

\bibitem[{{Ferrara}(2024{\natexlab{b}})}]{Ferrara2024a}
---. 2024{\natexlab{b}}, \aap, 689, A310, \dodoi{10.1051/0004-6361/202450944}

\bibitem[{{Ferrara} {et~al.}(2023){Ferrara}, {Pallottini}, \& {Dayal}}]{Ferrara2023}
{Ferrara}, A., {Pallottini}, A., \& {Dayal}, P. 2023, \mnras, 522, 3986, \dodoi{10.1093/mnras/stad1095}

\bibitem[{{Fialkov} {et~al.}(2013){Fialkov}, {Barkana}, {Visbal}, {Tseliakhovich}, \& {Hirata}}]{Fialkov2013}
{Fialkov}, A., {Barkana}, R., {Visbal}, E., {Tseliakhovich}, D., \& {Hirata}, C.~M. 2013, \mnras, 432, 2909, \dodoi{10.1093/mnras/stt650}

\bibitem[{{Finkelstein} {et~al.}(2022){Finkelstein}, {Bagley}, {Arrabal Haro}, {Dickinson}, {Ferguson}, {Kartaltepe}, {Papovich}, {Burgarella}, {Kocevski}, {Huertas-Company}, {Iyer}, {Koekemoer}, {Larson}, {P{\'e}rez-Gonz{\'a}lez}, {Rose}, {Tacchella}, {Wilkins}, {Chworowsky}, {Medrano}, {Morales}, {Somerville}, {Yung}, {Fontana}, {Giavalisco}, {Grazian}, {Grogin}, {Kewley}, {Kirkpatrick}, {Kurczynski}, {Lotz}, {Pentericci}, {Pirzkal}, {Ravindranath}, {Ryan}, {Trump}, {Yang}, {Almaini}, {Amor{\'\i}n}, {Annunziatella}, {Backhaus}, {Barro}, {Behroozi}, {Bell}, {Bhatawdekar}, {Bisigello}, {Bromm}, {Buat}, {Buitrago}, {Calabr{\`o}}, {Casey}, {Castellano}, {Ch{\'a}vez Ortiz}, {Ciesla}, {Cleri}, {Cohen}, {Cole}, {Cooke}, {Cooper}, {Cooray}, {Costantin}, {Cox}, {Croton}, {Daddi}, {Dav{\'e}}, {de La Vega}, {Dekel}, {Elbaz}, {Estrada-Carpenter}, {Faber}, {Fern{\'a}ndez}, {Finkelstein}, {Freundlich}, {Fujimoto}, {Garc{\'\i}a-Argum{\'a}nez}, {Gardner}, {Gawiser}, {G{\'o}mez-Guijarro}, {Guo}, {Hamblin}, {Hamilton},
  {Hathi}, {Holwerda}, {Hirschmann}, {Hutchison}, {Jaskot}, {Jha}, {Jogee}, {Juneau}, {Jung}, {Kassin}, {Le Bail}, {Leung}, {Lucas}, {Magnelli}, {Mantha}, {Matharu}, {McGrath}, {McIntosh}, {Merlin}, {Mobasher}, {Newman}, {Nicholls}, {Pandya}, {Rafelski}, {Ronayne}, {Santini}, {Seill{\'e}}, {Shah}, {Shen}, {Simons}, {Snyder}, {Stanway}, {Straughn}, {Teplitz}, {Vanderhoof}, {Vega-Ferrero}, {Wang}, {Weiner}, {Willmer}, {Wuyts}, {Zavala}, \& {Ceers Team}}]{Finkelstein2022}
{Finkelstein}, S.~L., {Bagley}, M.~B., {Arrabal Haro}, P., {et~al.} 2022, \apjl, 940, L55, \dodoi{10.3847/2041-8213/ac966e}

\bibitem[{{Finkelstein} {et~al.}(2023){Finkelstein}, {Bagley}, {Ferguson}, {Wilkins}, {Kartaltepe}, {Papovich}, {Yung}, {Arrabal Haro}, {Behroozi}, {Dickinson}, {Kocevski}, {Koekemoer}, {Larson}, {Le Bail}, {Morales}, {P{\'e}rez-Gonz{\'a}lez}, {Burgarella}, {Dav{\'e}}, {Hirschmann}, {Somerville}, {Wuyts}, {Bromm}, {Casey}, {Fontana}, {Fujimoto}, {Gardner}, {Giavalisco}, {Grazian}, {Grogin}, {Hathi}, {Hutchison}, {Jha}, {Jogee}, {Kewley}, {Kirkpatrick}, {Long}, {Lotz}, {Pentericci}, {Pierel}, {Pirzkal}, {Ravindranath}, {Ryan}, {Trump}, {Yang}, {Bhatawdekar}, {Bisigello}, {Buat}, {Calabr{\`o}}, {Castellano}, {Cleri}, {Cooper}, {Croton}, {Daddi}, {Dekel}, {Elbaz}, {Franco}, {Gawiser}, {Holwerda}, {Huertas-Company}, {Jaskot}, {Leung}, {Lucas}, {Mobasher}, {Pandya}, {Tacchella}, {Weiner}, \& {Zavala}}]{Finkelstein2023}
{Finkelstein}, S.~L., {Bagley}, M.~B., {Ferguson}, H.~C., {et~al.} 2023, \apjl, 946, L13, \dodoi{10.3847/2041-8213/acade4}

\bibitem[{{Finkelstein} {et~al.}(2024){Finkelstein}, {Leung}, {Bagley}, {Dickinson}, {Ferguson}, {Papovich}, {Akins}, {Arrabal Haro}, {Dav{\'e}}, {Dekel}, {Kartaltepe}, {Kocevski}, {Koekemoer}, {Pirzkal}, {Somerville}, {Yung}, {Amor{\'\i}n}, {Backhaus}, {Behroozi}, {Bisigello}, {Bromm}, {Casey}, {Ch{\'a}vez Ortiz}, {Cheng}, {Chworowsky}, {Cleri}, {Cooper}, {Davis}, {de la Vega}, {Elbaz}, {Franco}, {Fontana}, {Fujimoto}, {Giavalisco}, {Grogin}, {Holwerda}, {Huertas-Company}, {Hirschmann}, {Iyer}, {Jogee}, {Jung}, {Larson}, {Lucas}, {Mobasher}, {Morales}, {Morley}, {Mukherjee}, {P{\'e}rez-Gonz{\'a}lez}, {Ravindranath}, {Rodighiero}, {Rowland}, {Tacchella}, {Taylor}, {Trump}, \& {Wilkins}}]{Finkelstein2024}
{Finkelstein}, S.~L., {Leung}, G. C.~K., {Bagley}, M.~B., {et~al.} 2024, \apjl, 969, L2, \dodoi{10.3847/2041-8213/ad4495}

\bibitem[{{Furlanetto} \& {Mirocha}(2022)}]{Furlanetto2022}
{Furlanetto}, S.~R., \& {Mirocha}, J. 2022, \mnras, 511, 3895, \dodoi{10.1093/mnras/stac310}

\bibitem[{{Furtak} {et~al.}(2023){Furtak}, {Shuntov}, {Atek}, {Zitrin}, {Richard}, {Lehnert}, \& {Chevallard}}]{Furtak2023}
{Furtak}, L.~J., {Shuntov}, M., {Atek}, H., {et~al.} 2023, \mnras, 519, 3064, \dodoi{10.1093/mnras/stac3717}

\bibitem[{{Gelli} {et~al.}(2024){Gelli}, {Mason}, \& {Hayward}}]{Gelli2024}
{Gelli}, V., {Mason}, C., \& {Hayward}, C.~C. 2024, \apj, 975, 192, \dodoi{10.3847/1538-4357/ad7b36}

\bibitem[{{Greif} {et~al.}(2009){Greif}, {Glover}, {Bromm}, \& {Klessen}}]{Greif2009}
{Greif}, T.~H., {Glover}, S. C.~O., {Bromm}, V., \& {Klessen}, R.~S. 2009, \mnras, 392, 1381, \dodoi{10.1111/j.1365-2966.2008.14169.x}

\bibitem[{{Grudi{\'c}} {et~al.}(2019){Grudi{\'c}}, {Hopkins}, {Lee}, {Murray}, {Faucher-Gigu{\`e}re}, \& {Johnson}}]{Gaudic2019}
{Grudi{\'c}}, M.~Y., {Hopkins}, P.~F., {Lee}, E.~J., {et~al.} 2019, \mnras, 488, 1501, \dodoi{10.1093/mnras/stz1758}

\bibitem[{Hahn \& Abel(2011)}]{Hahn_2011}
Hahn, O., \& Abel, T. 2011, Monthly Notices of the Royal Astronomical Society, 415, 2101–2121, \dodoi{10.1111/j.1365-2966.2011.18820.x}

\bibitem[{{Hainline} {et~al.}(2024){Hainline}, {D'Eugenio}, {Jakobsen}, {Chevallard}, {Carniani}, {Witstok}, {Ji}, {Curtis-Lake}, {Johnson}, {Robertson}, {Tacchella}, {Curti}, {Charlot}, {Helton}, {Arribas}, {Bhatawdekar}, {Bunker}, {Cameron}, {Egami}, {Eisenstein}, {Hausen}, {Kumari}, {Maiolino}, {Perez-Gonzalez}, {Rieke}, {Saxena}, {Scholtz}, {Smit}, {Sun}, {Williams}, {Willmer}, \& {Willott}}]{Hainline2024}
{Hainline}, K.~N., {D'Eugenio}, F., {Jakobsen}, P., {et~al.} 2024, arXiv e-prints, arXiv:2404.04325, \dodoi{10.48550/arXiv.2404.04325}

\bibitem[{{Harikane} {et~al.}(2024){Harikane}, {Nakajima}, {Ouchi}, {Umeda}, {Isobe}, {Ono}, {Xu}, \& {Zhang}}]{Harikane2024}
{Harikane}, Y., {Nakajima}, K., {Ouchi}, M., {et~al.} 2024, \apj, 960, 56, \dodoi{10.3847/1538-4357/ad0b7e}

\bibitem[{{Harikane} {et~al.}(2022){Harikane}, {Ono}, {Ouchi}, {Liu}, {Sawicki}, {Shibuya}, {Behroozi}, {He}, {Shimasaku}, {Arnouts}, {Coupon}, {Fujimoto}, {Gwyn}, {Huang}, {Inoue}, {Kashikawa}, {Komiyama}, {Matsuoka}, \& {Willott}}]{Harikane2022}
{Harikane}, Y., {Ono}, Y., {Ouchi}, M., {et~al.} 2022, \apjs, 259, 20, \dodoi{10.3847/1538-4365/ac3dfc}

\bibitem[{{Harikane} {et~al.}(2023){Harikane}, {Ouchi}, {Oguri}, {Ono}, {Nakajima}, {Isobe}, {Umeda}, {Mawatari}, \& {Zhang}}]{Harikane2023}
{Harikane}, Y., {Ouchi}, M., {Oguri}, M., {et~al.} 2023, \apjs, 265, 5, \dodoi{10.3847/1538-4365/acaaa9}

\bibitem[{{Hartwig} {et~al.}(2022){Hartwig}, {Magg}, {Chen}, {Tarumi}, {Bromm}, {Glover}, {Ji}, {Klessen}, {Latif}, {Volonteri}, \& {Yoshida}}]{Hartwig2022}
{Hartwig}, T., {Magg}, M., {Chen}, L.-H., {et~al.} 2022, \apj, 936, 45, \dodoi{10.3847/1538-4357/ac7150}

\bibitem[{{Harvey} {et~al.}(2024){Harvey}, {Conselice}, {Adams}, {Austin}, {Juodzbalis}, {Trussler}, {Li}, {Ormerod}, {Ferreira}, {Duan}, {Westcott}, {Harris}, {Bhatawdekar}, {Coe}, {Cohen}, {Caruana}, {Cheng}, {Driver}, {Frye}, {Furtak}, {Grogin}, {Hathi}, {Holwerda}, {Jansen}, {Koekemoer}, {Lovell}, {Marshall}, {Nonino}, {Smail}, {Vijayan}, {Wilkins}, {Windhorst}, {Willmer}, {Yan}, \& {Zitrin}}]{Harvey2024}
{Harvey}, T., {Conselice}, C., {Adams}, N.~J., {et~al.} 2024, arXiv e-prints, arXiv:2403.03908, \dodoi{10.48550/arXiv.2403.03908}

\bibitem[{{Haslbauer} {et~al.}(2024){Haslbauer}, {Yan}, {Jerabkova}, {Gjergo}, {Kroupa}, \& {Hasani Zonoozi}}]{Haslbauer2024}
{Haslbauer}, M., {Yan}, Z., {Jerabkova}, T., {et~al.} 2024, \aap, 689, A221, \dodoi{10.1051/0004-6361/202347928}

\bibitem[{{Hegde} {et~al.}(2024){Hegde}, {Wyatt}, \& {Furlanetto}}]{Hegde2024}
{Hegde}, S., {Wyatt}, M.~M., \& {Furlanetto}, S.~R. 2024, \jcap, 2024, 025, \dodoi{10.1088/1475-7516/2024/08/025}

\bibitem[{{Heger} \& {Woosley}(2002)}]{Heger2002}
{Heger}, A., \& {Woosley}, S.~E. 2002, \apj, 567, 532, \dodoi{10.1086/338487}

\bibitem[{{Heger} \& {Woosley}(2010)}]{Heger2010}
---. 2010, \apj, 724, 341, \dodoi{10.1088/0004-637X/724/1/341}

\bibitem[{{Hirano} {et~al.}(2015){Hirano}, {Hosokawa}, {Yoshida}, {Omukai}, \& {Yorke}}]{Hirano2015}
{Hirano}, S., {Hosokawa}, T., {Yoshida}, N., {Omukai}, K., \& {Yorke}, H.~W. 2015, \mnras, 448, 568, \dodoi{10.1093/mnras/stv044}

\bibitem[{{Hirano} \& {Yoshida}(2024)}]{Hirano2024}
{Hirano}, S., \& {Yoshida}, N. 2024, \apj, 963, 2, \dodoi{10.3847/1538-4357/ad22e0}

\bibitem[{{Hosokawa} {et~al.}(2016){Hosokawa}, {Hirano}, {Kuiper}, {Yorke}, {Omukai}, \& {Yoshida}}]{Hosokawa2016}
{Hosokawa}, T., {Hirano}, S., {Kuiper}, R., {et~al.} 2016, \apj, 824, 119, \dodoi{10.3847/0004-637X/824/2/119}

\bibitem[{{Hsiao} {et~al.}(2023){Hsiao}, {Abdurro'uf}, {Coe}, {Larson}, {Jung}, {Mingozzi}, {Dayal}, {Kumari}, {Kokorev}, {Vikaeus}, {Brammer}, {Furtak}, {Adamo}, {Andrade-Santos}, {Antwi-Danso}, {Bradac}, {Bradley}, {Broadhurst}, {Carnall}, {Conselice}, {Diego}, {Donahue}, {Eldridge}, {Fujimoto}, {Henry}, {Hernandez}, {Hutchison}, {James}, {Norman}, {Park}, {Pirzkal}, {Postman}, {Ricotti}, {Rigby}, {Vanzella}, {Welch}, {Wilkins}, {Windhorst}, {Xu}, {Zackrisson}, \& {Zitrin}}]{Hsiao2023}
{Hsiao}, T. Y.-Y., {Abdurro'uf}, {Coe}, D., {et~al.} 2023, arXiv e-prints, arXiv:2305.03042, \dodoi{10.48550/arXiv.2305.03042}

\bibitem[{{Ibrahim} \& {Kobayashi}(2024)}]{Ibrahim2024}
{Ibrahim}, D., \& {Kobayashi}, C. 2024, \mnras, 527, 3276, \dodoi{10.1093/mnras/stad3313}

\bibitem[{{Inayoshi} {et~al.}(2022){Inayoshi}, {Harikane}, {Inoue}, {Li}, \& {Ho}}]{Inayoshi2022}
{Inayoshi}, K., {Harikane}, Y., {Inoue}, A.~K., {Li}, W., \& {Ho}, L.~C. 2022, \apjl, 938, L10, \dodoi{10.3847/2041-8213/ac9310}

\bibitem[{{Jaacks} {et~al.}(2018){Jaacks}, {Finkelstein}, \& {Bromm}}]{Jaacks2018}
{Jaacks}, J., {Finkelstein}, S.~L., \& {Bromm}, V. 2018, \mnras, 475, 3883, \dodoi{10.1093/mnras/sty049}

\bibitem[{{Jaacks} {et~al.}(2019){Jaacks}, {Finkelstein}, \& {Bromm}}]{Jaacks2019}
---. 2019, \mnras, 488, 2202, \dodoi{10.1093/mnras/stz1529}

\bibitem[{{Jeon} {et~al.}(2017){Jeon}, {Besla}, \& {Bromm}}]{Jeon_2017}
{Jeon}, M., {Besla}, G., \& {Bromm}, V. 2017, \apj, 848, 85, \dodoi{10.3847/1538-4357/aa8c80}

\bibitem[{{Jeon} \& {Bromm}(2019)}]{Jeon_2019}
{Jeon}, M., \& {Bromm}, V. 2019, \mnras, 485, 5939, \dodoi{10.1093/mnras/stz863}

\bibitem[{{Jeon} {et~al.}(2014){Jeon}, {Pawlik}, {Bromm}, \& {Milosavljevi{\'c}}}]{Jeon_2014}
{Jeon}, M., {Pawlik}, A.~H., {Bromm}, V., \& {Milosavljevi{\'c}}, M. 2014, \mnras, 444, 3288, \dodoi{10.1093/mnras/stu1980}

\bibitem[{{Je{\v{r}}{\'a}bkov{\'a}} {et~al.}(2018){Je{\v{r}}{\'a}bkov{\'a}}, {Hasani Zonoozi}, {Kroupa}, {Beccari}, {Yan}, {Vazdekis}, \& {Zhang}}]{Jerabkova2018}
{Je{\v{r}}{\'a}bkov{\'a}}, T., {Hasani Zonoozi}, A., {Kroupa}, P., {et~al.} 2018, \aap, 620, A39, \dodoi{10.1051/0004-6361/201833055}

\bibitem[{{Ji} {et~al.}(2014){Ji}, {Frebel}, \& {Bromm}}]{Ji2014}
{Ji}, A.~P., {Frebel}, A., \& {Bromm}, V. 2014, \apj, 782, 95, \dodoi{10.1088/0004-637X/782/2/95}

\bibitem[{Johnson {et~al.}(2024)Johnson, Foreman-Mackey, Sick, Leja, Walmsley, Tollerud, Leung, Scott, \& Park}]{ben_johnson_2024_12447779}
Johnson, B., Foreman-Mackey, D., Sick, J., {et~al.} 2024, dfm/python-fsps: v0.4.7, v0.4.7,  Zenodo, \dodoi{10.5281/zenodo.12447779}

\bibitem[{{Jones} \& {Nuth}(2011)}]{Jones2011}
{Jones}, A.~P., \& {Nuth}, J.~A. 2011, \aap, 530, A44, \dodoi{10.1051/0004-6361/201014440}

\bibitem[{{Jones} {et~al.}(2024){Jones}, {Smith}, {Dav{\'e}}, {Narayanan}, \& {Li}}]{Jones2024}
{Jones}, E., {Smith}, B., {Dav{\'e}}, R., {Narayanan}, D., \& {Li}, Q. 2024, arXiv e-prints, arXiv:2402.06728, \dodoi{10.48550/arXiv.2402.06728}

\bibitem[{{Kang} {et~al.}(2024){Kang}, {Kimm}, {Han}, {Katz}, {Devriendt}, {Slyz}, \& {Teyssier}}]{Kang2024}
{Kang}, C., {Kimm}, T., {Han}, D., {et~al.} 2024, arXiv e-prints, arXiv:2407.12090, \dodoi{10.48550/arXiv.2407.12090}

\bibitem[{{Kannan} {et~al.}(2022){Kannan}, {Garaldi}, {Smith}, {Pakmor}, {Springel}, {Vogelsberger}, \& {Hernquist}}]{Kannan2022}
{Kannan}, R., {Garaldi}, E., {Smith}, A., {et~al.} 2022, \mnras, 511, 4005, \dodoi{10.1093/mnras/stab3710}

\bibitem[{{Katz} {et~al.}(2023){Katz}, {Kimm}, {Ellis}, {Devriendt}, \& {Slyz}}]{Katz2023}
{Katz}, H., {Kimm}, T., {Ellis}, R.~S., {Devriendt}, J., \& {Slyz}, A. 2023, \mnras, 524, 351, \dodoi{10.1093/mnras/stad1903}

\bibitem[{{Keller} {et~al.}(2023){Keller}, {Munshi}, {Trebitsch}, \& {Tremmel}}]{Keller2023}
{Keller}, B.~W., {Munshi}, F., {Trebitsch}, M., \& {Tremmel}, M. 2023, \apjl, 943, L28, \dodoi{10.3847/2041-8213/acb148}

\bibitem[{{Kim} {et~al.}(2023){Kim}, {Jeon}, {Choi}, {Richstein}, {Sacchi}, \& {Kallivayalil}}]{Kim2023}
{Kim}, J., {Jeon}, M., {Choi}, Y., {et~al.} 2023, \apj, 959, 31, \dodoi{10.3847/1538-4357/acfe08}

\bibitem[{{Klessen} \& {Glover}(2023)}]{Klessen2023}
{Klessen}, R.~S., \& {Glover}, S. C.~O. 2023, \araa, 61, 65, \dodoi{10.1146/annurev-astro-071221-053453}

\bibitem[{{Kokorev} {et~al.}(2024{\natexlab{a}}){Kokorev}, {Caputi}, {Greene}, {Dayal}, {Trebitsch}, {Cutler}, {Fujimoto}, {Labb{\'e}}, {Miller}, {Iani}, {Navarro-Carrera}, \& {Rinaldi}}]{Kokorev2024a}
{Kokorev}, V., {Caputi}, K.~I., {Greene}, J.~E., {et~al.} 2024{\natexlab{a}}, \apj, 968, 38, \dodoi{10.3847/1538-4357/ad4265}

\bibitem[{{Kokorev} {et~al.}(2024{\natexlab{b}}){Kokorev}, {Chisholm}, {Endsley}, {Finkelstein}, {Greene}, {Akins}, {Bromm}, {Casey}, {Fujimoto}, {Labb{\'e}}, \& {Larson}}]{Kokorev2024b}
{Kokorev}, V., {Chisholm}, J., {Endsley}, R., {et~al.} 2024{\natexlab{b}}, arXiv e-prints, arXiv:2407.20320, \dodoi{10.48550/arXiv.2407.20320}

\bibitem[{{Kroupa}(2001)}]{Kroupa2001}
{Kroupa}, P. 2001, \mnras, 322, 231, \dodoi{10.1046/j.1365-8711.2001.04022.x}

\bibitem[{{Kroupa} {et~al.}(2024){Kroupa}, {Gjergo}, {Jerabkova}, \& {Yan}}]{Kroupa2024}
{Kroupa}, P., {Gjergo}, E., {Jerabkova}, T., \& {Yan}, Z. 2024, arXiv e-prints, arXiv:2410.07311, \dodoi{10.48550/arXiv.2410.07311}

\bibitem[{{Kulkarni} {et~al.}(2021){Kulkarni}, {Visbal}, \& {Bryan}}]{Kulkarni2021}
{Kulkarni}, M., {Visbal}, E., \& {Bryan}, G.~L. 2021, \apj, 917, 40, \dodoi{10.3847/1538-4357/ac08a3}

\bibitem[{{Larson}(1998)}]{Larson1998}
{Larson}, R.~B. 1998, \mnras, 301, 569, \dodoi{10.1046/j.1365-8711.1998.02045.x}

\bibitem[{{Latif} {et~al.}(2022){Latif}, {Whalen}, \& {Khochfar}}]{Latif2022}
{Latif}, M.~A., {Whalen}, D., \& {Khochfar}, S. 2022, \apj, 925, 28, \dodoi{10.3847/1538-4357/ac3916}

\bibitem[{{Lazar} \& {Bromm}(2022)}]{Lazar2022}
{Lazar}, A., \& {Bromm}, V. 2022, \mnras, 511, 2505, \dodoi{10.1093/mnras/stac176}

\bibitem[{{Lee} {et~al.}(2024){Lee}, {Jeon}, \& {Bromm}}]{Lee2024}
{Lee}, T., {Jeon}, M., \& {Bromm}, V. 2024, \mnras, 527, 1257, \dodoi{10.1093/mnras/stad3198}

\bibitem[{{Leitherer} {et~al.}(1999){Leitherer}, {Schaerer}, {Goldader}, {Delgado}, {Robert}, {Kune}, {de Mello}, {Devost}, \& {Heckman}}]{Starburst99}
{Leitherer}, C., {Schaerer}, D., {Goldader}, J.~D., {et~al.} 1999, \apjs, 123, 3, \dodoi{10.1086/313233}

\bibitem[{{Leroy} {et~al.}(2008){Leroy}, {Walter}, {Brinks}, {Bigiel}, {de Blok}, {Madore}, \& {Thornley}}]{Leroy2008}
{Leroy}, A.~K., {Walter}, F., {Brinks}, E., {et~al.} 2008, \aj, 136, 2782, \dodoi{10.1088/0004-6256/136/6/2782}

\bibitem[{{Li} {et~al.}(2023){Li}, {Dekel}, {Sarkar}, {Aung}, {Giavalisco}, {Mandelker}, \& {Tacchella}}]{Li2023}
{Li}, Z., {Dekel}, A., {Sarkar}, K.~C., {et~al.} 2023, arXiv e-prints, arXiv:2311.14662, \dodoi{10.48550/arXiv.2311.14662}

\bibitem[{{Liu} \& {Bromm}(2020)}]{Liu2020MNRAS}
{Liu}, B., \& {Bromm}, V. 2020, \mnras, 497, 2839, \dodoi{10.1093/mnras/staa2143}

\bibitem[{{Liu} \& {Bromm}(2022)}]{Liu2022}
---. 2022, \apjl, 937, L30, \dodoi{10.3847/2041-8213/ac927f}

\bibitem[{{Liu} \& {Bromm}(2023)}]{Liu2023}
---. 2023, arXiv e-prints, arXiv:2312.04085, \dodoi{10.48550/arXiv.2312.04085}

\bibitem[{{Liu} {et~al.}(2024){Liu}, {Gurian}, {Inayoshi}, {Hirano}, {Hosokawa}, {Bromm}, \& {Yoshida}}]{Liu2024}
{Liu}, B., {Gurian}, J., {Inayoshi}, K., {et~al.} 2024, arXiv e-prints, arXiv:2407.14294, \dodoi{10.48550/arXiv.2407.14294}

\bibitem[{{Lu} {et~al.}(2024){Lu}, {Frenk}, {Bose}, {Lacey}, {Cole}, {Baugh}, \& {Helly}}]{Lu2024}
{Lu}, S., {Frenk}, C.~S., {Bose}, S., {et~al.} 2024, arXiv e-prints, arXiv:2406.02672, \dodoi{10.48550/arXiv.2406.02672}

\bibitem[{{Ma} {et~al.}(2016){Ma}, {Hopkins}, {Faucher-Gigu{\`e}re}, {Zolman}, {Muratov}, {Kere{\v{s}}}, \& {Quataert}}]{Ma2016}
{Ma}, X., {Hopkins}, P.~F., {Faucher-Gigu{\`e}re}, C.-A., {et~al.} 2016, \mnras, 456, 2140, \dodoi{10.1093/mnras/stv2659}

\bibitem[{{Ma} {et~al.}(2018{\natexlab{a}}){Ma}, {Hopkins}, {Boylan-Kolchin}, {Faucher-Gigu{\`e}re}, {Quataert}, {Feldmann}, {Garrison-Kimmel}, {Hayward}, {Kere{\v{s}}}, \& {Wetzel}}]{Ma2018a}
{Ma}, X., {Hopkins}, P.~F., {Boylan-Kolchin}, M., {et~al.} 2018{\natexlab{a}}, \mnras, 477, 219, \dodoi{10.1093/mnras/sty684}

\bibitem[{{Ma} {et~al.}(2018{\natexlab{b}}){Ma}, {Hopkins}, {Garrison-Kimmel}, {Faucher-Gigu{\`e}re}, {Quataert}, {Boylan-Kolchin}, {Hayward}, {Feldmann}, \& {Kere{\v{s}}}}]{Ma2018b}
{Ma}, X., {Hopkins}, P.~F., {Garrison-Kimmel}, S., {et~al.} 2018{\natexlab{b}}, \mnras, 478, 1694, \dodoi{10.1093/mnras/sty1024}

\bibitem[{{Ma} {et~al.}(2019){Ma}, {Hayward}, {Casey}, {Hopkins}, {Quataert}, {Liang}, {Faucher-Gigu{\`e}re}, {Feldmann}, \& {Kere{\v{s}}}}]{Ma2019}
{Ma}, X., {Hayward}, C.~C., {Casey}, C.~M., {et~al.} 2019, \mnras, 487, 1844, \dodoi{10.1093/mnras/stz1324}

\bibitem[{{Ma} {et~al.}(2020){Ma}, {Grudi{\'c}}, {Quataert}, {Hopkins}, {Faucher-Gigu{\`e}re}, {Boylan-Kolchin}, {Wetzel}, {Kim}, {Murray}, \& {Kere{\v{s}}}}]{Ma2020}
{Ma}, X., {Grudi{\'c}}, M.~Y., {Quataert}, E., {et~al.} 2020, \mnras, 493, 4315, \dodoi{10.1093/mnras/staa527}

\bibitem[{{Madau} \& {Dickinson}(2014)}]{Madau2014}
{Madau}, P., \& {Dickinson}, M. 2014, \araa, 52, 415, \dodoi{10.1146/annurev-astro-081811-125615}

\bibitem[{{Maiolino} {et~al.}(2024){Maiolino}, {Scholtz}, {Witstok}, {Carniani}, {D'Eugenio}, {de Graaff}, {{\"U}bler}, {Tacchella}, {Curtis-Lake}, {Arribas}, {Bunker}, {Charlot}, {Chevallard}, {Curti}, {Looser}, {Maseda}, {Rawle}, {Rodr{\'\i}guez del Pino}, {Willott}, {Egami}, {Eisenstein}, {Hainline}, {Robertson}, {Williams}, {Willmer}, {Baker}, {Boyett}, {DeCoursey}, {Fabian}, {Helton}, {Ji}, {Jones}, {Kumari}, {Laporte}, {Nelson}, {Perna}, {Sandles}, {Shivaei}, \& {Sun}}]{Maiolino2024}
{Maiolino}, R., {Scholtz}, J., {Witstok}, J., {et~al.} 2024, \nat, 627, 59, \dodoi{10.1038/s41586-024-07052-5}

\bibitem[{{Marigo}(2001)}]{Marigo2001}
{Marigo}, P. 2001, \aap, 370, 194, \dodoi{10.1051/0004-6361:20000247}

\bibitem[{{Marszewski} {et~al.}(2024){Marszewski}, {Sun}, {Faucher-Gigu{\`e}re}, {Hayward}, \& {Feldmann}}]{Marszewski2024}
{Marszewski}, A., {Sun}, G., {Faucher-Gigu{\`e}re}, C.-A., {Hayward}, C.~C., \& {Feldmann}, R. 2024, \apjl, 967, L41, \dodoi{10.3847/2041-8213/ad4cee}

\bibitem[{{Martin-Alvarez} {et~al.}(2023){Martin-Alvarez}, {Sijacki}, {Haehnelt}, {Farcy}, {Dubois}, {Belokurov}, {Rosdahl}, \& {Lopez-Rodriguez}}]{Martin-Alvarez2023}
{Martin-Alvarez}, S., {Sijacki}, D., {Haehnelt}, M.~G., {et~al.} 2023, \mnras, 525, 3806, \dodoi{10.1093/mnras/stad2559}

\bibitem[{{Mason} {et~al.}(2015){Mason}, {Trenti}, \& {Treu}}]{Mason2015}
{Mason}, C.~A., {Trenti}, M., \& {Treu}, T. 2015, \apj, 813, 21, \dodoi{10.1088/0004-637X/813/1/21}

\bibitem[{{Mason} {et~al.}(2023){Mason}, {Trenti}, \& {Treu}}]{Mason2023}
---. 2023, \mnras, 521, 497, \dodoi{10.1093/mnras/stad035}

\bibitem[{{Matthee} {et~al.}(2024){Matthee}, {Naidu}, {Brammer}, {Chisholm}, {Eilers}, {Goulding}, {Greene}, {Kashino}, {Labbe}, {Lilly}, {Mackenzie}, {Oesch}, {Weibel}, {Wuyts}, {Xiao}, {Bordoloi}, {Bouwens}, {van Dokkum}, {Illingworth}, {Kramarenko}, {Maseda}, {Mason}, {Meyer}, {Nelson}, {Reddy}, {Shivaei}, {Simcoe}, \& {Yue}}]{Matthee2024ApJ}
{Matthee}, J., {Naidu}, R.~P., {Brammer}, G., {et~al.} 2024, \apj, 963, 129, \dodoi{10.3847/1538-4357/ad2345}

\bibitem[{{McCaffrey} {et~al.}(2023){McCaffrey}, {Hardin}, {Wise}, \& {Regan}}]{McCaffrey2023}
{McCaffrey}, J., {Hardin}, S., {Wise}, J.~H., \& {Regan}, J.~A. 2023, The Open Journal of Astrophysics, 6, 47, \dodoi{10.21105/astro.2304.13755}

\bibitem[{{Mirocha} \& {Furlanetto}(2023)}]{Mirocha2023}
{Mirocha}, J., \& {Furlanetto}, S.~R. 2023, \mnras, 519, 843, \dodoi{10.1093/mnras/stac3578}

\bibitem[{{Morales} {et~al.}(2024){Morales}, {Finkelstein}, {Leung}, {Bagley}, {Cleri}, {Dave}, {Dickinson}, {Ferguson}, {Hathi}, {Jones}, {Koekemoer}, {Papovich}, {P{\'e}rez-Gonz{\'a}lez}, {Pirzkal}, {Smith}, {Wilkins}, \& {Yung}}]{Morales2024}
{Morales}, A.~M., {Finkelstein}, S.~L., {Leung}, G. C.~K., {et~al.} 2024, \apjl, 964, L24, \dodoi{10.3847/2041-8213/ad2de4}

\bibitem[{{Naidu} {et~al.}(2022){Naidu}, {Oesch}, {van Dokkum}, {Nelson}, {Suess}, {Brammer}, {Whitaker}, {Illingworth}, {Bouwens}, {Tacchella}, {Matthee}, {Allen}, {Bezanson}, {Conroy}, {Labbe}, {Leja}, {Leonova}, {Magee}, {Price}, {Setton}, {Strait}, {Stefanon}, {Toft}, {Weaver}, \& {Weibel}}]{Naidu2022}
{Naidu}, R.~P., {Oesch}, P.~A., {van Dokkum}, P., {et~al.} 2022, \apjl, 940, L14, \dodoi{10.3847/2041-8213/ac9b22}

\bibitem[{{Nakajima} \& {Maiolino}(2022)}]{Nakajima2022}
{Nakajima}, K., \& {Maiolino}, R. 2022, \mnras, 513, 5134, \dodoi{10.1093/mnras/stac1242}

\bibitem[{{Nelson} {et~al.}(2019){Nelson}, {Springel}, {Pillepich}, {Rodriguez-Gomez}, {Torrey}, {Genel}, {Vogelsberger}, {Pakmor}, {Marinacci}, {Weinberger}, {Kelley}, {Lovell}, {Diemer}, \& {Hernquist}}]{Nelson2019}
{Nelson}, D., {Springel}, V., {Pillepich}, A., {et~al.} 2019, Computational Astrophysics and Cosmology, 6, 2, \dodoi{10.1186/s40668-019-0028-x}

\bibitem[{{Oesch} {et~al.}(2016){Oesch}, {Brammer}, {van Dokkum}, {Illingworth}, {Bouwens}, {Labb{\'e}}, {Franx}, {Momcheva}, {Ashby}, {Fazio}, {Gonzalez}, {Holden}, {Magee}, {Skelton}, {Smit}, {Spitler}, {Trenti}, \& {Willner}}]{Oesch2016}
{Oesch}, P.~A., {Brammer}, G., {van Dokkum}, P.~G., {et~al.} 2016, \apj, 819, 129, \dodoi{10.3847/0004-637X/819/2/129}

\bibitem[{{Omukai}(2000)}]{Omukai2000}
{Omukai}, K. 2000, \apj, 534, 809, \dodoi{10.1086/308776}

\bibitem[{{Ono} {et~al.}(2023){Ono}, {Harikane}, {Ouchi}, {Yajima}, {Abe}, {Isobe}, {Shibuya}, {Wise}, {Zhang}, {Nakajima}, \& {Umeda}}]{Ono2023}
{Ono}, Y., {Harikane}, Y., {Ouchi}, M., {et~al.} 2023, \apj, 951, 72, \dodoi{10.3847/1538-4357/acd44a}

\bibitem[{{O'Shea} {et~al.}(2015){O'Shea}, {Wise}, {Xu}, \& {Norman}}]{O'Shea2015}
{O'Shea}, B.~W., {Wise}, J.~H., {Xu}, H., \& {Norman}, M.~L. 2015, \apjl, 807, L12, \dodoi{10.1088/2041-8205/807/1/L12}

\bibitem[{{Padmanabhan} \& {Loeb}(2023)}]{Padmanabhan2023}
{Padmanabhan}, H., \& {Loeb}, A. 2023, \apjl, 953, L4, \dodoi{10.3847/2041-8213/acea7a}

\bibitem[{{Pallottini} \& {Ferrara}(2023)}]{Pallottini2023}
{Pallottini}, A., \& {Ferrara}, A. 2023, \aap, 677, L4, \dodoi{10.1051/0004-6361/202347384}

\bibitem[{{Pallottini} {et~al.}(2022){Pallottini}, {Ferrara}, {Gallerani}, {Behrens}, {Kohandel}, {Carniani}, {Vallini}, {Salvadori}, {Gelli}, {Sommovigo}, {D'Odorico}, {Di Mascia}, \& {Pizzati}}]{Pallottini2022}
{Pallottini}, A., {Ferrara}, A., {Gallerani}, S., {et~al.} 2022, \mnras, 513, 5621, \dodoi{10.1093/mnras/stac1281}

\bibitem[{Pawlik {et~al.}(2011)Pawlik, Milosavljević, \& Bromm}]{Pawlik_2011}
Pawlik, A.~H., Milosavljević, M., \& Bromm, V. 2011, The Astrophysical Journal, 731, 54, \dodoi{10.1088/0004-637x/731/1/54}

\bibitem[{Pawlik \& Schaye(2008)}]{Pawlik_2008}
Pawlik, A.~H., \& Schaye, J. 2008, Monthly Notices of the Royal Astronomical Society, 389, 651–677, \dodoi{10.1111/j.1365-2966.2008.13601.x}

\bibitem[{{Pawlik} \& {Schaye}(2011)}]{Pawlik2011TRAPHHIC}
{Pawlik}, A.~H., \& {Schaye}, J. 2011, \mnras, 412, 1943, \dodoi{10.1111/j.1365-2966.2010.18032.x}

\bibitem[{{Planck Collaboration} {et~al.}(2016){Planck Collaboration}, {Ade}, {Aghanim}, {Arnaud}, {Ashdown}, {Aumont}, {Baccigalupi}, {Banday}, {Barreiro}, {Bartlett}, {Bartolo}, {Battaner}, {Battye}, {Benabed}, {Beno{\^\i}t}, {Benoit-L{\'e}vy}, {Bernard}, {Bersanelli}, {Bielewicz}, {Bock}, {Bonaldi}, {Bonavera}, {Bond}, {Borrill}, {Bouchet}, {Boulanger}, {Bucher}, {Burigana}, {Butler}, {Calabrese}, {Cardoso}, {Catalano}, {Challinor}, {Chamballu}, {Chary}, {Chiang}, {Chluba}, {Christensen}, {Church}, {Clements}, {Colombi}, {Colombo}, {Combet}, {Coulais}, {Crill}, {Curto}, {Cuttaia}, {Danese}, {Davies}, {Davis}, {de Bernardis}, {de Rosa}, {de Zotti}, {Delabrouille}, {D{\'e}sert}, {Di Valentino}, {Dickinson}, {Diego}, {Dolag}, {Dole}, {Donzelli}, {Dor{\'e}}, {Douspis}, {Ducout}, {Dunkley}, {Dupac}, {Efstathiou}, {Elsner}, {En{\ss}lin}, {Eriksen}, {Farhang}, {Fergusson}, {Finelli}, {Forni}, {Frailis}, {Fraisse}, {Franceschi}, {Frejsel}, {Galeotta}, {Galli}, {Ganga}, {Gauthier}, {Gerbino}, {Ghosh}, {Giard},
  {Giraud-H{\'e}raud}, {Giusarma}, {Gjerl{\o}w}, {Gonz{\'a}lez-Nuevo}, {G{\'o}rski}, {Gratton}, {Gregorio}, {Gruppuso}, {Gudmundsson}, {Hamann}, {Hansen}, {Hanson}, {Harrison}, {Helou}, {Henrot-Versill{\'e}}, {Hern{\'a}ndez-Monteagudo}, {Herranz}, {Hildebrandt}, {Hivon}, {Hobson}, {Holmes}, {Hornstrup}, {Hovest}, {Huang}, {Huffenberger}, {Hurier}, {Jaffe}, {Jaffe}, {Jones}, {Juvela}, {Keih{\"a}nen}, {Keskitalo}, {Kisner}, {Kneissl}, {Knoche}, {Knox}, {Kunz}, {Kurki-Suonio}, {Lagache}, {L{\"a}hteenm{\"a}ki}, {Lamarre}, {Lasenby}, {Lattanzi}, {Lawrence}, {Leahy}, {Leonardi}, {Lesgourgues}, {Levrier}, {Lewis}, {Liguori}, {Lilje}, {Linden-V{\o}rnle}, {L{\'o}pez-Caniego}, {Lubin}, {Mac{\'\i}as-P{\'e}rez}, {Maggio}, {Maino}, {Mandolesi}, {Mangilli}, {Marchini}, {Maris}, {Martin}, {Martinelli}, {Mart{\'\i}nez-Gonz{\'a}lez}, {Masi}, {Matarrese}, {McGehee}, {Meinhold}, {Melchiorri}, {Melin}, {Mendes}, {Mennella}, {Migliaccio}, {Millea}, {Mitra}, {Miville-Desch{\^e}nes}, {Moneti}, {Montier}, {Morgante}, {Mortlock},
  {Moss}, {Munshi}, {Murphy}, {Naselsky}, {Nati}, {Natoli}, {Netterfield}, {N{\o}rgaard-Nielsen}, {Noviello}, {Novikov}, {Novikov}, {Oxborrow}, {Paci}, {Pagano}, {Pajot}, {Paladini}, {Paoletti}, {Partridge}, {Pasian}, {Patanchon}, {Pearson}, {Perdereau}, {Perotto}, {Perrotta}, {Pettorino}, {Piacentini}, {Piat}, {Pierpaoli}, {Pietrobon}, {Plaszczynski}, {Pointecouteau}, {Polenta}, {Popa}, {Pratt}, {Pr{\'e}zeau}, {Prunet}, {Puget}, {Rachen}, {Reach}, {Rebolo}, {Reinecke}, {Remazeilles}, {Renault}, {Renzi}, {Ristorcelli}, {Rocha}, {Rosset}, {Rossetti}, {Roudier}, {Rouill{\'e} d'Orfeuil}, {Rowan-Robinson}, {Rubi{\~n}o-Mart{\'\i}n}, {Rusholme}, {Said}, {Salvatelli}, {Salvati}, {Sandri}, {Santos}, {Savelainen}, {Savini}, {Scott}, {Seiffert}, {Serra}, {Shellard}, {Spencer}, {Spinelli}, {Stolyarov}, {Stompor}, {Sudiwala}, {Sunyaev}, {Sutton}, {Suur-Uski}, {Sygnet}, {Tauber}, {Terenzi}, {Toffolatti}, {Tomasi}, {Tristram}, {Trombetti}, {Tucci}, {Tuovinen}, {T{\"u}rler}, {Umana}, {Valenziano}, {Valiviita}, {Van Tent},
  {Vielva}, {Villa}, {Wade}, {Wandelt}, {Wehus}, {White}, {White}, {Wilkinson}, {Yvon}, {Zacchei}, \& {Zonca}}]{Planck2016}
{Planck Collaboration}, {Ade}, P.~A.~R., {Aghanim}, N., {et~al.} 2016, \aap, 594, A13, \dodoi{10.1051/0004-6361/201525830}

\bibitem[{{Portinari} {et~al.}(1998){Portinari}, {Chiosi}, \& {Bressan}}]{Portinari1998}
{Portinari}, L., {Chiosi}, C., \& {Bressan}, A. 1998, \aap, 334, 505, \dodoi{10.48550/arXiv.astro-ph/9711337}

\bibitem[{{Prole} {et~al.}(2023){Prole}, {Schauer}, {Clark}, {Glover}, {Priestley}, \& {Klessen}}]{Prole2023}
{Prole}, L.~R., {Schauer}, A. T.~P., {Clark}, P.~C., {et~al.} 2023, \mnras, 520, 2081, \dodoi{10.1093/mnras/stad188}

\bibitem[{{Rhoads} {et~al.}(2023){Rhoads}, {Wold}, {Harish}, {Kim}, {Pharo}, {Malhotra}, {Gabrielpillai}, {Jiang}, \& {Yang}}]{Rhoads2023}
{Rhoads}, J.~E., {Wold}, I. G.~B., {Harish}, S., {et~al.} 2023, \apjl, 942, L14, \dodoi{10.3847/2041-8213/acaaaf}

\bibitem[{{Ritter} {et~al.}(2012){Ritter}, {Safranek-Shrader}, {Gnat}, {Milosavljevi{\'c}}, \& {Bromm}}]{Ritter2012}
{Ritter}, J.~S., {Safranek-Shrader}, C., {Gnat}, O., {Milosavljevi{\'c}}, M., \& {Bromm}, V. 2012, \apj, 761, 56, \dodoi{10.1088/0004-637X/761/1/56}

\bibitem[{{Robertson} {et~al.}(2024){Robertson}, {Johnson}, {Tacchella}, {Eisenstein}, {Hainline}, {Arribas}, {Baker}, {Bunker}, {Carniani}, {Cargile}, {Carreira}, {Charlot}, {Chevallard}, {Curti}, {Curtis-Lake}, {D'Eugenio}, {Egami}, {Hausen}, {Helton}, {Jakobsen}, {Ji}, {Jones}, {Maiolino}, {Maseda}, {Nelson}, {P{\'e}rez-Gonz{\'a}lez}, {Pusk{\'a}s}, {Rieke}, {Smit}, {Sun}, {{\"U}bler}, {Whitler}, {Williams}, {Willmer}, {Willott}, \& {Witstok}}]{Robertson2024}
{Robertson}, B., {Johnson}, B.~D., {Tacchella}, S., {et~al.} 2024, \apj, 970, 31, \dodoi{10.3847/1538-4357/ad463d}

\bibitem[{{Robertson} {et~al.}(2023){Robertson}, {Tacchella}, {Johnson}, {Hainline}, {Whitler}, {Eisenstein}, {Endsley}, {Rieke}, {Stark}, {Alberts}, {Dressler}, {Egami}, {Hausen}, {Rieke}, {Shivaei}, {Williams}, {Willmer}, {Arribas}, {Bonaventura}, {Bunker}, {Cameron}, {Carniani}, {Charlot}, {Chevallard}, {Curti}, {Curtis-Lake}, {D'Eugenio}, {Jakobsen}, {Looser}, {L{\"u}tzgendorf}, {Maiolino}, {Maseda}, {Rawle}, {Rix}, {Smit}, {{\"U}bler}, {Willott}, {Witstok}, {Baum}, {Bhatawdekar}, {Boyett}, {Chen}, {de Graaff}, {Florian}, {Helton}, {Hviding}, {Ji}, {Kumari}, {Lyu}, {Nelson}, {Sandles}, {Saxena}, {Suess}, {Sun}, {Topping}, \& {Wallace}}]{Robertson2023}
{Robertson}, B.~E., {Tacchella}, S., {Johnson}, B.~D., {et~al.} 2023, Nature Astronomy, 7, 611, \dodoi{10.1038/s41550-023-01921-1}

\bibitem[{{Robitaille}(2011)}]{Robitaille2011}
{Robitaille}, T.~P. 2011, \aap, 536, A79, \dodoi{10.1051/0004-6361/201117150}

\bibitem[{{Roca-F{\`a}brega} {et~al.}(2024){Roca-F{\`a}brega}, {Kim}, {Primack}, {Jung}, {Genina}, {Hausammann}, {Kim}, {Lupi}, {Nagamine}, {Powell}, {Revaz}, {Shimizu}, {Strawn}, {Vel{\'a}zquez}, {Abel}, {Ceverino}, {Dong}, {Quinn}, {Shin}, {Segovia-Otero}, {Agertz}, {Barrow}, {Cadiou}, {Dekel}, {Hummels}, {Oh}, {Teyssier}, \& {AGORA Collaboration}}]{AGORA2024}
{Roca-F{\`a}brega}, S., {Kim}, J.-H., {Primack}, J.~R., {et~al.} 2024, \apj, 968, 125, \dodoi{10.3847/1538-4357/ad43de}

\bibitem[{{Safranek-Shrader} {et~al.}(2014){Safranek-Shrader}, {Milosavljevic}, \& {Bromm}}]{Safranek-Shrader2014}
{Safranek-Shrader}, C., {Milosavljevic}, M., \& {Bromm}, V. 2014, \mnras, 440, L76, \dodoi{10.1093/mnrasl/slu027}

\bibitem[{{Safranek-Shrader} {et~al.}(2016){Safranek-Shrader}, {Montgomery}, {Milosavljevi{\'c}}, \& {Bromm}}]{Safranek-Shrader2016}
{Safranek-Shrader}, C., {Montgomery}, M.~H., {Milosavljevi{\'c}}, M., \& {Bromm}, V. 2016, \mnras, 455, 3288, \dodoi{10.1093/mnras/stv2545}

\bibitem[{{Salim} \& {Narayanan}(2020)}]{Salim2020}
{Salim}, S., \& {Narayanan}, D. 2020, \araa, 58, 529, \dodoi{10.1146/annurev-astro-032620-021933}

\bibitem[{{Salpeter}(1955)}]{Salpeter1955}
{Salpeter}, E.~E. 1955, \apj, 121, 161, \dodoi{10.1086/145971}

\bibitem[{{Schaerer}(2003)}]{Schaerer2003}
{Schaerer}, D. 2003, \aap, 397, 527, \dodoi{10.1051/0004-6361:20021525}

\bibitem[{{Schauer} {et~al.}(2021){Schauer}, {Glover}, {Klessen}, \& {Clark}}]{Schauer2021}
{Schauer}, A. T.~P., {Glover}, S. C.~O., {Klessen}, R.~S., \& {Clark}, P. 2021, \mnras, 507, 1775, \dodoi{10.1093/mnras/stab1953}

\bibitem[{{Schmidt}(1959)}]{Schmidt1959}
{Schmidt}, M. 1959, \apj, 129, 243, \dodoi{10.1086/146614}

\bibitem[{{Schneider} \& {Omukai}(2010)}]{Schneider2010}
{Schneider}, R., \& {Omukai}, K. 2010, \mnras, 402, 429, \dodoi{10.1111/j.1365-2966.2009.15891.x}

\bibitem[{{S{\'e}rsic}(1963)}]{Sersic1963}
{S{\'e}rsic}, J.~L. 1963, Boletin de la Asociacion Argentina de Astronomia La Plata Argentina, 6, 41

\bibitem[{{Shen} {et~al.}(2023){Shen}, {Vogelsberger}, {Boylan-Kolchin}, {Tacchella}, \& {Kannan}}]{Shen2023}
{Shen}, X., {Vogelsberger}, M., {Boylan-Kolchin}, M., {Tacchella}, S., \& {Kannan}, R. 2023, \mnras, 525, 3254, \dodoi{10.1093/mnras/stad2508}

\bibitem[{Springel(2005)}]{Springel_2005}
Springel, V. 2005, Monthly Notices of the Royal Astronomical Society, 364, 1105–1134, \dodoi{10.1111/j.1365-2966.2005.09655.x}

\bibitem[{{Springel} {et~al.}(2001){Springel}, {Yoshida}, \& {White}}]{2001NewA....6...79S}
{Springel}, V., {Yoshida}, N., \& {White}, S. D.~M. 2001, \na, 6, 79, \dodoi{10.1016/S1384-1076(01)00042-2}

\bibitem[{Stacy {et~al.}(2016)Stacy, Bromm, \& Lee}]{Stacy2016}
Stacy, A., Bromm, V., \& Lee, A.~T. 2016, Monthly Notices of the Royal Astronomical Society, 462, 1307, \dodoi{10.1093/mnras/stw1728}

\bibitem[{{Steinhardt} {et~al.}(2023){Steinhardt}, {Kokorev}, {Rusakov}, {Garcia}, \& {Sneppen}}]{Steinhardt2023}
{Steinhardt}, C.~L., {Kokorev}, V., {Rusakov}, V., {Garcia}, E., \& {Sneppen}, A. 2023, \apjl, 951, L40, \dodoi{10.3847/2041-8213/acdef6}

\bibitem[{{Sugimura} {et~al.}(2020){Sugimura}, {Matsumoto}, {Hosokawa}, {Hirano}, \& {Omukai}}]{Sugimura2020}
{Sugimura}, K., {Matsumoto}, T., {Hosokawa}, T., {Hirano}, S., \& {Omukai}, K. 2020, \apjl, 892, L14, \dodoi{10.3847/2041-8213/ab7d37}

\bibitem[{{Sun} {et~al.}(2023{\natexlab{a}}){Sun}, {Faucher-Gigu{\`e}re}, {Hayward}, \& {Shen}}]{Sun2023a}
{Sun}, G., {Faucher-Gigu{\`e}re}, C.-A., {Hayward}, C.~C., \& {Shen}, X. 2023{\natexlab{a}}, \mnras, 526, 2665, \dodoi{10.1093/mnras/stad2902}

\bibitem[{{Sun} {et~al.}(2023{\natexlab{b}}){Sun}, {Faucher-Gigu{\`e}re}, {Hayward}, {Shen}, {Wetzel}, \& {Cochrane}}]{Sun2023b}
{Sun}, G., {Faucher-Gigu{\`e}re}, C.-A., {Hayward}, C.~C., {et~al.} 2023{\natexlab{b}}, \apjl, 955, L35, \dodoi{10.3847/2041-8213/acf85a}

\bibitem[{{Sun} \& {Furlanetto}(2016)}]{Sun&Furlanetto2016}
{Sun}, G., \& {Furlanetto}, S.~R. 2016, \mnras, 460, 417, \dodoi{10.1093/mnras/stw980}

\bibitem[{{Susa} {et~al.}(2014){Susa}, {Hasegawa}, \& {Tominaga}}]{Susa2014}
{Susa}, H., {Hasegawa}, K., \& {Tominaga}, N. 2014, \apj, 792, 32, \dodoi{10.1088/0004-637X/792/1/32}

\bibitem[{{Tacchella} {et~al.}(2018){Tacchella}, {Bose}, {Conroy}, {Eisenstein}, \& {Johnson}}]{Tacchella2018}
{Tacchella}, S., {Bose}, S., {Conroy}, C., {Eisenstein}, D.~J., \& {Johnson}, B.~D. 2018, \apj, 868, 92, \dodoi{10.3847/1538-4357/aae8e0}

\bibitem[{{Tacchella} {et~al.}(2022){Tacchella}, {Finkelstein}, {Bagley}, {Dickinson}, {Ferguson}, {Giavalisco}, {Graziani}, {Grogin}, {Hathi}, {Hutchison}, {Jung}, {Koekemoer}, {Larson}, {Papovich}, {Pirzkal}, {Rojas-Ruiz}, {Song}, {Schneider}, {Somerville}, {Wilkins}, \& {Yung}}]{Tacchella2022}
{Tacchella}, S., {Finkelstein}, S.~L., {Bagley}, M., {et~al.} 2022, \apj, 927, 170, \dodoi{10.3847/1538-4357/ac4cad}

\bibitem[{{Tacchella} {et~al.}(2023{\natexlab{a}}){Tacchella}, {Eisenstein}, {Hainline}, {Johnson}, {Baker}, {Helton}, {Robertson}, {Suess}, {Chen}, {Nelson}, {Pusk{\'a}s}, {Sun}, {Alberts}, {Egami}, {Hausen}, {Rieke}, {Rieke}, {Shivaei}, {Williams}, {Willmer}, {Bunker}, {Cameron}, {Carniani}, {Charlot}, {Curti}, {Curtis-Lake}, {Looser}, {Maiolino}, {Maseda}, {Rawle}, {Rix}, {Smit}, {{\"U}bler}, {Willott}, {Witstok}, {Baum}, {Bhatawdekar}, {Boyett}, {Danhaive}, {de Graaff}, {Endsley}, {Ji}, {Lyu}, {Sandles}, {Saxena}, {Scholtz}, {Topping}, \& {Whitler}}]{Tacchella2023}
{Tacchella}, S., {Eisenstein}, D.~J., {Hainline}, K., {et~al.} 2023{\natexlab{a}}, \apj, 952, 74, \dodoi{10.3847/1538-4357/acdbc6}

\bibitem[{{Tacchella} {et~al.}(2023{\natexlab{b}}){Tacchella}, {Johnson}, {Robertson}, {Carniani}, {D'Eugenio}, {Kumari}, {Maiolino}, {Nelson}, {Suess}, {{\"U}bler}, {Williams}, {Adebusola}, {Alberts}, {Arribas}, {Bhatawdekar}, {Bonaventura}, {Bowler}, {Bunker}, {Cameron}, {Curti}, {Egami}, {Eisenstein}, {Frye}, {Hainline}, {Helton}, {Ji}, {Looser}, {Lyu}, {Perna}, {Rawle}, {Rieke}, {Rieke}, {Saxena}, {Sandles}, {Shivaei}, {Simmonds}, {Sun}, {Willmer}, {Willott}, \& {Witstok}}]{Tacchella2023a}
{Tacchella}, S., {Johnson}, B.~D., {Robertson}, B.~E., {et~al.} 2023{\natexlab{b}}, \mnras, 522, 6236, \dodoi{10.1093/mnras/stad1408}

\bibitem[{{Tonry}(2006)}]{Tonry2006}
{Tonry}, J. 2006, {Revisiting the Supernova Ia Rate at z>1}, HST Proposal ID 10980. Cycle 15

\bibitem[{{Trinca} {et~al.}(2023){Trinca}, {Schneider}, {Valiante}, {Graziani}, {Ferrotti}, {Omukai}, \& {Chon}}]{Trinca2023}
{Trinca}, A., {Schneider}, R., {Valiante}, R., {et~al.} 2023, arXiv e-prints, arXiv:2305.04944, \dodoi{10.48550/arXiv.2305.04944}

\bibitem[{{Trussler} {et~al.}(2023){Trussler}, {Conselice}, {Adams}, {Maiolino}, {Nakajima}, {Zackrisson}, {Austin}, {Ferreira}, \& {Harvey}}]{Trussler2023}
{Trussler}, J. A.~A., {Conselice}, C.~J., {Adams}, N.~J., {et~al.} 2023, \mnras, 525, 5328, \dodoi{10.1093/mnras/stad2553}

\bibitem[{{Tsuna} {et~al.}(2023){Tsuna}, {Nakazato}, \& {Hartwig}}]{Tsuna2023}
{Tsuna}, D., {Nakazato}, Y., \& {Hartwig}, T. 2023, \mnras, 526, 4801, \dodoi{10.1093/mnras/stad3043}

\bibitem[{{Venditti} {et~al.}(2024{\natexlab{a}}){Venditti}, {Bromm}, {Finkelstein}, {Calabr{\`o}}, {Napolitano}, {Graziani}, \& {Schneider}}]{Venditti2024}
{Venditti}, A., {Bromm}, V., {Finkelstein}, S.~L., {et~al.} 2024{\natexlab{a}}, arXiv e-prints, arXiv:2405.10940, \dodoi{10.48550/arXiv.2405.10940}

\bibitem[{{Venditti} {et~al.}(2024{\natexlab{b}}){Venditti}, {Bromm}, {Finkelstein}, {Graziani}, \& {Schneider}}]{Venditti2024_PISN}
{Venditti}, A., {Bromm}, V., {Finkelstein}, S.~L., {Graziani}, L., \& {Schneider}, R. 2024{\natexlab{b}}, \mnras, 527, 5102, \dodoi{10.1093/mnras/stad3513}

\bibitem[{{Ventura} {et~al.}(2024){Ventura}, {Qin}, {Balu}, \& {Wyithe}}]{Ventura2024MNRAS}
{Ventura}, E.~M., {Qin}, Y., {Balu}, S., \& {Wyithe}, J. S.~B. 2024, \mnras, 529, 628, \dodoi{10.1093/mnras/stae567}

\bibitem[{{Wang} {et~al.}(2023){Wang}, {Fujimoto}, {Labb{\'e}}, {Furtak}, {Miller}, {Setton}, {Zitrin}, {Atek}, {Bezanson}, {Brammer}, {Leja}, {Oesch}, {Price}, {Chemerynska}, {Cutler}, {Dayal}, {van Dokkum}, {Goulding}, {Greene}, {Fudamoto}, {Khullar}, {Kokorev}, {Marchesini}, {Pan}, {Weaver}, {Whitaker}, \& {Williams}}]{Wang2023}
{Wang}, B., {Fujimoto}, S., {Labb{\'e}}, I., {et~al.} 2023, \apjl, 957, L34, \dodoi{10.3847/2041-8213/acfe07}

\bibitem[{{Wang} {et~al.}(2012){Wang}, {Bromm}, {Greif}, {Stacy}, {Dai}, {Loeb}, \& {Cheng}}]{Wang2012}
{Wang}, F.~Y., {Bromm}, V., {Greif}, T.~H., {et~al.} 2012, \apj, 760, 27, \dodoi{10.1088/0004-637X/760/1/27}

\bibitem[{{Weingartner} \& {Draine}(2001)}]{LMCSMC}
{Weingartner}, J.~C., \& {Draine}, B.~T. 2001, \apj, 548, 296, \dodoi{10.1086/318651}

\bibitem[{Welch(1967)}]{Welch1967}
Welch, P. 1967, IEEE Transactions on Audio and Electroacoustics, 15, 70, \dodoi{10.1109/TAU.1967.1161901}

\bibitem[{{Wiersma} {et~al.}(2009){Wiersma}, {Schaye}, {Theuns}, {Dalla Vecchia}, \& {Tornatore}}]{Wiersma2009}
{Wiersma}, R. P.~C., {Schaye}, J., {Theuns}, T., {Dalla Vecchia}, C., \& {Tornatore}, L. 2009, \mnras, 399, 574, \dodoi{10.1111/j.1365-2966.2009.15331.x}

\bibitem[{{Williams} {et~al.}(2018){Williams}, {Curtis-Lake}, {Hainline}, {Chevallard}, {Robertson}, {Charlot}, {Endsley}, {Stark}, {Willmer}, {Alberts}, {Amorin}, {Arribas}, {Baum}, {Bunker}, {Carniani}, {Crandall}, {Egami}, {Eisenstein}, {Ferruit}, {Husemann}, {Maseda}, {Maiolino}, {Rawle}, {Rieke}, {Smit}, {Tacchella}, \& {Willott}}]{Williams2018}
{Williams}, C.~C., {Curtis-Lake}, E., {Hainline}, K.~N., {et~al.} 2018, \apjs, 236, 33, \dodoi{10.3847/1538-4365/aabcbb}

\bibitem[{{Wise} {et~al.}(2012){Wise}, {Turk}, {Norman}, \& {Abel}}]{Wise2012}
{Wise}, J.~H., {Turk}, M.~J., {Norman}, M.~L., \& {Abel}, T. 2012, \apj, 745, 50, \dodoi{10.1088/0004-637X/745/1/50}

\bibitem[{{Yajima} {et~al.}(2023){Yajima}, {Abe}, {Fukushima}, {Ono}, {Harikane}, {Ouchi}, {Hashimoto}, \& {Khochfar}}]{Yajima2023}
{Yajima}, H., {Abe}, M., {Fukushima}, H., {et~al.} 2023, \mnras, 525, 4832, \dodoi{10.1093/mnras/stad2497}

\bibitem[{{Yajima} {et~al.}(2022){Yajima}, {Abe}, {Khochfar}, {Nagamine}, {Inoue}, {Kodama}, {Arata}, {Dalla Vecchia}, {Fukushima}, {Hashimoto}, {Kashikawa}, {Kubo}, {Li}, {Matsuda}, {Mawatari}, {Ouchi}, \& {Umehata}}]{Yajima2022}
{Yajima}, H., {Abe}, M., {Khochfar}, S., {et~al.} 2022, \mnras, 509, 4037, \dodoi{10.1093/mnras/stab3092}

\bibitem[{Yung {et~al.}(2023)Yung, Somerville, Finkelstein, Wilkins, \& Gardner}]{Yung_2023}
Yung, L. Y.~A., Somerville, R.~S., Finkelstein, S.~L., Wilkins, S.~M., \& Gardner, J.~P. 2023, Monthly Notices of the Royal Astronomical Society, 527, 5929–5948, \dodoi{10.1093/mnras/stad3484}

\bibitem[{{Yung} {et~al.}(2019){Yung}, {Somerville}, {Popping}, {Finkelstein}, {Ferguson}, \& {Dav{\'e}}}]{Yung2019}
{Yung}, L.~Y.~A., {Somerville}, R.~S., {Popping}, G., {et~al.} 2019, \mnras, 490, 2855, \dodoi{10.1093/mnras/stz2755}

\bibitem[{{Zackrisson} {et~al.}(2011){Zackrisson}, {Rydberg}, {Schaerer}, {{\"O}stlin}, \& {Tuli}}]{Zackrisson2011}
{Zackrisson}, E., {Rydberg}, C.-E., {Schaerer}, D., {{\"O}stlin}, G., \& {Tuli}, M. 2011, \apj, 740, 13, \dodoi{10.1088/0004-637X/740/1/13}

\bibitem[{{Zavala} {et~al.}(2024){Zavala}, {Castellano}, {Akins}, {Bakx}, {Burgarella}, {Casey}, {Ch{\'a}vez Ortiz}, {Dickinson}, {Finkelstein}, {Mitsuhashi}, {Nakajima}, {P{\'e}rez-Gonz{\'a}lez}, {Arrabal Haro}, {Buat}, {Backhaus}, {Calabr{\`o}}, {Cleri}, {Fern{\'a}ndez-Arenas}, {Fontana}, {Franco}, {Giavalisco}, {Grogin}, {Hathi}, {Hirschmann}, {Ikeda}, {Jung}, {Kartaltepe}, {Koekemoer}, {Larson}, {McKinney}, {Papovich}, {Saito}, {Santini}, {Terlevich}, {Terlevich}, {Treu}, \& {Yung}}]{Zavala2024}
{Zavala}, J.~A., {Castellano}, M., {Akins}, H.~B., {et~al.} 2024, arXiv e-prints, arXiv:2403.10491, \dodoi{10.48550/arXiv.2403.10491}

\bibitem[{{Zhang} {et~al.}(2024){Zhang}, {Bromm}, \& {Liu}}]{Zhang2024}
{Zhang}, S., {Bromm}, V., \& {Liu}, B. 2024, arXiv e-prints, arXiv:2405.11381, \dodoi{10.48550/arXiv.2405.11381}

\bibitem[{{Ziparo} {et~al.}(2023){Ziparo}, {Ferrara}, {Sommovigo}, \& {Kohandel}}]{Ziparo2023}
{Ziparo}, F., {Ferrara}, A., {Sommovigo}, L., \& {Kohandel}, M. 2023, \mnras, 520, 2445, \dodoi{10.1093/mnras/stad125}

\end{thebibliography}

\bibliographystyle{aasjournal}

\end{document}